\documentclass[a4paper,11pt]{article}

\usepackage{subcaption, caption}
\usepackage{amssymb,amsmath,amsfonts,makeidx,placeins,pbox,multirow}
\usepackage{graphicx,rotate,subcaption,color,slashed,cite,caption,epstopdf,verbatim}
\usepackage{longtable,tabu}
\usepackage{array}
\usepackage{caption}
\usepackage{subcaption}
\usepackage{url}
\usepackage{xcolor}
\usepackage[norule]{footmisc}
\usepackage{academicons}
\usepackage[colorlinks=true,
            linkcolor=magenta,
            urlcolor=blue,
            citecolor=blue]{hyperref}
\usepackage{orcidlink}

\evensidemargin=\oddsidemargin
\begin{document}

\newcommand{\eg}      {\textit{e.g.}}
\newcommand{\gte}     {\geq} 

\def\hmath#1{\text{\scalebox{1.6}{$#1$}}}
\def\lmath#1{\text{\scalebox{1.4}{$#1$}}}
\def\mmath#1{\text{\scalebox{1.2}{$#1$}}}
\def\smath#1{\text{\scalebox{.8}{$#1$}}}
\def\ssmath#1{\text{\scalebox{.6}{$#1$}}}

\def\hfrac#1#2{\hmath{\frac{#1}{#2}}}
\def\lfrac#1#2{\lmath{\frac{#1}{#2}}}
\def\mfrac#1#2{\mmath{\frac{#1}{#2}}}
\def\sfrac#1#2{\smath{\frac{#1}{#2}}}
\def\ssfrac#1#2{\smath{\frac{#1}{#2}}}

\def\pow{^\mmath}

\newcommand{\ahalf}{\ensuremath{\tfrac{1}{2}}}
\newcommand{\shalf}{\ensuremath{\sfrac{1}{2}}}
\newcommand{\jhalf}{\ensuremath{\frac{1}{2}}}

\newcommand{\dzero}{\ensuremath{d_{0}}}
\newcommand{\dosig}{\ensuremath{d_{0}/\sigma_{d_{0}}}}
\newcommand{\dxy}{\ensuremath{d_{xy}}}
\newcommand{\dxysig}{\ensuremath{d_{xy}/\sigma_{d_{xy}}}}

\def\MET{\ensuremath{E_{\mathrm{T}}^{\mathrm{miss}}}} 
\def\met{\ensuremath{E_{\mathrm{T}}^{\mathrm{miss}}}} 
\def\ptmiss{\ensuremath{p_{\mathrm{T}}^{\mathrm{miss}}}} 
\newcommand{\Etmiss}{\met}
\newcommand{\metrel}{\ensuremath{E_{\mathrm{T,Rel}}^{\mathrm{miss}}}}
\newcommand{\DeltaR}{\ensuremath{\Delta R}}
\newcommand{\dr}{\DeltaR}

\newcommand{\mll}{\ensuremath{m_{ll}}}


\newcommand{\alpgen}{{\sc alpgen}}
\newcommand{\jimmy}{{\sc jimmy}}
\newcommand{\mcatnlo}{{\sc mc@nlo}}
\newcommand{\sherpa}{{\sc sherpa}}
\newcommand{\powheg}{{\sc powheg}}
\newcommand{\herwig}{{\sc herwig}}
\newcommand{\pythia}{{\sc Pythia}}
\newcommand{\blackmax}{{\sc Blackmax}}
\newcommand{\acer}{{\sc AcerMC}}
\newcommand{\madg}{{\sc MadGraph}}
\newcommand{\delphes}{{\sc Delphes}}
\newcommand{\fastjet}{{\sc FastJet}}
\newcommand{\tensorflow}{{\textsc{TensorFlow v2.16.2}}}
\newcommand{\keras}{{\textsc{Keras v3.10.0}}}

\newcommand{\ie}{{\it i.e.}}

\renewcommand{\thefootnote}{\fnsymbol{footnote}}

%
%
\def\ra{\ensuremath{\rightarrow}}
\def\la{\ensuremath{\leftarrow}}
\let\rarrow=\ra
\let\larrow=\la
\def\lapprox{\ensuremath{\sim\kern-1em\raise 0.65ex\hbox{$<$}}}
\def\rapprox{\ensuremath{\sim\kern-1em\raise 0.65ex\hbox{$>$}}}
\def\gam{\ensuremath{\gamma}}
\def\rts {\ensuremath{\sqrt{s}}}
\def\stat{\mbox{$\;$(stat.)}}
\def\syst{\mbox{$\;$(syst.)}}
%
%
\def\Mtau{\ensuremath{m_{\tau}}}
\def\swsq{\ensuremath{\sin^2\!\theta_{W}}}
\def\swel{\ensuremath{\sin^2\!\theta_{\mathrm{eff}}^{\mathrm{lept}}}}
\def\swsqb{\ensuremath{\sin^2\!\overline{\theta}_{W}}}
\def\swsqon{\ensuremath{\swsq\equiv 1-\mW^2/\mZ^2}} 
\def\gv{\ensuremath{g_{\mathrm{V}}}} 
\def\ga{\ensuremath{g_{\mathrm{A}}}} 
\def\gvbar{\ensuremath{\bar{g}_\mathrm{V}}} 
\def\gabar{\ensuremath{\bar{g}_\mathrm{A}}} 
%
%
\def\antibar#1{\ensuremath{#1\bar{#1}}}
\def\tbar{\ensuremath{\bar{t}}}
\def\ttbar{\antibar{t}}
\def\bbar{\ensuremath{\bar{b}}}
\def\bbbar{\antibar{b}}
\def\cbar{\ensuremath{\bar{c}}}
\def\ccbar{\antibar{c}}
\def\sbar{\ensuremath{\bar{s}}}
\def\ssbar{\antibar{s}}
\def\ubar{\ensuremath{\bar{u}}}
\def\uubar{\antibar{u}}
\def\dbar{\ensuremath{\bar{d}}}
\def\ddbar{\antibar{d}}
\def\fbar{\ensuremath{\bar{f}}}
\def\ffbar{\antibar{f}}
\def\qbar{\ensuremath{\bar{q}}}
\def\qqbar{\antibar{q}}
\def\nbar{\ensuremath{\bar{\nu}}}
\def\nnbar{\antibar{\nu}}
%
%
\def\ee{\ensuremath{e^+ e^-}}%
\def\epm{\ensuremath{e^{\pm}}}%
\def\epem{\ensuremath{e^+ e^-}}%
\def\mumu{\ensuremath{\mathrm{\mu^+ \mu^-}}}%
\def\tautau{\ensuremath{\mathrm{\tau^+ \tau^-}}}%
\let\muchless=\ll
\def\ll{\ensuremath{\ell^+ \ell^-}}%
\def\lnu{\ensuremath{\ell \nu}}%
%

\newcommand{\zgs}{\ensuremath{Z/\gamma^{*}}}

\def\ZZ{\ensuremath{Z Z}}
\def\WZ{\ensuremath{W Z}}
\newcommand{\wjets}{\ensuremath{\mathrm{W+jets}}}
\newcommand{\gamjets}{\ensuremath{\mathrm{\gamma+jets}}}

\def\Zzero{\ensuremath{Z}}
\def\Zboson{\ensuremath{Z}}
\def\Wplus{\ensuremath{W^+}}
\def\Wminus{\ensuremath{W^-}}
\def\Wboson{\ensuremath{W}}%
\def\Wpm{\ensuremath{W^{\pm}}}%
\def\Wmp{\ensuremath{W^{\mp}}}%
\def\Zzv{\ensuremath{\Zzero^{\textstyle *}}}
\def\Ginv{\ensuremath{\Gamma_{\mathrm{inv}}}}
%
\def\Azero{\ensuremath{A^0}}%
\def\hzero{\ensuremath{h^0}}%
\def\Hzero{\ensuremath{H^0}}%
\def\Hboson{\ensuremath{H}}%
\def\Hplus{\ensuremath{H^+}}%
\def\Hminus{\ensuremath{H^-}}%
\def\Hpm{\ensuremath{H^{\pm}}}%
\def\Hmp{\ensuremath{H^{\mp}}}%
\def\susy#1{\ensuremath{\tilde{#1}}}%
\def\ellell{\ensuremath{\mathrm{\ell^+ \ell^-}}}%
\def\ggino{\ensuremath{\mathchoice%
      {\displaystyle\raise.4ex\hbox{$\displaystyle\tilde\chi$}}%
         {\textstyle\raise.4ex\hbox{$\textstyle\tilde\chi$}}%
       {\scriptstyle\raise.3ex\hbox{$\scriptstyle\tilde\chi$}}%
 {\scriptscriptstyle\raise.3ex\hbox{$\scriptscriptstyle\tilde\chi$}}}}

\def\chinop{\ensuremath{\mathchoice%
      {\displaystyle\raise.4ex\hbox{$\displaystyle\tilde\chi^+$}}%
         {\textstyle\raise.4ex\hbox{$\textstyle\tilde\chi^+$}}%
       {\scriptstyle\raise.3ex\hbox{$\scriptstyle\tilde\chi^+$}}%
 {\scriptscriptstyle\raise.3ex\hbox{$\scriptscriptstyle\tilde\chi^+$}}}}
\def\chinom{\ensuremath{\mathchoice%
      {\displaystyle\raise.4ex\hbox{$\displaystyle\tilde\chi^-$}}%
         {\textstyle\raise.4ex\hbox{$\textstyle\tilde\chi^-$}}%
       {\scriptstyle\raise.3ex\hbox{$\scriptstyle\tilde\chi^-$}}%
 {\scriptscriptstyle\raise.3ex\hbox{$\scriptscriptstyle\tilde\chi^-$}}}}
\def\chinopm{\ensuremath{\mathchoice%
      {\displaystyle\raise.4ex\hbox{$\displaystyle\tilde\chi^\pm$}}%
         {\textstyle\raise.4ex\hbox{$\textstyle\tilde\chi^\pm$}}%
       {\scriptstyle\raise.3ex\hbox{$\scriptstyle\tilde\chi^\pm$}}%
 {\scriptscriptstyle\raise.3ex\hbox{$\scriptscriptstyle\tilde\chi^\pm$}}}}
\def\chinomp{\ensuremath{\mathchoice%
      {\displaystyle\raise.4ex\hbox{$\displaystyle\tilde\chi^\mp$}}%
         {\textstyle\raise.4ex\hbox{$\textstyle\tilde\chi^\mp$}}%
       {\scriptstyle\raise.3ex\hbox{$\scriptstyle\tilde\chi^\mp$}}%
 {\scriptscriptstyle\raise.3ex\hbox{$\scriptscriptstyle\tilde\chi^\mp$}}}}

\def\chinoonep{\ensuremath{\mathchoice%
      {\displaystyle\raise.4ex\hbox{$\displaystyle\tilde\chi^+_1$}}%
         {\textstyle\raise.4ex\hbox{$\textstyle\tilde\chi^+_1$}}%
       {\scriptstyle\raise.3ex\hbox{$\scriptstyle\tilde\chi^+_1$}}%
 {\scriptscriptstyle\raise.3ex\hbox{$\scriptscriptstyle\tilde\chi^+_1$}}}}
\def\chinoonem{\ensuremath{\mathchoice%
      {\displaystyle\raise.4ex\hbox{$\displaystyle\tilde\chi^-_1$}}%
         {\textstyle\raise.4ex\hbox{$\textstyle\tilde\chi^-_1$}}%
       {\scriptstyle\raise.3ex\hbox{$\scriptstyle\tilde\chi^-_1$}}%
 {\scriptscriptstyle\raise.3ex\hbox{$\scriptscriptstyle\tilde\chi^-_1$}}}}
\def\chinoonepm{\ensuremath{\mathchoice%
      {\displaystyle\raise.4ex\hbox{$\displaystyle\tilde\chi^\pm_1$}}%
         {\textstyle\raise.4ex\hbox{$\textstyle\tilde\chi^\pm_1$}}%
       {\scriptstyle\raise.3ex\hbox{$\scriptstyle\tilde\chi^\pm_1$}}%
 {\scriptscriptstyle\raise.3ex\hbox{$\scriptscriptstyle\tilde\chi^\pm_1$}}}}

\def\chinotwop{\ensuremath{\mathchoice%
      {\displaystyle\raise.4ex\hbox{$\displaystyle\tilde\chi^+_2$}}%
         {\textstyle\raise.4ex\hbox{$\textstyle\tilde\chi^+_2$}}%
       {\scriptstyle\raise.3ex\hbox{$\scriptstyle\tilde\chi^+_2$}}%
 {\scriptscriptstyle\raise.3ex\hbox{$\scriptscriptstyle\tilde\chi^+_2$}}}}
\def\chinotwom{\ensuremath{\mathchoice%
      {\displaystyle\raise.4ex\hbox{$\displaystyle\tilde\chi^-_2$}}%
         {\textstyle\raise.4ex\hbox{$\textstyle\tilde\chi^-_2$}}%
       {\scriptstyle\raise.3ex\hbox{$\scriptstyle\tilde\chi^-_2$}}%
 {\scriptscriptstyle\raise.3ex\hbox{$\scriptscriptstyle\tilde\chi^-_2$}}}}
\def\chinotwopm{\ensuremath{\mathchoice%
      {\displaystyle\raise.4ex\hbox{$\displaystyle\tilde\chi^\pm_2$}}%
         {\textstyle\raise.4ex\hbox{$\textstyle\tilde\chi^\pm_2$}}%
       {\scriptstyle\raise.3ex\hbox{$\scriptstyle\tilde\chi^\pm_2$}}%
 {\scriptscriptstyle\raise.3ex\hbox{$\scriptscriptstyle\tilde\chi^\pm_2$}}}}

\def\nino{\ensuremath{\mathchoice%
      {\displaystyle\raise.4ex\hbox{$\displaystyle\tilde\chi^0$}}%
         {\textstyle\raise.4ex\hbox{$\textstyle\tilde\chi^0$}}%
       {\scriptstyle\raise.3ex\hbox{$\scriptstyle\tilde\chi^0$}}%
 {\scriptscriptstyle\raise.3ex\hbox{$\scriptscriptstyle\tilde\chi^0$}}}}

\def\ninoone{\ensuremath{\mathchoice%
      {\displaystyle\raise.4ex\hbox{$\displaystyle\tilde\chi^0_1$}}%
         {\textstyle\raise.4ex\hbox{$\textstyle\tilde\chi^0_1$}}%
       {\scriptstyle\raise.3ex\hbox{$\scriptstyle\tilde\chi^0_1$}}%
 {\scriptscriptstyle\raise.3ex\hbox{$\scriptscriptstyle\tilde\chi^0_1$}}}}
\def\ninotwo{\ensuremath{\mathchoice%
      {\displaystyle\raise.4ex\hbox{$\displaystyle\tilde\chi^0_2$}}%
         {\textstyle\raise.4ex\hbox{$\textstyle\tilde\chi^0_2$}}%
       {\scriptstyle\raise.3ex\hbox{$\scriptstyle\tilde\chi^0_2$}}%
 {\scriptscriptstyle\raise.3ex\hbox{$\scriptscriptstyle\tilde\chi^0_2$}}}}
\def\ninothree{\ensuremath{\mathchoice%
      {\displaystyle\raise.4ex\hbox{$\displaystyle\tilde\chi^0_3$}}%
         {\textstyle\raise.4ex\hbox{$\textstyle\tilde\chi^0_3$}}%
       {\scriptstyle\raise.3ex\hbox{$\scriptstyle\tilde\chi^0_3$}}%
 {\scriptscriptstyle\raise.3ex\hbox{$\scriptscriptstyle\tilde\chi^0_3$}}}}
\def\ninofour{\ensuremath{\mathchoice%
      {\displaystyle\raise.4ex\hbox{$\displaystyle\tilde\chi^0_4$}}%
         {\textstyle\raise.4ex\hbox{$\textstyle\tilde\chi^0_4$}}%
       {\scriptstyle\raise.3ex\hbox{$\scriptstyle\tilde\chi^0_4$}}%
 {\scriptscriptstyle\raise.3ex\hbox{$\scriptscriptstyle\tilde\chi^0_4$}}}}

\def\gravino{\ensuremath{\tilde{G}}}%
\def\Zprime{\ensuremath{Z^\prime}}
\def\Zstar{\ensuremath{Z^{*}}}
\def\squark{\ensuremath{\tilde{q}}}
\def\squarkL{\ensuremath{\tilde{q}_{\mathrm{L}}}} 
\def\squarkR{\ensuremath{\tilde{q}_{\mathrm{R}}}} 
\def\gluino{\ensuremath{\tilde{g}}}
\def\stop{\ensuremath{\tilde{t}}}
\def\stopone{\ensuremath{\tilde{t}_1}}
\def\stoptwo{\ensuremath{\tilde{t}_2}}
\def\stopL{\ensuremath{\tilde{t}_{\mathrm{L}}}} 
\def\stopR{\ensuremath{\tilde{t}_{\mathrm{R}}}} 
\def\sbottom{\ensuremath{\tilde{b}}}
\def\sbottomone{\ensuremath{\tilde{b}_1}}
\def\sbottomtwo{\ensuremath{\tilde{b}_2}}
\def\sbottomL{\ensuremath{\tilde{b}_{\mathrm{L}}}} 
\def\sbottomR{\ensuremath{\tilde{b}_{\mathrm{R}}}} 
\def\slepton{\ensuremath{\tilde{\ell}}}
\def\sleptonL{\ensuremath{\tilde{\ell}_{\mathrm{L}}}} 
\def\sleptonR{\ensuremath{\tilde{\ell}_{\mathrm{R}}}} 
\def\sel{\ensuremath{\tilde{e}}}
\def\selL{\ensuremath{\tilde{e}_{\mathrm{L}}}} 
\def\selR{\ensuremath{\tilde{e}_{\mathrm{R}}}} 
\def\smu{\ensuremath{\tilde{\mu}}}
\def\smuL{\ensuremath{\tilde{\mu}_{\mathrm{L}}}} 
\def\smuR{\ensuremath{\tilde{\mu}_{\mathrm{R}}}} 
\def\stau{\ensuremath{\tilde{\tau}}}
\def\stauL{\ensuremath{\tilde{\tau}_{\mathrm{L}}}} 
\def\stauR{\ensuremath{\tilde{\tau}_{\mathrm{R}}}} 
\def\stauone{\ensuremath{\tilde{\tau}_1}}
\def\stautwo{\ensuremath{\tilde{\tau}_2}}
\def\snu{\ensuremath{\tilde{\nu}}}
%
%
\let\pii=\pi
\def\pi{\ensuremath{\pii}}%
\def\pizero{\ensuremath{\pii^0}}%
\def\piplus{\ensuremath{\pii^+}}%
\def\piminus{\ensuremath{\pii^-}}%
\def\pipm{\ensuremath{\pii^{\pm}}}%
\def\pimp{\ensuremath{\pii^{\mp}}}%
\let\etaa=\eta
\def\eta{\ensuremath{\etaa}}%
\def\etaprime{\ensuremath{\eta^{\sst\prime}}}%
%
%
\def\kzero{\ensuremath{K^0}}%
\def\kzerobar{\ensuremath{\overline{K}\vphantom{K}^0}}%
\def\kaon{\ensuremath{K}}%
\def\kplus{\ensuremath{K^+}}%
\def\kminus{\ensuremath{K^-}}%
\def\kzeroL{\ensuremath{K^0_{\mathrm{L}}}} 
\def\kzerol{\ensuremath{K^0_{\mathrm{L}}}} 
\def\klong{\ensuremath{K^0_{\mathrm{L}}}} 
\def\kzeroS{\ensuremath{K^0_{\mathrm{S}}}} 
\def\kzeros{\ensuremath{K^0_{\mathrm{S}}}} 
\def\kshort{\ensuremath{K^0_{\mathrm{S}}}} 
%
%
\def\Ups{\ensuremath{\mit\Upsilon}} 
\def\Upsp{\ensuremath{\mit\Upsilon^{\sst\prime}}} 
\def\Upspp{\ensuremath{\mit\Upsilon^{\sst\prime\prime}}} 
\def\Upsppp{\ensuremath{\mit\Upsilon^{\sst\prime\prime\prime}}} 
\def\Upspppp{\ensuremath{\mit\Upsilon^{\sst\prime\prime\prime\prime}}} 
\def\itUpsp{\ensuremath{\mit\Upsilon^{\sst\prime}}}%
\def\UoneS{\ensuremath{\Upsilon(\mathrm{1S})}}%

%
%
\def\ups#1{\ensuremath{\mit{\Upsilon}(\mathrm{#1S})}} 
%
%
\def\nsPj#1#2#3{\ensuremath{#1\,^{#2}\!P_{#3}}}%
\let\nspj=\nsPj
\def\nsSj#1#2#3{\ensuremath{#1\,^{#2}\!S_{#3}}}%
\let\nssj=\nsSj
%
%
\def\pt{\ensuremath{p_{\mathrm{T}}}} 
\def\pT{\ensuremath{p_{\mathrm{T}}}} 
\def\et{\ensuremath{E_{\mathrm{T}}}} 
\def\eT{\ensuremath{E_{\mathrm{T}}}} 
\def\ET{\ensuremath{E_{\mathrm{T}}}} 
\def\HT{\ensuremath{H_{\mathrm{T}}}} 

\newcommand{\St}      {\ensuremath{S_{\mathrm{T}}}}
\newcommand{\meff}{\ensuremath{m_{\mathrm{eff}}}}
\newcommand{\st}      {\St}
\newcommand{\htlep}   {\ensuremath{H_{\mathrm{T}}^{\mathrm{leptons}}}}
\newcommand{\Htlep}   {\htlep}
\newcommand{\mt}      {\ensuremath{m_{\mathrm{T}}}}
\newcommand{\mo}      {\ensuremath{^{-1}}}

\newcommand{\Kt}      {\ensuremath{\mathrm{k_{T}}}}

\def\ptsq{\ensuremath{p^2_{\mathrm{T}}}} 

\def\degr{\ensuremath{^\circ}} 
\def\abseta{\ensuremath{|\eta|}}
\def\Hgg{\ensuremath{H\to\gamma\gamma}}
\def\mh{\ensuremath{m_h}}
\def\mW{\ensuremath{m_W}}
\def\mZ{\ensuremath{m_Z}}
\def\mH{\ensuremath{m_H}}
\def\mA{\ensuremath{m_A}}
\def\Wjj{\ensuremath{W \rightarrow jj}}
\def\tjjb{\ensuremath{t \rightarrow jjb}}
\def\Hbb{\ensuremath{H \rightarrow b\bar b}}
\def\Zmm{\ensuremath{Z \rightarrow \mu\mu}}
\def\Zee{\ensuremath{Z \rightarrow ee}}
\def\Zll{\ensuremath{Z \rightarrow \ell\ell}}
\def\Wln{\ensuremath{W \rightarrow \ell\nu}}
\def\Wen{\ensuremath{W \rightarrow e\nu}}
\def\Wmn{\ensuremath{W \rightarrow \mu\nu}}
\def\Hllll{\ensuremath{H \rightarrow ZZ^{(*)} \rightarrow \mu\mu\mu\mu}}
\def\Hmmmm{\ensuremath{H \rightarrow \mu\mu\mu\mu}}
\def\Heeee{\ensuremath{H \rightarrow eeee}}
\def\Amm{\ensuremath{A \rightarrow \mu\mu}}
\def\Ztau{\ensuremath{Z \rightarrow \tau\tau}}
\def\Wtau{\ensuremath{W \rightarrow \tau\nu}}
\def\Atau{\ensuremath{A \rightarrow \tau\tau}}
\def\Htau{\ensuremath{H \rightarrow \tau\tau}}
%
%
\def\TeV{\ifmmode {\mathrm{\ Te\kern -0.1em V}}\else
                   \textrm{Te\kern -0.1em V}\fi}%
\def\GeV{\ifmmode {\mathrm{\ Ge\kern -0.1em V}}\else
                   \textrm{Ge\kern -0.1em V}\fi}%
\def\MeV{\ifmmode {\mathrm{\ Me\kern -0.1em V}}\else
                   \textrm{Me\kern -0.1em V}\fi}%
\def\keV{\ifmmode {\mathrm{\ ke\kern -0.1em V}}\else
                   \textrm{ke\kern -0.1em V}\fi}%
\def\eV{\ifmmode  {\mathrm{\ e\kern -0.1em V}}\else
                   \textrm{e\kern -0.1em V}\fi}%
\let\tev=\TeV
\let\gev=\GeV
\let\mev=\MeV
\let\kev=\keV
\let\ev=\eV

\def\TeVc{\ifmmode {\mathrm{\ Te\kern -0.1em V}/c}\else
                   {\textrm{Te\kern -0.1em V}/$c$}\fi}%
\def\GeVc{\ifmmode {\mathrm{\ Ge\kern -0.1em V}/c}\else
                   {\textrm{Ge\kern -0.1em V}/$c$}\fi}%
\def\MeVc{\ifmmode {\mathrm{\ Me\kern -0.1em V}/c}\else
                   {\textrm{Me\kern -0.1em V}/$c$}\fi}%
\def\keVc{\ifmmode {\mathrm{\ ke\kern -0.1em V}/c}\else
                   {\textrm{ke\kern -0.1em V}/$c$}\fi}%
\def\eVc{\ifmmode  {\mathrm{\ e\kern -0.1em V}/c}\else
                   {\textrm{e\kern -0.1em V}/$c$}\fi}%
\let\tevc=\TeVc
\let\gevc=\GeVc
\let\mevc=\MeVc
\let\kevc=\keVc
\let\evc=\eVc

\def\TeVcc{\ifmmode {\mathrm{\ Te\kern -0.1em V}/c^2}\else
                   {\textrm{Te\kern -0.1em V}/$c^2$}\fi}%
\def\GeVcc{\ifmmode {\mathrm{\ Ge\kern -0.1em V}/c^2}\else
                   {\textrm{Ge\kern -0.1em V}/$c^2$}\fi}%
\def\MeVcc{\ifmmode {\mathrm{\ Me\kern -0.1em V}/c^2}\else
                   {\textrm{Me\kern -0.1em V}/$c^2$}\fi}%
\def\keVcc{\ifmmode {\mathrm{\ ke\kern -0.1em V}/c^2}\else
                   {\textrm{ke\kern -0.1em V}/$c^2$}\fi}%
\def\eVcc{\ifmmode  {\mathrm{\ e\kern -0.1em V}/c^2}\else
                   {\textrm{e\kern -0.1em V}/$c^2$}\fi}%
\let\tevcc=\TeVcc
\let\gevcc=\GeVcc
\let\mevcc=\MeVcc
\let\kevcc=\keVcc
\let\evcc=\eVcc

\def\cm{\ifmmode  {\mathrm{\ cm}}\else
                   \textrm{~cm}\fi}%
\def\ifb{\mbox{fb$^{-1}$}}
\def\ipb{\mbox{pb$^{-1}$}}
\def\inb{\mbox{nb$^{-1}$}}
\def\mass#1{\ensuremath{m_{#1#1}}}
\def\twomass#1#2{\ensuremath{m_{#1#2}}}%
\def\Ecm{\ensuremath{E_{\mathrm{cm}}}} 
%
%
\newbox\boxsqbox
\newdimen\boxsize\boxsize=1.2ex%
\def\boxop{%
\setbox\boxsqbox=\vbox{\hrule depth0.8pt width0.8\boxsize height0pt%
                       \kern0.8\boxsize
                       \hrule height0.8pt width0.8\boxsize depth0pt}%
           \hbox{%
           \vrule height1.0\boxsize width0.8pt depth0pt%
           \copy\boxsqbox
           \vrule height1.0\boxsize width0.8pt depth0pt\kern1.5pt}}%
\def\boxsq{\ensuremath{\boxop^2}}%
%
\def\spinor#1{\ensuremath{\left(\matrix{#1_1\cr#1_2\cr#1_3\cr#1_4\cr}\right)}} 
\def\pmb#1{\setbox0=\hbox{$#1$}
  \kern-.025em\copy0\kern-1.0\wd0%
  \kern.05em\copy0\kern-1.0\wd0%
  \kern-.025em\raise.0433em\box0}%
\def\grad{\pmb{\nabla}}%
%
%
\newdimen\dkwidth
\def\dk{%
   \dkwidth=\baselineskip
   {\def\to{\rightarrow}
   \kern 3pt%
   \hbox{%
      \raise 3pt%
      \hbox{%
         \vrule height 0.8\dkwidth width 0.7pt depth0pt%
      }%
      \kern-0.4pt%
      \hbox to 1.5\dkwidth{%
         \rightarrowfill
      }%
   \kern0.6em%
   }}%
}%
%
%
\def\eqalign#1{%
   \,
   \vcenter{%
      \openup\jot\m@th
      \ialign{%
         \strut\hfil$\displaystyle{##}$&&$%
         \displaystyle{{}##}$\hfil\crcr#1\crcr%
      }%
   }%
   \,
}%
%
%
\newcommand {\AcPA}   {Acta Phys. Austriaca{} }
\newcommand {\ARevNS} {Ann.{} Rev.{} Nucl.{} Sci.{} }
\newcommand {\CPC}    {Comp.{} Phys.{} Comm.{} }
\newcommand {\FortP}  {Fortschr.{} Phys.{} }
\newcommand {\IJMP}   {Int.{} J.{} Mod.{} Phys.{} }
\newcommand {\JETP}   {Sov.{} Phys.{} JETP{} }
\newcommand {\JETPL}  {JETP Lett.{} }
\newcommand {\JaFi}   {Jad.{} Fiz.{} }
\newcommand {\JMP}    {J.{} Math.{} Phys.{} }
\newcommand {\MPL}    {Mod.{} Phys.{} Lett.{} }
\newcommand {\NCim}   {Nuovo Cimento{} }
\newcommand {\NIM}    {Nucl.{} Instrum.{} Meth.{} }
\newcommand {\NP}     {Nucl.{} Phys.{} }
\newcommand {\PL}     {Phys.{} Lett.{} }
\newcommand {\PR}     {Phys.{} Rev.{} }
\newcommand {\PRL}    {Phys.{} Rev.{} Lett.{} }
\newcommand {\PRep}   {Phys.{} Rep.{} }   
\newcommand {\RMP}    {Rev.{} Mod.{} Phys.{} }
\newcommand {\ZfP}    {Z.{} Phys.{} }
\newcommand {\EPJ}    {Eur.{} Phys.{} J.{} }
        
%

\newcommand{\HRule}{\rule[20pt]{\linewidth}{0.3mm}}
\newcommand{\HRulesmall}{\rule[10pt]{0.5\linewidth}{0.3mm}}
\newcommand{\atlas}{{\sc Atlas} }
\newcommand{\cms}{{\sc CMS} }
\newcommand{\cdf}{ {\sc CDF} }
\newcommand{\brapsi} {\ensuremath{|\psi\rangle} }
\newcommand{\ketpsi} {\ensuremath{\langle\psi|} }

\newcommand{\lvp}{\mathrm{\textbf{p}}}


\def\smspn#1#2{\ensuremath{\left( \begin{smallmatrix} #1\\ #2 \end{smallmatrix}\right)}}%
\def\spn#1#2{\ensuremath{\left( \begin{matrix} #1\\ #2 \end{matrix}\right)}}%
\def\spnop#1#2#3#4{\ensuremath{\begin{pmatrix} #1& #2\\ #3& #4 \end{pmatrix}}}%

\def\nucl#1#2#3{\ensuremath{{}^{#1}_{#2}\textrm{#3} }}%

\newcommand{\gmma}{\spnop{1}{0}{0}{-1}}
\newcommand{\gmmi}{\spnop{0}{\sigma^i}{-\sigma^i}{0}}
\newcommand{\gmme}{\spnop{0}{1}{1}{0}}

\newcommand{\gma}{\ensuremath{\gamma^0}}
\newcommand{\gmb}{\ensuremath{\gamma^1}}
\newcommand{\gmc}{\ensuremath{\gamma^2}}
\newcommand{\gmd}{\ensuremath{\gamma^3}}
\newcommand{\gme}{\ensuremath{\gamma^5}}
\newcommand{\gmi}{\ensuremath{\gamma^i}}
\newcommand{\gmmua}{\ensuremath{\gamma^\mu}}
\newcommand{\gmnua}{\ensuremath{\gamma^\nu}}
\newcommand{\gmmub}{\ensuremath{\gamma_\mu}}
\newcommand{\gmnub}{\ensuremath{\gamma_\nu}}

\newcommand{\Amua}{\ensuremath{A^\mu}}
\newcommand{\Amub}{\ensuremath{A_\mu}}
\newcommand{\Anua}{\ensuremath{A^\nu}}
\newcommand{\Anub}{\ensuremath{A_\nu}}
\newcommand{\gmunu}{\ensuremath{g_{\mu\nu}}}
\newcommand{\nuL}{\ensuremath{\nu_L}}
\newcommand{\eL}{\ensuremath{e_L}}
\newcommand{\nuR}{\ensuremath{\nu_R}}
\newcommand{\eR}{\ensuremath{e_R}}
\newcommand{\chiL}{\ensuremath{\chi_L}}
\newcommand{\chiR}{\ensuremath{\chi_R}}

\newcommand\Zdis{Z_{\rm dis}}
\newcommand\Zexc{Z_{\rm exc}}

\def\LT{\ensuremath{L_{\mathrm{T}}}} 
\def\ljpt{\ensuremath{p_{\mathrm{T0}}j_{\mathrm{0p_{T}}}}} 

\begin{center}{\Large \textbf{
Applying self-organizing maps to nonresonant searches\\
}}\end{center}

\begin{center}\textbf{
Vaidehi Tikhe\textsuperscript{1}\footnote[1]{Corresponding author: vaidehitikhe@gmail.com} \orcidlink{0009-0006-2540-488X},
N. Kirutheeka\textsuperscript{2} \orcidlink{0009-0007-8390-0168} 
Sourabh Dube\textsuperscript{2}\orcidlink{0000-0002-5145-3777}
}\end{center}

\begin{center}

{\bf 1} Department of Physics, Savitribai Phule Pune University, Pune, India
\\
{\bf 2} Indian Institute of Science Education and Research (IISER), Pune, India
\end{center}

\section*{Abstract}
\textbf{\boldmath{%
    An excess of events seen in a nonresonant search at a collider can be difficult to ascribe
    to a specific hypothesis from amongst several competing ones. We explore a novel approach of
applying self-organizing maps to such an excess to uniquely identify a hypothesis. We use a search for vector-like leptons in a
trilepton final state to outline the method. We define an approach combining the inherent
clustering of these maps and elements of supervised learning. We compare the performance of
this approach with a multiclassifying neural network. We find that
the method using self-organizing maps competes well (despite not using any standard model processes
in the training), and provides additional tools that would help characterize any observed excesses in searches.
}}

\vspace{\baselineskip}

\vspace{10pt}
\noindent\rule{\textwidth}{1pt}
\tableofcontents
\noindent\rule{\textwidth}{1pt}
\vspace{10pt}

\section{Introduction}
\label{sec:intro}
The standard model (SM) of particle physics is one of the most successful theories across all of science~\cite{ParticleDataGroup:2024cfk}.
Well-tested as it is, there are several experimental results that the SM does not explain, such as the particle nature of dark matter,
or the observed matter-antimatter
asymmetry in the universe. Explaining these results requires either a modification of the SM, or the presence of beyond standard model (BSM)
phenomena. There exist a large number of theoretical proposals that aim to address the inadequacies of the standard model. These can be
organized by paradigms, such as models of supersymmetry or models of extra dimensions, or by specific phenomenological predictions, such
as predictions of sterile neutrinos or predictions of nonchiral fermions. Particle physics experiments also have vibrant programs that
search across the board for these BSM phenomena~\cite{cmspublic,atlaspublic}. These searches can be model-dependent (looking for a specific hypothetical particle) or
model-independent (looking for deviations from the SM).

A particular inverse problem in particle physics is to uniquely identify a specific theory or parameters, given a set of experimental results.
Approaches to tackle this can vary from aiming to generally map observed distributions to
theoretically predicted ones (such as~\cite{Shmakov:2023kjj}) to choosing a very large number of experimental observables and
performing a ``fit'' to narrow down the theoretical parameters (such as~\cite{Bornhauser:2012iy}).
The problem can sometimes be straightforward to address. For example, consider a resonant search for an arbitrary particle $X$ which
decays to charged leptons ($X\rightarrow \ell \ell$). If this search observes an excess over the predicted background, then
a study of the invariant mass of the lepton pair, and the relative abundances of events where $\ell = e, \mu, \tau$,
would enable one to uniquely identify several properties of the particle $X$.

However in a nonresonant search, with cascade decays of the hypothetical particles, uniquely identifying a signal hypothesis
from a set of possible choices is a
non-trivial problem. Consider the search conducted by the {\sc CMS} collaboration in the multilepton final state~\cite{CMS:2022nty}.
The search is conducted in seven final states with a varying number of charged leptons as well as hadronically decaying $\tau$-leptons.
{\sc CMS} searches for four different kinds of BSM models and
the searches are carried out as counting experiments in signal regions defined using boosted decision trees, or in signal regions defined by placing
selections on different kinematic quantities. In either case, the searches are nonresonant, and if an excess had been seen, then further
detailed steps would have been needed to uniquely identify the model.

Consider the situation where a nonresonant search as in Ref.~\cite{CMS:2022nty} observes a small excess. The observed events could
consist of a mixture of SM and BSM processes. For a small number of events, distributions of traditional experimental
variables such as sum of lepton momenta, or sum of jet momenta may be insufficient to draw conclusions. 

In this paper we outline a procedure that could help in uniquely identifying the origin of a small excess. We construct our
problem as follows: a search for an example BSM process (vector-like leptons), in a final state with three leptons ($\ell = e, \mu$),
observes an excess. This excess is of the order of 10 events. We define a novel approach using
self-organizing maps (SOMs)~\cite{Kohonen2004SelforganizedFO} to identify the most likely mass of the vector-like leptons which
has given rise to the observed excess. SOMs are an example of an unsupervised machine learning algorithm which clusters inputs in
a latent space. A seemingly homogenous collection of events
may thus also be split into subparts based on the interplay between the properties of those events.
Here we use the SOMs in a `supervised' way, while keeping its clustering properties intact.
Rather than frame the problem as a regression problem (``estimate the vector-like lepton mass''), we
define the procedure to tackle a categorization problem; thus our approach could be useful even when the identification of the
hypothesis does not depend on identifying a numerical value. A natural alternative method is the use of deep neural networks (DNNs) to solve
the categorization problem. We compare the DNN approach to the SOM approach. In the problem of the small excess that we consider,
it is also possible that the excess stems from a standard model process that is unconsidered in the background estimate. To illustrate
our approach, we exclude all SM processes from our SOM training. Despite this, we find that the clustering property of SOMs can
be exploited to identify the signal category.


\section{Setup of the problem}
\label{sec:setup}

Vector-like leptons (VLLs) arise in a wide variety of models, and have been searched for by both the {\sc CMS}~\cite{CMS:2022nty} and {\sc ATLAS}
collaborations~\cite{ATLAS:2024mrr} in a multilepton final state.
Here, we consider the doublet model of VLLs~\cite{Kumar:2015tna,Bhattiprolu:2019vdu}.
The model proposes two new particles: $L$ (charged) and $N$ (neutral) which follow the SM lepton flavors. In this paper we consider
VLLs of the $\mu$-flavor, without loss of generality. The VLLs decay to a SM boson (\Hboson\ or \Zboson\ or \Wpm) and a corresponding lepton.
The SM bosons may further decay leptonically or hadronically. The $L$ and the $N$ are assumed to be mass degenerate, and can be either
pair produced at the LHC, or produced in association with each other. Here are two examples of
decay chains that may result in a final state with three charged leptons ($3\ell$):
\begin{align*}
  L \bar{L} &\rightarrow W\nu_\mu\ \Zboson\mu \rightarrow q q'\ \nu_\mu\ \ \ell\ell\ \mu \\
  L N &\rightarrow \Zboson\mu\ W\mu\ \rightarrow \nu\nu\ \mu\ \ \ell\nu_{\ell}\ \mu \\
\end{align*}
We generate VLL simulation samples for $pp$ collisions at $13.6 \tev$ using \madg~\cite{Alwall_2014}. We generate both possible
processes, $L\bar{L}$ and $LN$. Events are hadronized with
\pythia~\cite{pythia83}. We use the \delphes~\cite{deFavereau2014} framework to obtain the output as {\sc Root}~\cite{Brun:1997pa}
files for further analysis. Additionally, the anti-$k_t$ jet clustering algorithm~\cite{Cacciari:2008gp} implemented in the
\textsc{FastJet}~\cite{Cacciari:2011ma} package is used with the radius parameter set to $0.4$ to cluster hadrons into jets.
We generate simulation samples for six different mass hypotheses of the VLL: $m_L = 500, 600, 750, 1000, 1500, 2500 \gev$.
Additionally, we follow the same simulation chain to generate simulation samples
of the SM $WZ$ production, and the $t\bar{t}Z$ production. The $WZ$ and $t\bar{t}Z$ processes serve as examples of SM processes that
would populate the $3\ell$ final state.

We generate $200,000$ events for $m_L = 500, 1000, 1500 \gev$. For $m_L = 600, 750, 2500 \gev$, and the $WZ$ process, we generate $50,000$ events, while for
$t\bar{t}Z$ we generate $100,000$ events.

We select events with leptons using criteria applied at the particle level to mimic an experimental search in the multilepton final state.
We select light leptons ($e$ or $\mu$) that satisfy $\pt > 10 \gev$ and $|\eta|<2.4$. The leading light lepton is required to satisfy $\pt > 30 \gev$ to
mimic an experimental trigger requirement.
We construct a lepton isolation variable by considering all stable particles (excluding neutrinos) that satisfy $\Delta R<0.4$ from
the lepton, where $\Delta R = \sqrt{\Delta \eta^2 + \Delta \phi^2}$. The scalar sum of \pt\ of these particles is the lepton isolation. We require
that leptons satisy  isolation$/\pt^\ell < 0.15$.
Hadronically decaying $\tau$ leptons are reconstructed from their stable hadronic daughters, and are selected if they satisfy $\pt > 20 \gev$ and $|\eta|<2.4$.
Jets are clustered using the anti-$k_t$ algorithm with radius parameter set to $0.4$, and are required to have $\pt > 30 \gev$. The transverse component
of the vector sum of momenta of all neutrinos is defined as the missing momentum \ptmiss\ of the event.
We require events to have at least three light leptons to be considered for further study.

Typical collider analyses will use different kinematic variables to separate signal from background, either by making direct requirements on these
quantities, or by using them as inputs for a multivariate algorithm. Following Refs.~\cite{CMS:2022nty,ATLAS:2024mrr},
we construct a set of eight kinematic variables defined using the three leptons of the final state. These kinematic variables are used as training variables
for both the DNN and the SOM. Figure~\ref{fig:VLL_inputs} shows the distribution of the kinematic variables for three different
VLL mass hypotheses.
\begin{itemize}
\item We define \LT\ as the scalar sum of \pt\ of the three light leptons. In case there are hadronically decaying taus in the event, we also add
  the \pt\ of the leading tau to the \LT. 
\item We define \HT\ as the scalar sum of \pt\ of all selected jets.
\item We define $m_{\ell\ell\ell}$ as the invariant mass of the three light leptons.
\item We define mos$^{\mathrm{high}}$ and mos$^{\mathrm{low}}$ as the higher and lower (respectively) invariant mass of any of the possible oppositely charged lepton pairs. 
\item We define $\mt^{\mathrm{high}}$ as the highest value of the transverse masses calculated for each lepton and the \ptmiss. The transverse mass for lepton $i$ is
  defined as $\mt = \sqrt{2\ \pt^i\ \ptmiss\ (1-\cos{\Delta\phi})}$, where $\Delta\phi$ is the azimuthal angle separation between the lepton direction and \ptmiss\ direction.
\item We define $\mt^{\mathrm{all lep}}$ as the transverse mass of the resultant vector of the sum of the three leptons and the \ptmiss.
\item We define $\pt^{\ell j}$ as the scalar sum of \pt\ of the leading lepton and the leading jet.
\end{itemize}

\begin{figure}[h]
 
  \centering
  \includegraphics[width=1.0\textwidth]{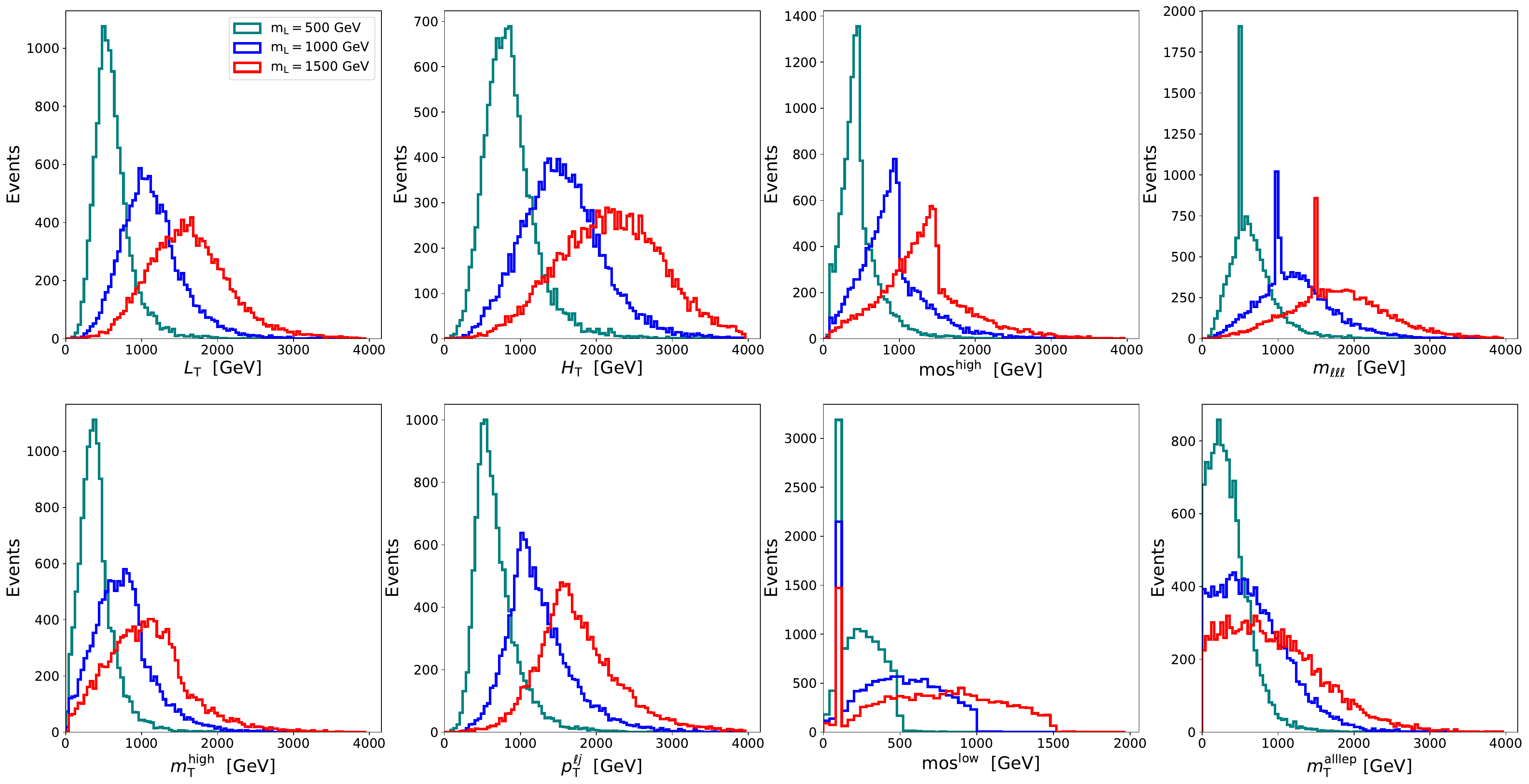}
  \caption[Input features]{The distributions of the kinematic variables are shown for the selected three-lepton events, for three different VLL mass hypotheses.
    The top row (from left to right) shows \LT, \HT, mos$^{\mathrm{high}}$ and $m_{\ell\ell\ell}$, and the bottom row (from left to right) shows
    $\mt^{\mathrm{high}}$, $\pt^{\ell j}$, mos$^{\mathrm{low}}$ and $\mt^{\mathrm{all lep}}$.}
  \label{fig:VLL_inputs}
\end{figure}

We train the algorithm (whether DNN or SOM) on events from three different VLL masses: $m_L = 500,$ or  $1000,$ or  $1500 \gev$, with the DNN
also being trained on SM processes. This mass range is chosen to mimic the range probed by collider searches for VLLs; experimental sensitivity
remains approximately constant over this range. The trained algorithms are then confronted with several example cases of what a particular counting
experiment of the search may observe. These cases are devised such that the excess events may be purely signal (i.e. VLL) or a mixture
of signal and SM processes. We define a procedure, and then for each case we apply our procedure to determine which mass hypothesis is supported by
that counting experiment.

The cases we consider are
\begin{itemize}
\item Case 1: A background free search, where the excess corresponds to one of the $m_L$ hypotheses we have trained on. Here we assume that
  the experiment has ``observed'' $10$ events. We take these $10$ events from the $m_L=1000\gev$ process.
\item Case 2: A background free search, where the excess is of an $m_L$ hypothesis we have {\emph {not}} trained on. Here we assume that the experiment
  has observed $10$ events and we take these events from the $m_L=2500\gev$ process.
\item Case 3: A search with mixed SM processes and an $m_L$ hypothesis we have trained on. Here we assume that the experiment has observed $20$ events.
  We construct these $20$ events as $5$ events of $WZ$ and $t\bar{t}Z$ each, and $10$ events corresponding to the $m_L=500\gev$ process. We note
  that even with merely statistical uncertainties considered, such an excess has a significance of less than $3\sigma$. In practice, such a low
  significance might not trigger an experiment to consider the inverse problem. However, the method outlined here can still be applied.
\item Case 4: A search with mixed SM processes and an $m_L$ hypothesis we have {\emph{not}} trained on. Here we assume that the experiment has observed $15$ events.
  We construct these as $2$ events from $WZ$, $3$ events from $t\bar{t}Z$, and $10$ events from the $m_L=750\gev$ process.
\item Case 5: This mirrors Case 4 above, but with the VLL considered to be from the $m_L=600\gev$ process.
\end{itemize}

As stated above, we specifically do not include the SM processes in our training for the SOM.
We aim to outline a method for the use of the SOM to
separate different classes of events, and thus we treat the SM processes as just another class, one which is unknown to the SOM during training.
By excluding the SM processes, we aim to cover cases where 
\begin{itemize}
\item The background for a search is estimated in a completely data-driven way, with no labelled background available for training.
\item The background yield is too low for training.
\item The search already employs a multivariate algorithm (such as a neural network) to suppress background - and thus the SM process events
  that survive selections are highly similar to the BSM signal hypothesis.
\end{itemize}


\section{Multiclassifying DNN}
\label{sec:dnn}

Deep neural networks are a supervised machine learning technique and are widely used in experimental particle physics. We implement a multiclassifier
using DNNs to infer the correct hypothesis given a set of input events.  We use the $\tensorflow$ library~\cite{tensorflow2015-whitepaper} in Python
using  $\keras$~\cite{keras_chollet2015} as the interface. The DNN is trained using the eight input variables described in Section~\ref{sec:setup}.
The DNN has three hidden layers with $32,16,8$ neurons respectively. The output layer uses a softmax activation function
and contains four neurons, representing three VLL mass hypotheses (500, 1000, 1500) and the SM. These output neurons are respectively labelled as
$n500$, $n1000$, $n1500$, and $nSM$ and their output score represents the probability that an input event belongs to that particular class.
A categorical cross-entropy loss function is used for the DNN with Adam \cite{adam_optimizer} chosen as the learning rate optimizer. The model has $988$ trainable parameters.
The neural network is trained using $9500$ events each for $m_L = 500, 1000, 1500$, and SM (for the SM, $4250$ events each of $WZ$ and $t\overline{t}Z$ are combined). The
DNN performance remains stable with as low as a $1000$ events used in the training.

For a binary classifier, one can use a receiver operating characteristic (ROC) curve to assess performance. For this multiclassifier we define a `one-versus-others'
ROC curve in the following way. To calculate the ROC curve for a given signal hypothesis, say $m_L=1000$, we choose the corresponding output neuron ($n1000$). The
distribution of output scores for $m_L=1000$ class of events on $n1000$ is treated as `signal'. The distribution of all other processes (viz. $m_L=500, 1500$ and SM) on $n1000$ is
treated as `background'. Using these signal and background distributions, we can calculate an ROC curve.

Figure~\ref{fig:DNNperformance} shows the one-versus-others ROC curves for each class of events, using a statistically independent (from training)
dataset of $2000$ events. We estimate an uncertainty on the ROC curves (and thus the AUC) using the bootstrapping method: the uncertainty is
the standard deviation of the true positive rate of 10 iterations of training done using randomly chosen input events.\\

\begin{figure}[h]
  \centering
  \includegraphics[width=0.45\textwidth]{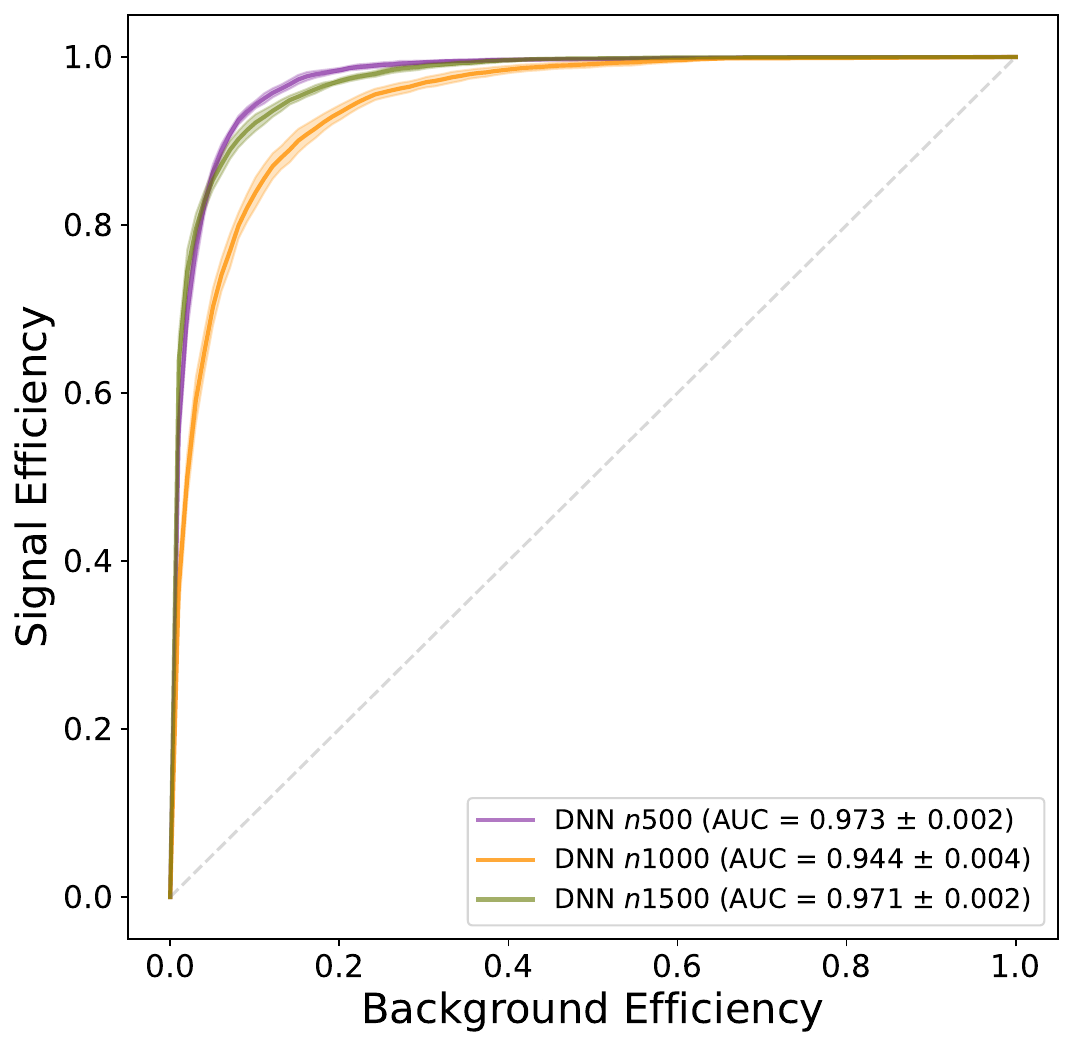}
  \caption[DNN performance]{The one-versus-others ROC curves along with uncertainties for the multiclassifying DNN for the three signal
    hypotheses that the DNN is trained to identify.}
  \label{fig:DNNperformance}
\end{figure}


We now take the various cases outlined at the end of Section~\ref{sec:setup} and for each case, use the trained DNN to infer the right $m_L$ value. We perform
a $100$ trials of each case, and illustratively show plots of one case.

~\\ \noindent {\bf Case 1:} The experiment has observed $10$ events (each event corresponds to $m_L=1000\gev$).

We pass the $10$ events through the DNN to obtain the scores for each of the output neurons. These scores are shown in Fig.~\ref{fig:DNNcase1} (left).
The legend of the figure also shows the median value for that distribution.
We compare the median of each distribution, and see that the median value for $n1000$ is the highest, at $0.8$. From the comparison of these
median values we conclude that these observed events correspond to $m_L=1000\gev$. For a $100$ independent trials, we get $m_L=1000\gev$ in all of them.

\begin{figure}[h!]
  \centering
  \includegraphics[width=0.40\textwidth]{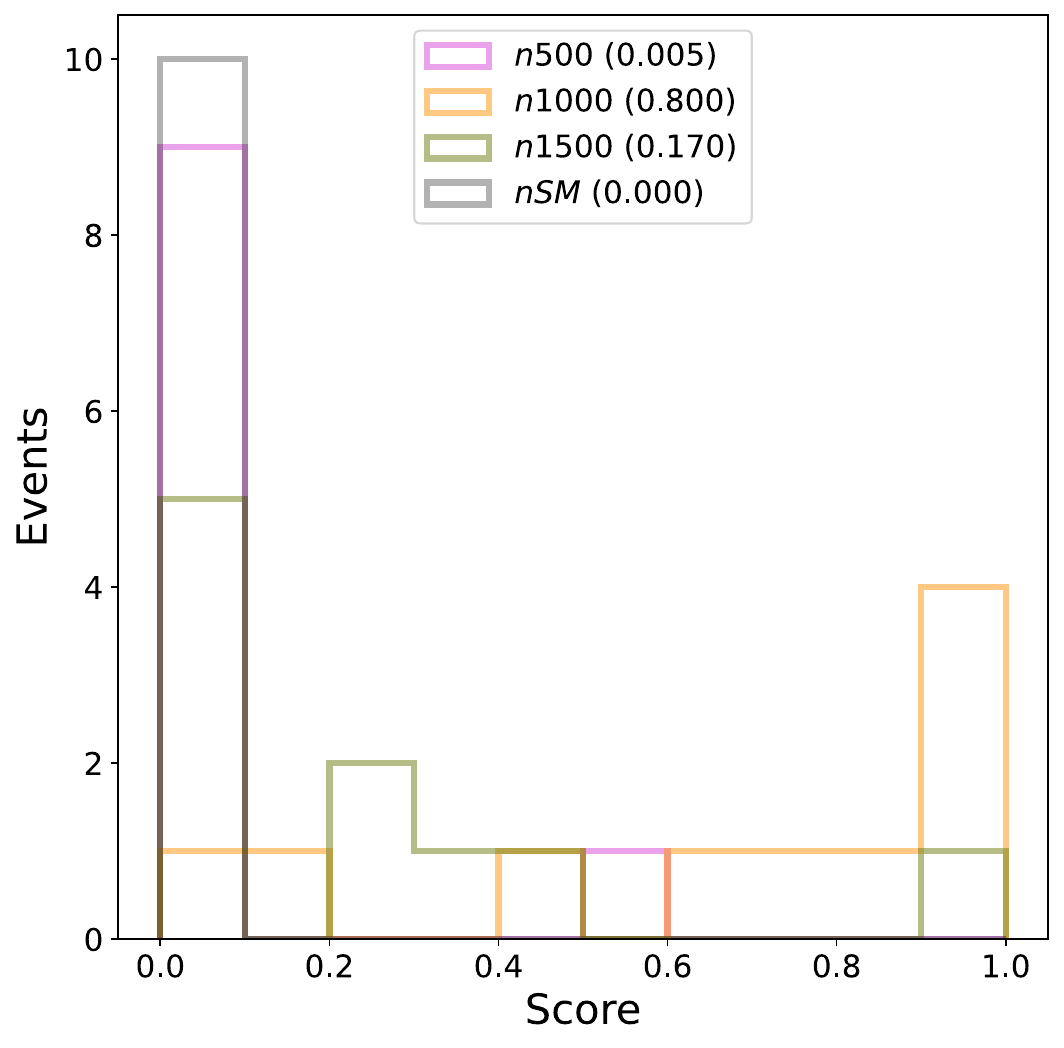}
  \includegraphics[width=0.40\textwidth]{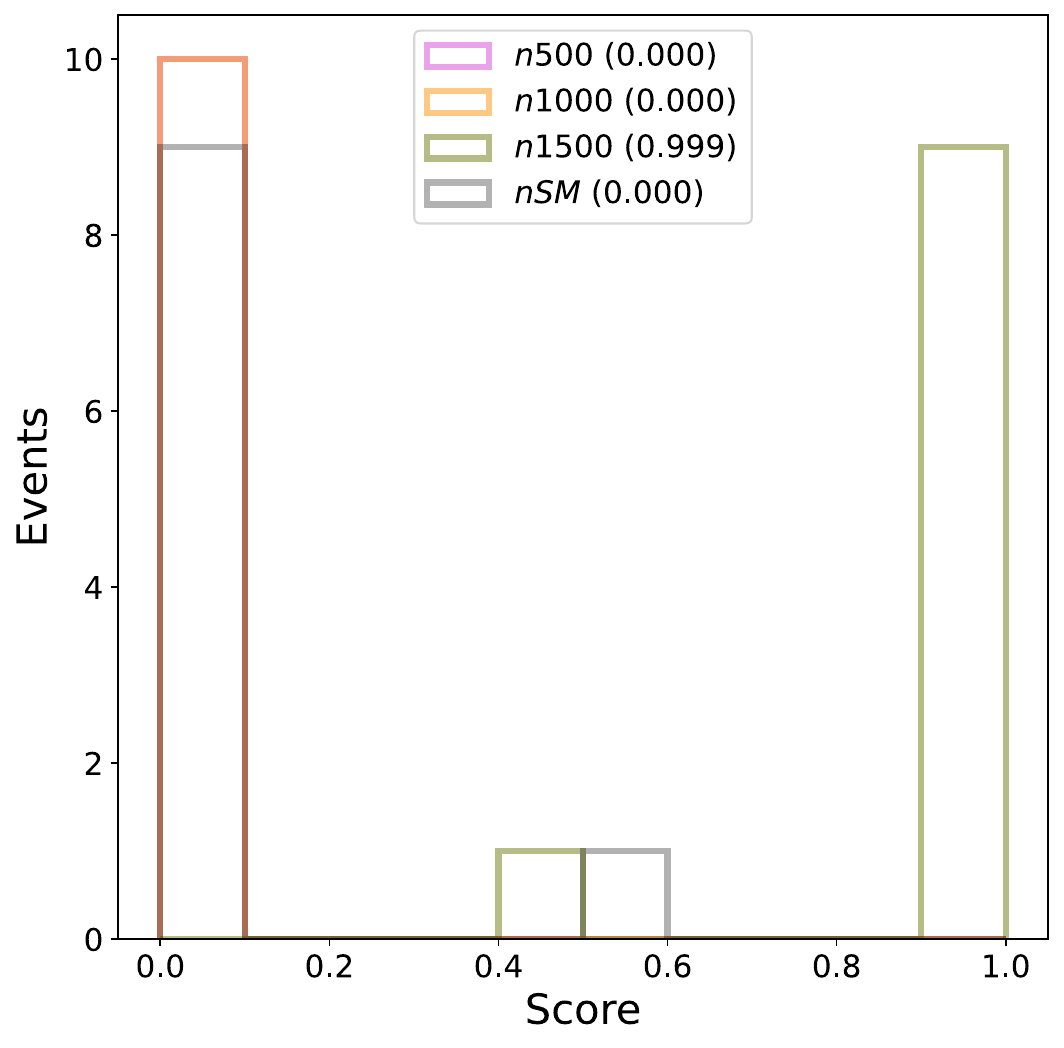}
  \caption[DNN Infering 1 and 2]{The distribution of the scores of the observed events for case 1 (left) and case 2 (right)
    on the four output neurons of the DNN. The median
    value for each distribution is shown in the legend. For case 1, the highest median is for the $n1000$ score,
    while for case 2, it is for the $n1500$ score.}
  \label{fig:DNNcase1}
\end{figure}

~\\ \noindent {\bf Case 2:} The experiment has observed $10$ events (each event corresponds to $m_L=2500\gev$).

We pass the $10$ events through the DNN; Fig.~\ref{fig:DNNcase1} (right) shows distribution of scores for each of the output neurons. Here the median value
for $n1500$ is highest, and thus one would conclude (erroneously) that the observed events correspond to $m_L=1500\gev$.
For a $100$ independent trials, we get $m_L=1500\gev$ in all trials.
This result is expected;
the properties of the events of the $m_L=2500\gev$ hypothesis most closely resemble those of the $m_L=1500\gev$, which the network is trained to identify.

~\\ \noindent {\bf Case 3:} The experiment has observed $20$ events, with an expected background of $10$ events. (The composition of the observed events is $10$ from the SM
processes, and $10$ from the $m_L=500\gev$ process.)

The distribution of the neuron scores for the $20$ events is shown in Fig.~\ref{fig:DNNcase3} (left). In this case, we see that several events have a high score
on $nSM$, thus indicating the presence of SM events. We make an additional requirement on the events that the $nSM$ score $< 0.8$. This selection is designed to reject
90\% of the SM events. We show the scores for the
$11$ events surviving this selection in Fig.~\ref{fig:DNNcase3} (right). For these $11$ events, the median score for $n500$ is the highest, and we conclude
that the observed excess corresponds to the $m_L=500\gev$ hypothesis.
Here as well, for a $100$ independent trials, we get $m_L=500\gev$ in all trials.

\begin{figure}[h!]
  \centering
  \includegraphics[width=0.40\textwidth]{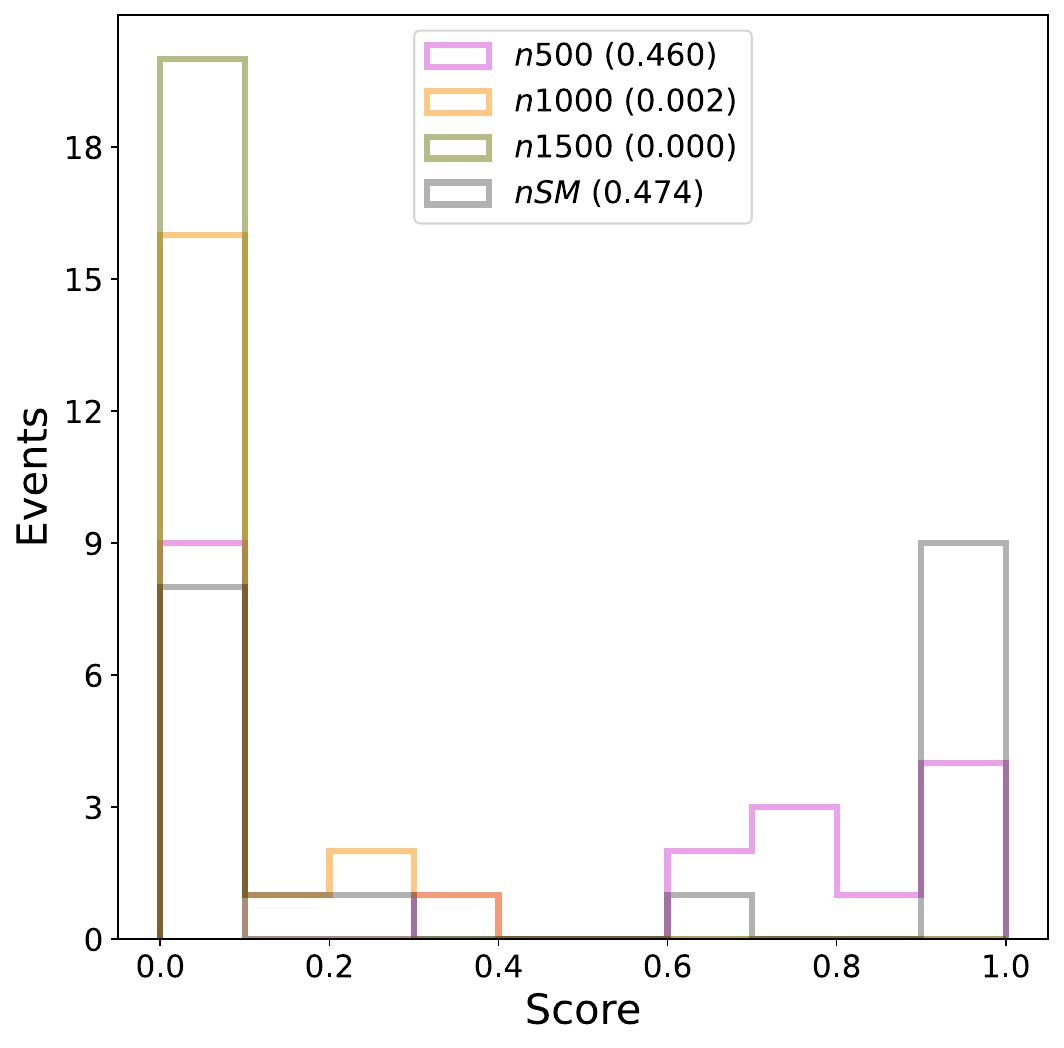}
  \includegraphics[width=0.40\textwidth]{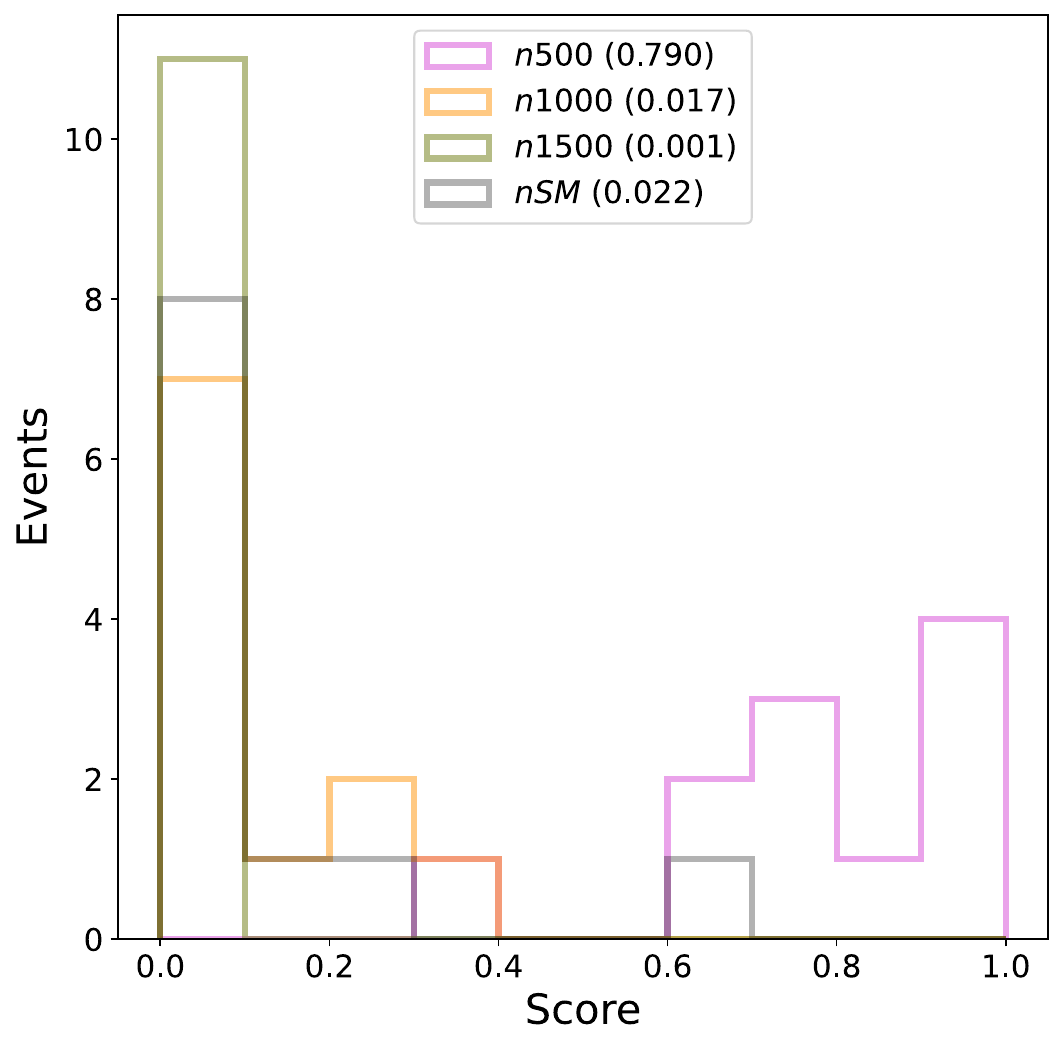}
  \caption[DNN Infering 3]{The distribution of the output neuron scores of the observed events for case 3. The left plot shows the distributions for all events,
    while the right plot shows the distribution after requiring that $nSM$ score $<0.8$. The median
    value for each distribution is shown in the legend. }
  \label{fig:DNNcase3}
\end{figure}

~\\ \noindent {\bf Case 4:} The experiment has observed $15$ events, with an expected background of $5$ events. (The composition of the observed events is $5$ from the SM
processes, and $10$ from the $m_L=750\gev$ process.)

The distribution of the neuron scores for the events is shown in Figure~\ref{fig:DNNcase4}, for both before and after a requirement that $nSM$ score $<0.8$.
We see that in this case, the median scores are high for both the $n500$ and $n1000$ output neurons - and thus a clear conclusion for the observed events
cannot be obtained. However, it is possible to rule out the $m_L=1500\gev$ hypothesis. In a $100$ independent trials, we get the highest median for
$m_L=1000\gev$ in $71$ trials, and $m_L=500\gev$ in $29$ trials.

\begin{figure}[h!]
  \centering
  \includegraphics[width=0.40\textwidth]{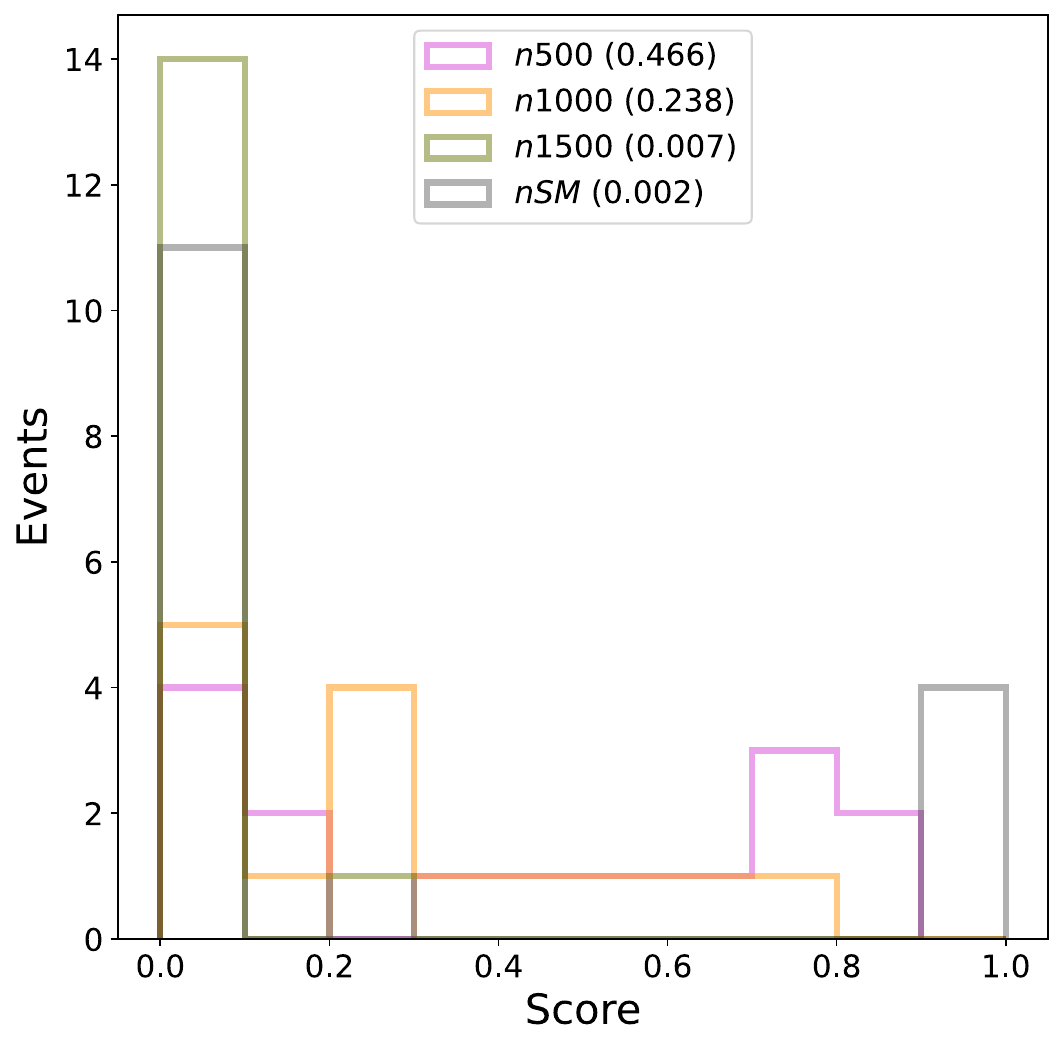}
  \includegraphics[width=0.40\textwidth]{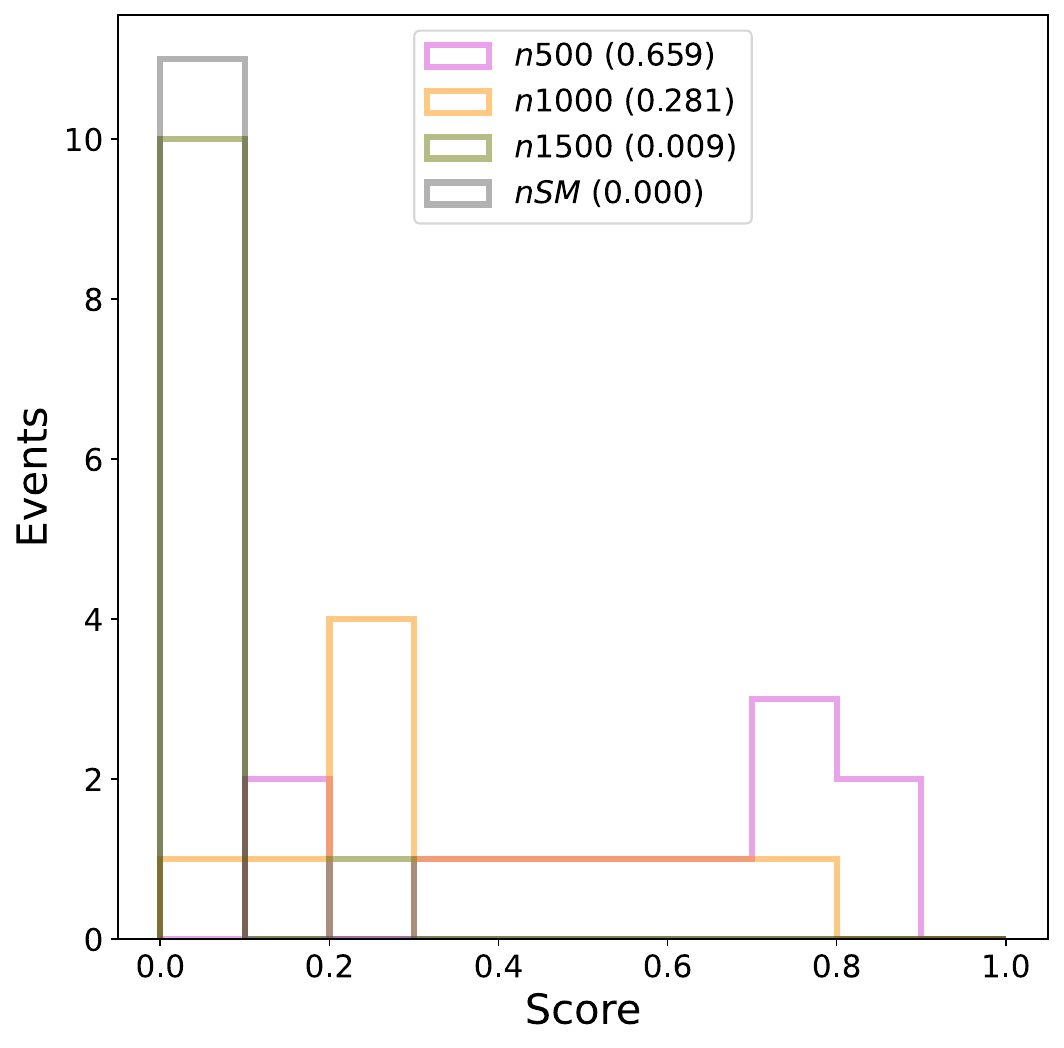}
  \caption[DNN Infering 4]{ The distribution of the output neuron scores of the observed events for case 4. The left plot shows the distributions for all events,
    while the right plot shows the distribution after requiring that $nSM$ score $<0.8$. The median
    value for each distribution is shown in the legend. }
  \label{fig:DNNcase4}
\end{figure}

~\\ \noindent {\bf Case 5:} The experiment has observed $15$ events, with an expected background of $5$ events. (The composition of the observed events
is $5$ from the SM processes, and $10$ from the $m_L=600\gev$ process.)

The distribution of the neuron scores for the events is shown in Figure~\ref{fig:DNNcase5}, for both before and after a requirement that $nSM$ score $<0.8$.
We see that in this case, the median scores are high for $n500$ after the requirement of $nSM$, and thus one would infer that the observed events
correspond to $m_L=500\gev$. While this conclusion is erroneous, as in the previous case, it is possible to rule out the  $m_L=1000$ and $1500\gev$ hypotheses.
In a $100$ independent trials, we get $m_L=500\gev$ in $99$ trials, and $m_L=1000\gev$ in $1$ trial.

\begin{figure}[h!]
  \centering
  \includegraphics[width=0.40\textwidth]{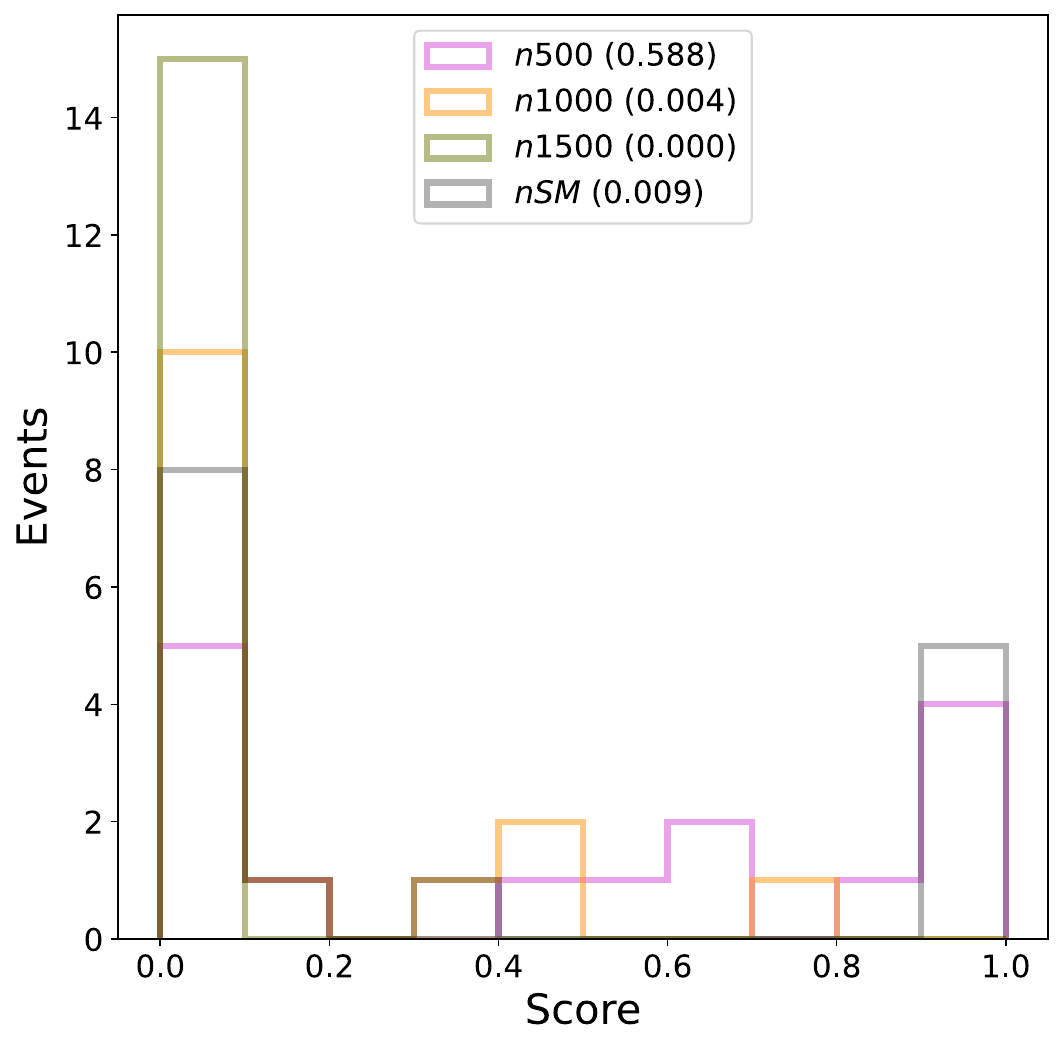}
  \includegraphics[width=0.40\textwidth]{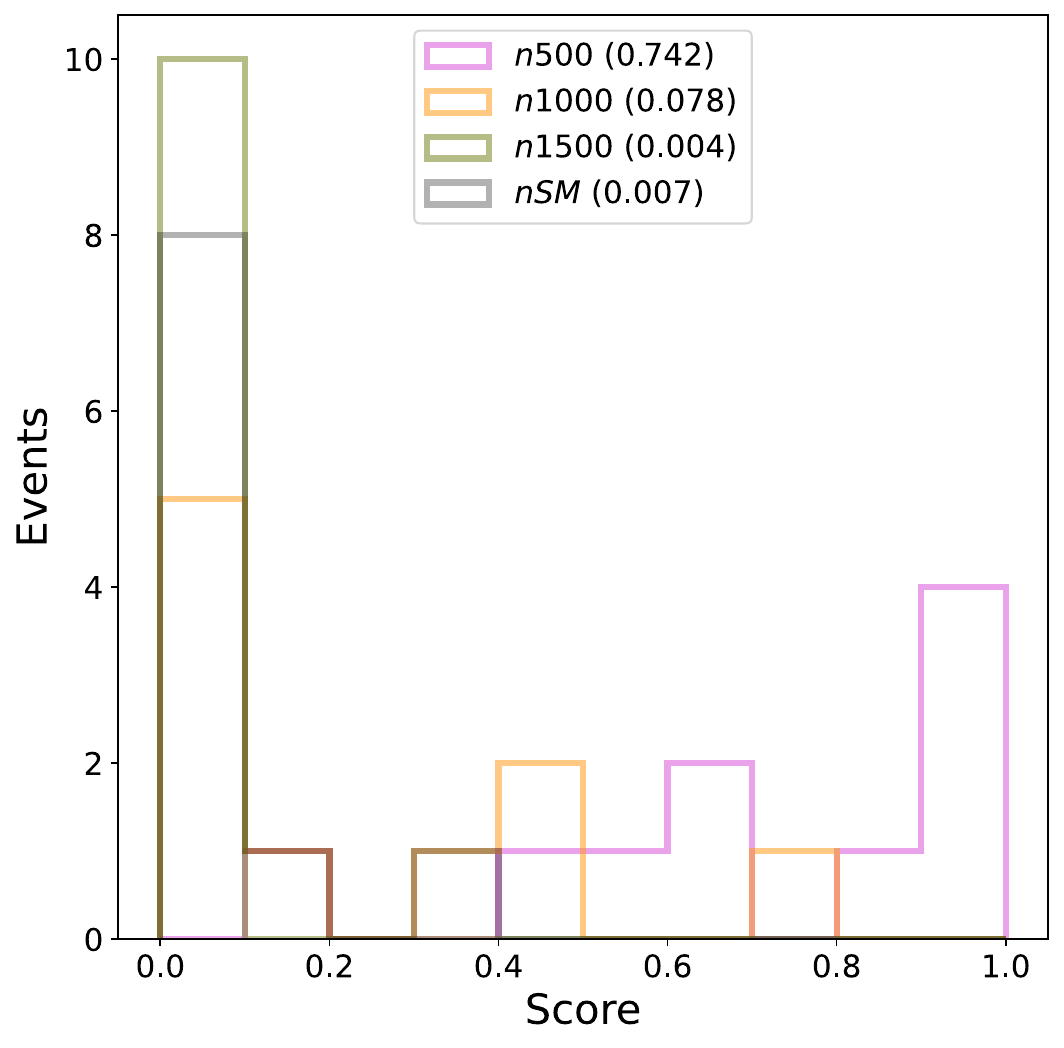}
  \caption[DNN Infering 5]{ The distribution of the output neuron scores of the observed events for case 5. The left plot shows the distributions for all events,
    while the right plot shows the distribution after requiring that $nSM$ score $<0.8$. The median
    value for each distribution is shown in the legend. }
  \label{fig:DNNcase5}
\end{figure}


\section{Self-Organizing Maps}
\label{sec:som}

Self-organizing maps are a two-dimensional representation of a multidimensional dataset.
So far, they have not seen significant use for experimental particle physics. Studies have been conducted to use SOMs as
anomaly finders at the LHC~\cite{Chowdhury:2025mul}, which is an example of using them in an unsupervised way.
Here we outline a method to use SOMs in a `supervised' way; we train the SOM on labelled datasets and provide
a recipe to use the trained SOM to infer `unlabelled' data as described in Section~\ref{sec:setup}. 

We use the `Minimalistic Implementation of the Self Organizing Maps' or MiniSOM~\cite{vettigliminisom} library to construct our SOM model.
The training of a SOM is an iterative process. The SOM includes a single layer of neurons arranged in a two-dimensional grid ($n \times q$).
We consider only square grids ($n=q$). Each neuron has a weight vector $w$ associated to it, with the dimensions of $w$ being the same as
the number of input features of the data. Each neuron thus represents a point in the multidimensional space of the input features; in the
same way, each instance of the input data (i.e. each event) is also a point in this space. During training, a Euclidean distance is calculated between
each neuron and that particular input event. The neuron closest to this event is flagged as the Best Matching Unit (BMU) for that input event.
A radius parameter $\sigma$ defines the neighborhood of the BMU. We choose the neighborhood function for our training to be a Gaussian distribution.
At each iteration $t$, the $w$'s for the BMU along with its neighborhood are updated to bring it closer to the input event.
The learning rate $L(t)$ decides the size of the change in $w$. The updated weights are given by
\begin{equation}
  w(t+1) = w(t) + L(t) \times (x - w(t))
  \label{eq:wtsUpdate}
\end{equation}
where $x$ stands for the input vector. The weights for the SOM neurons are initialized using PCA components of the input data.
The algorithm is repeated for all the input rows over a number of iterations resulting in the clustering of similar data points into a group.

We train the SOM on $9500$ events each for $m_L=500, 1000, 1500 \gev$. A well trained SOM is expected to separate the input classes into
separate clusters. Additionally, even within a class separation into subclusters is desirable. These subclusters will help inform the
likely properties and correlations between events when the SOM is used on the observed events. 
To assess performance, we define a separation score for each mass hypothesis. The separation score aims to assess the overall separation
of the classes, as well as the spreading out of each class in the grid.
The score for $m_L=500\gev$ is defined as
\begin{align} \mathrm{score}_{500}^i &= \frac{{\mathrm{N_{500} - N_{1000} - N_{1500}}}}{{\mathrm{N_{500} + N_{1000} + N_{1500}}}} \\
  \mathrm{SepScore}_{500} &= \frac{\Sigma_i\ {\mathrm{score}_{500}^{i}}}{{\mathrm{N_{non-empty}}}}
\end{align}
where $\mathrm{N_{500}, N_{1000}, N_{1500}}$ are the number of events of each class falling in neuron $i$, and $\mathrm{N_{non-empty}}$ is the
number of neurons which are populated by at least one event of any class. Similarly we calculate $\mathrm{SepScore}_{1000}$ and $\mathrm{SepScore}_{1500}$.
If the clustering is perfect, then any given neuron will be populated by only one class. Thus $\mathrm{score}^i$ for that class would be $1.0$.
If that particular class also then occupies a larger area in the grid, the resultant $\mathrm{SepScore}$ will tend to $1.0$. 
In case a class of events is absent from a neuron, then the corresponding $\mathrm{score}^i$
will be $-1.0$, and a larger fraction of such neurons will make the $\mathrm{SepScore}$ tend to -1.0. The $\mathrm{SepScore}$ is thus an
interplay between separating the classes and occupying a larger area in the grid; the former is independent for each class, the latter
is necessarily dependent.

\begin{figure}[h!]
  \centering
  \begin{subfigure}[b]{0.24\textwidth}
    \centering
    \includegraphics[width=\textwidth]{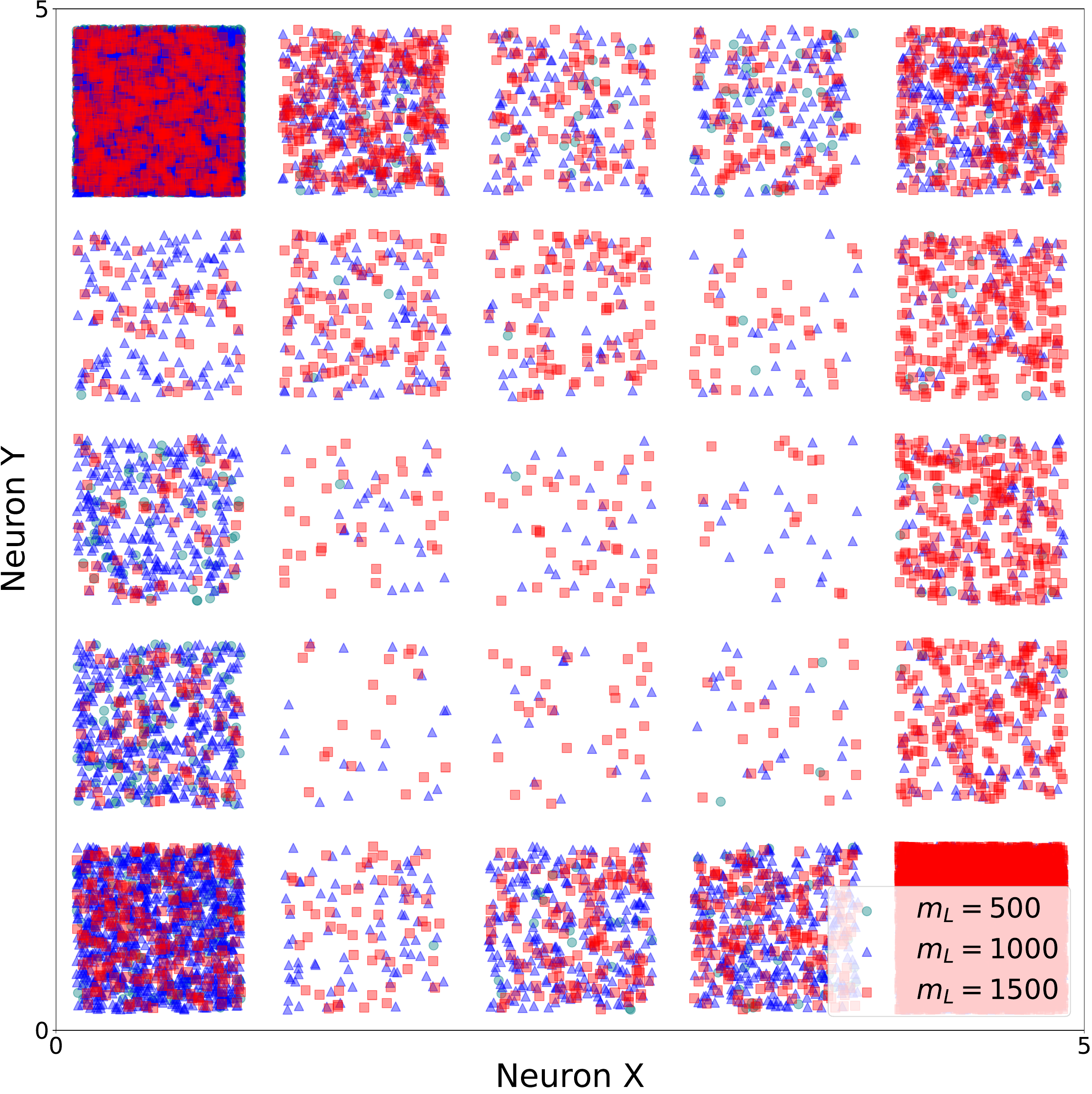}
    \caption{$n=5$}
    \label{subfig:(a)}
  \end{subfigure}
  \hfill
  \begin{subfigure}[b]{0.24\textwidth}
    \centering
    \includegraphics[width=\textwidth]{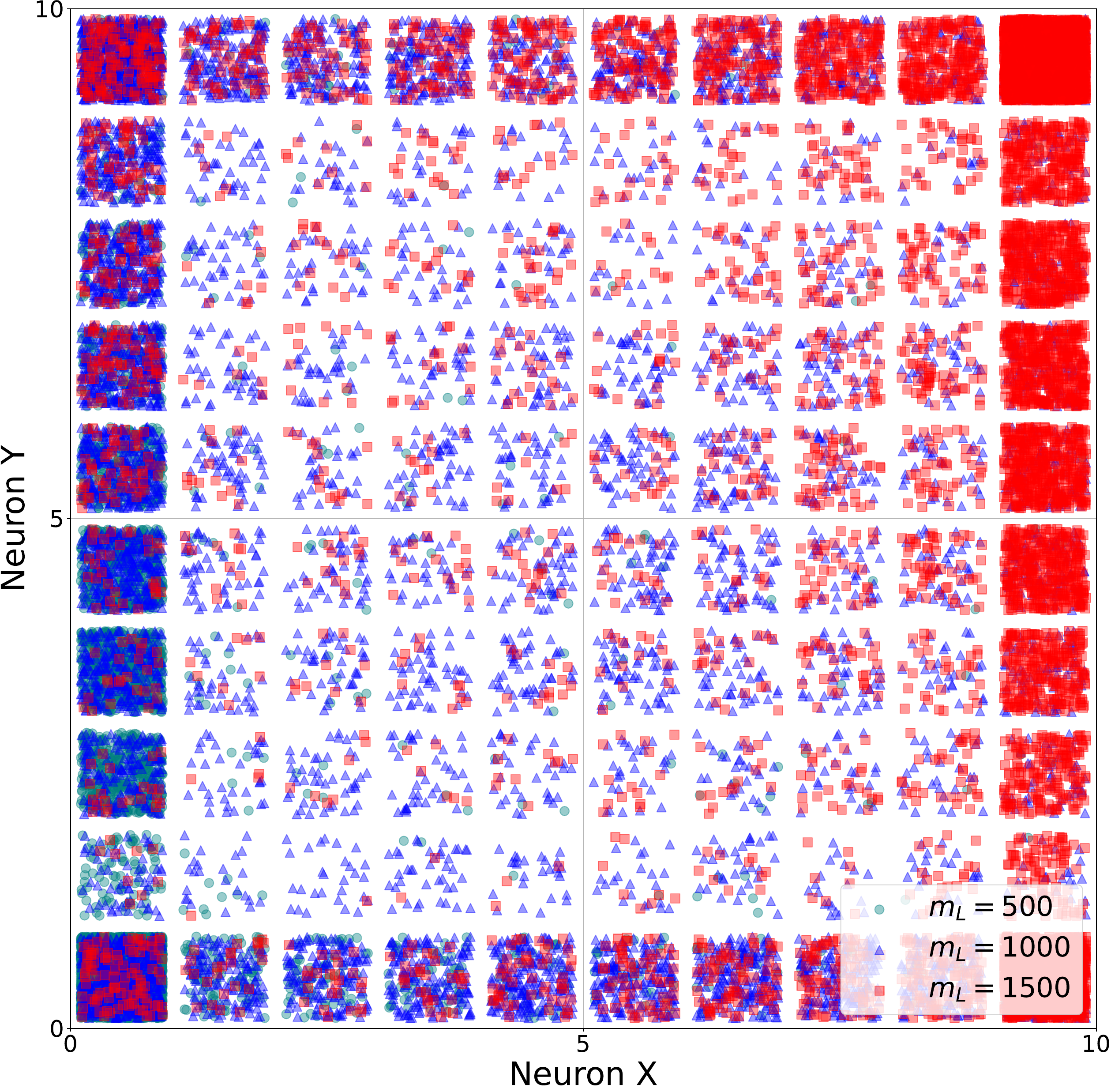}
    \caption{$n=10$}
    \label{subfig:(b)}
  \end{subfigure}
   \hfill
  \begin{subfigure}[b]{0.24\textwidth}
    \centering
    \includegraphics[width=\textwidth]{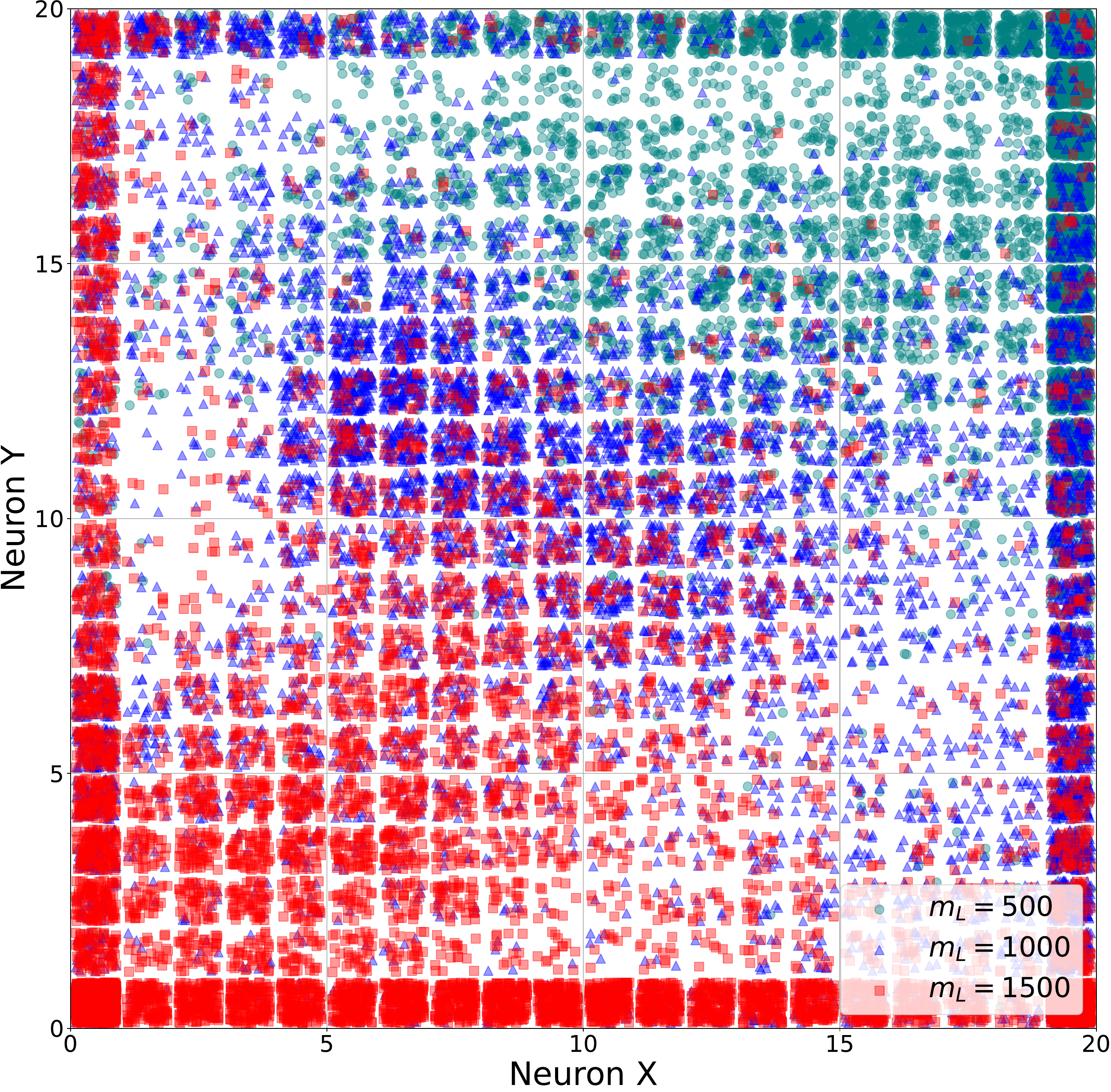}
    \caption{$n=20$}
    \label{subfig:(c)}
  \end{subfigure}
   \hfill
  \begin{subfigure}[b]{0.24\textwidth}
    \centering
    \includegraphics[width=\textwidth]{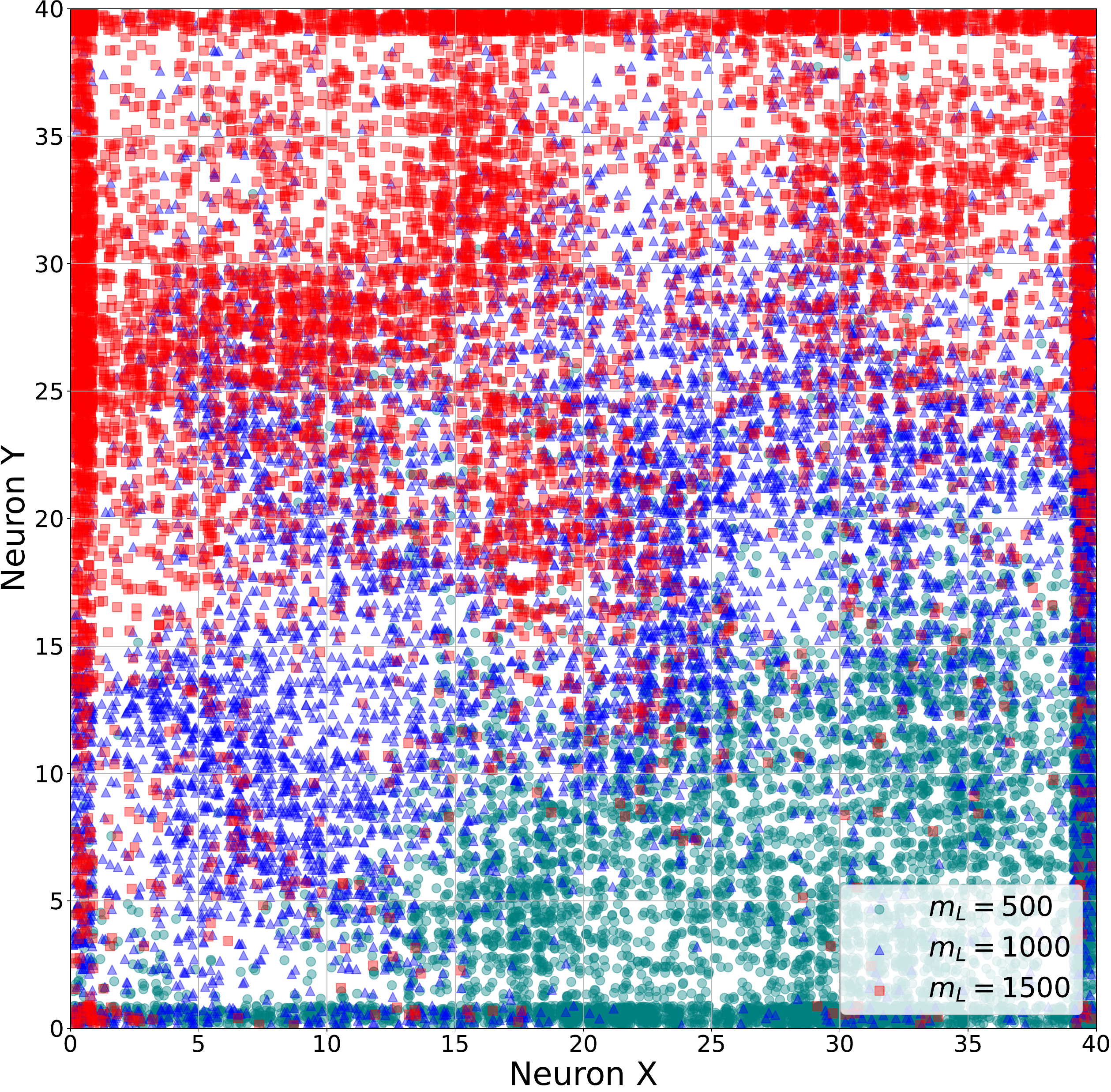}
    \caption{$n=40$}
    \label{subfig:(d)}
  \end{subfigure}
   
  \begin{subfigure}[b]{0.24\textwidth}
    \centering
    \includegraphics[width=\textwidth]{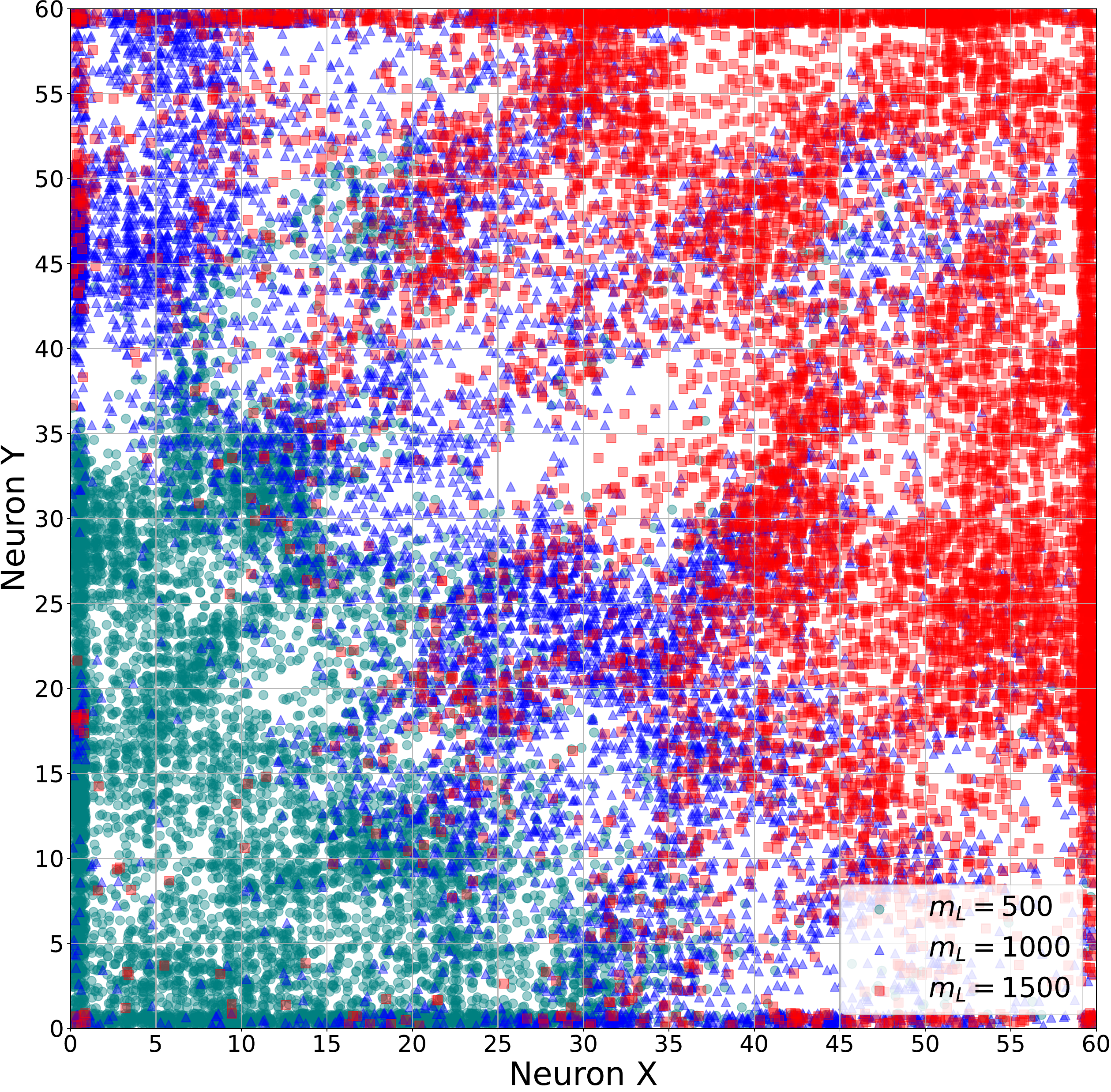}
    \caption{$n=60$}
    \label{subfig:(e)}
  \end{subfigure}
   \hfill
  \begin{subfigure}[b]{0.24\textwidth}
    \centering
    \includegraphics[width=\textwidth]{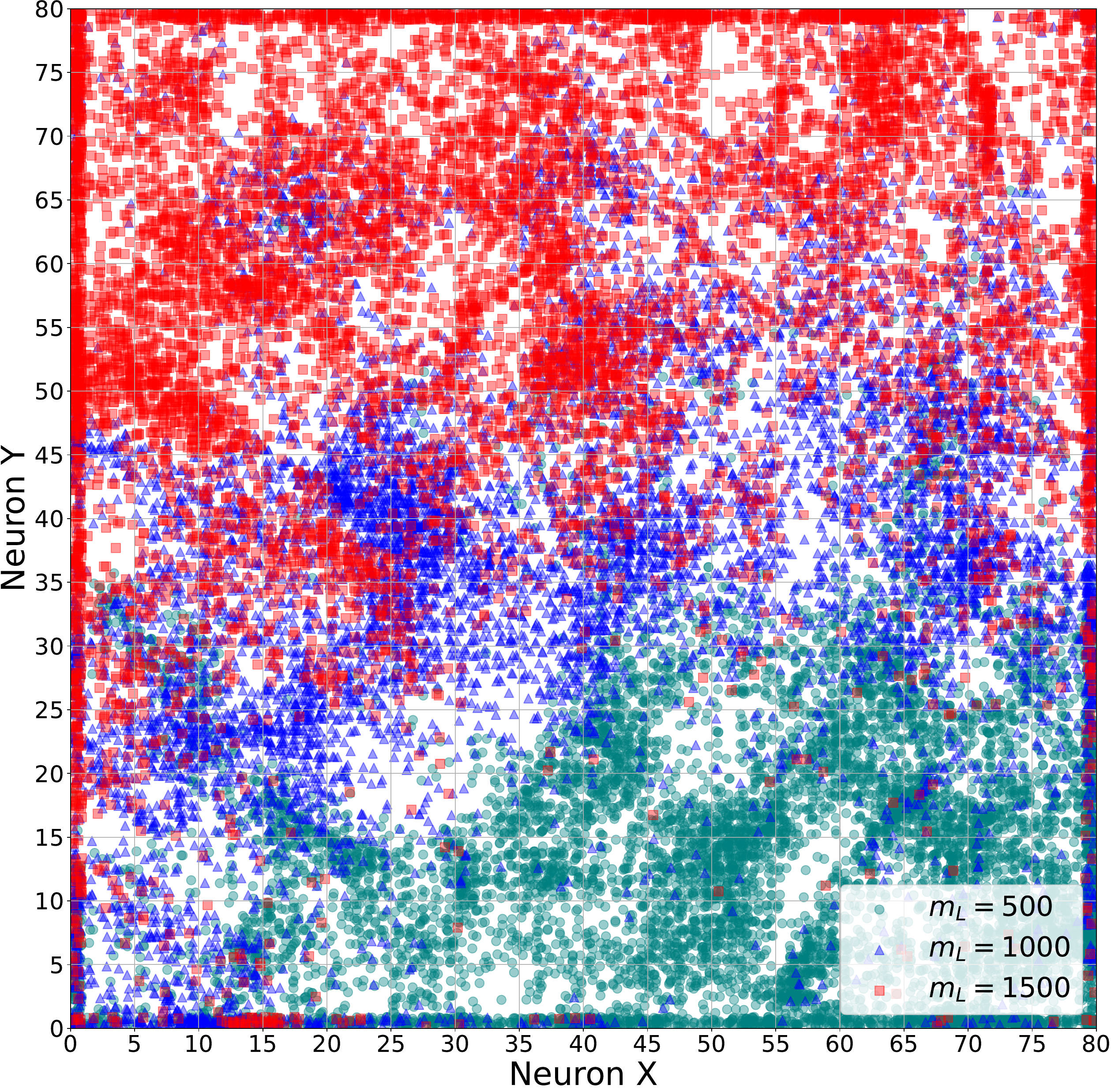}
    \caption{$n=80$}
    \label{subfig:(f)}
  \end{subfigure}
   \hfill
  \begin{subfigure}[b]{0.24\textwidth}
    \centering
    \includegraphics[width=\textwidth]{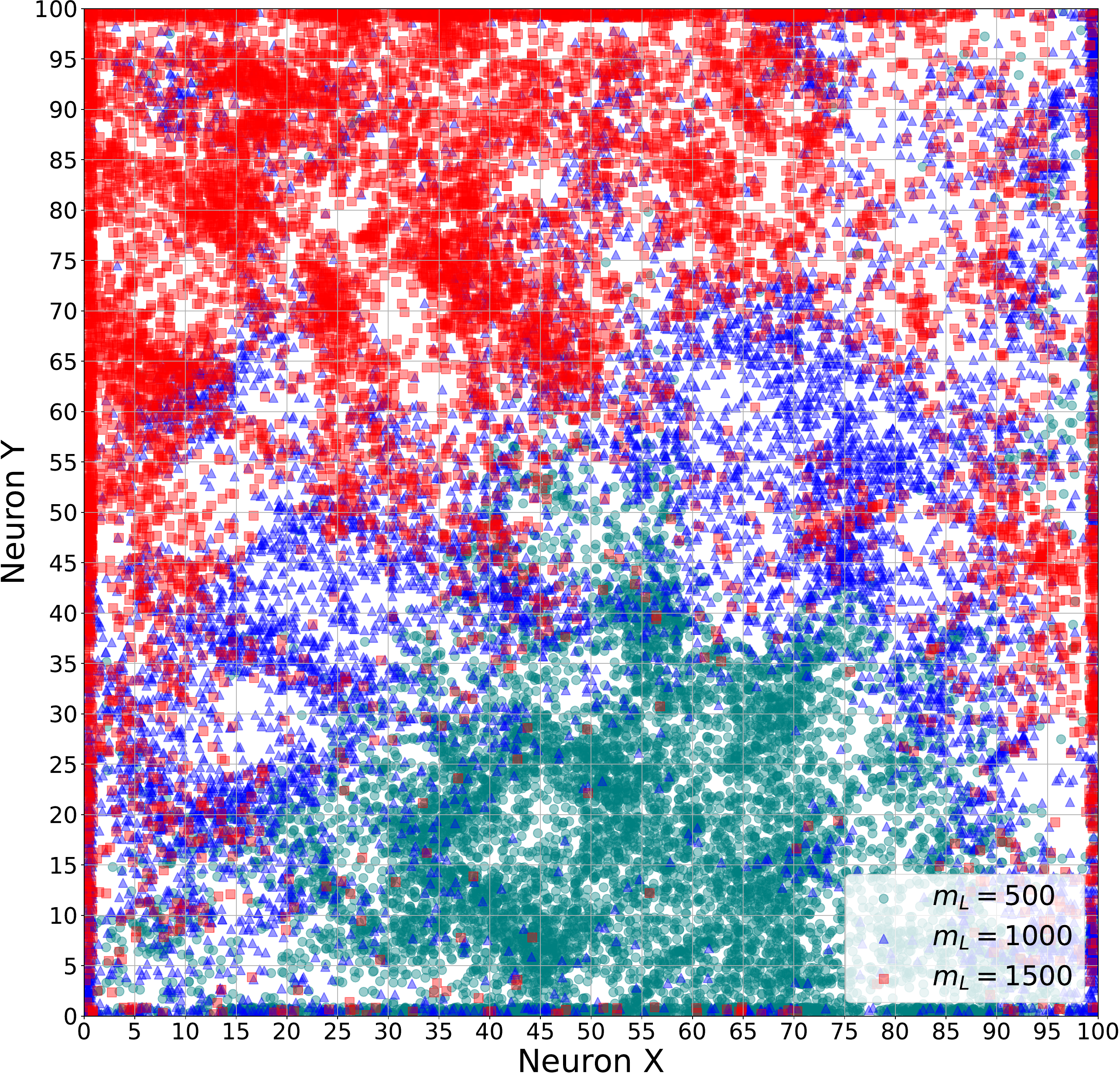}
    \caption{$n=100$}
    \label{subfig:(g)}
  \end{subfigure}
   \hfill
  \begin{subfigure}[b]{0.24\textwidth}
    \centering
    \includegraphics[width=\textwidth]{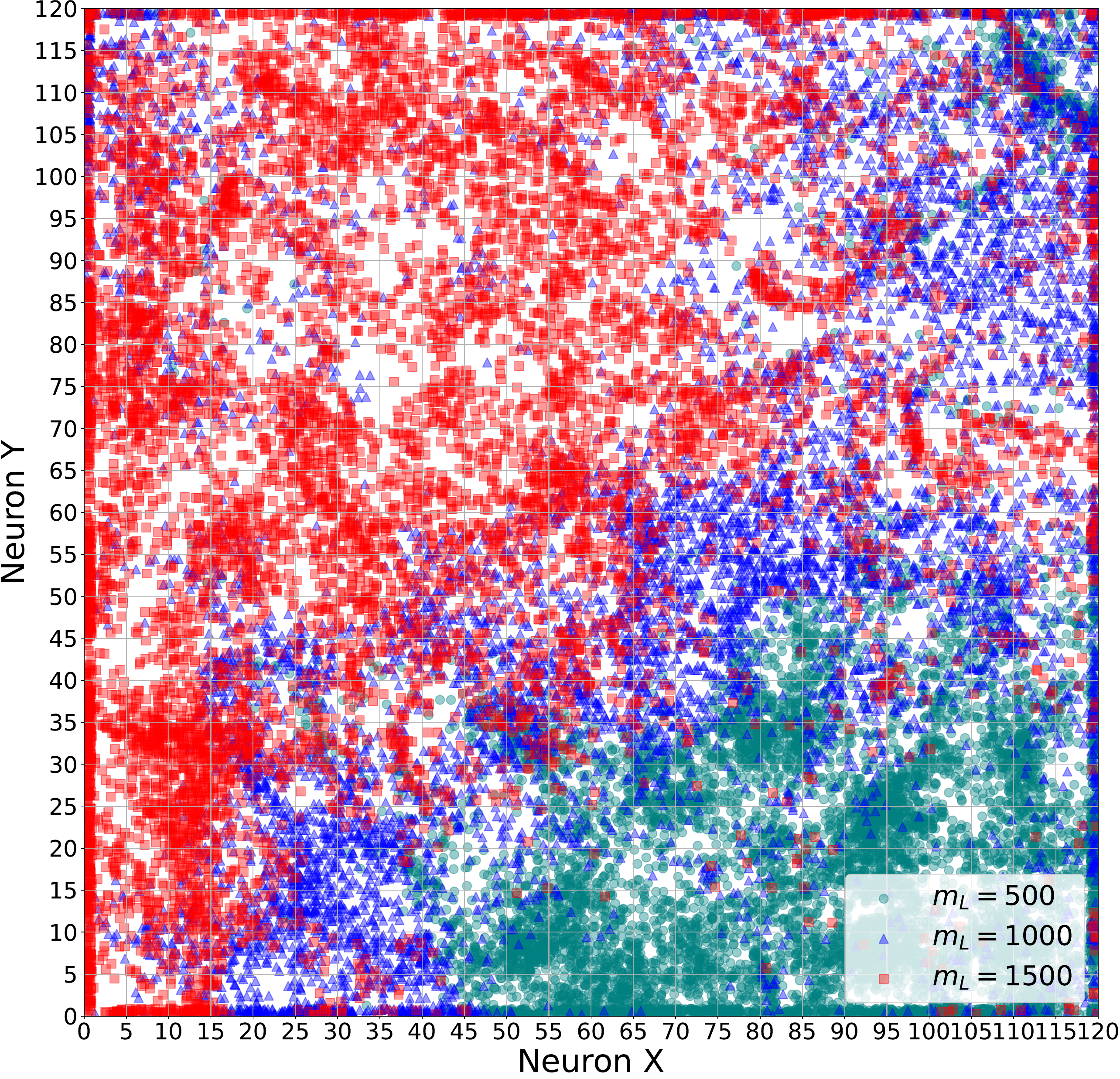}
    \caption{$n=120$}
    \label{subfig:(h)}
  \end{subfigure}
  \caption[SOM grids with different sizes]{Different SOM models with varying grid size $n$.
    The models are populated by events of the training dataset.}
  \label{fig:Somgrids1}
\end{figure}

\begin{table}[!h]
  \centering
\begin{tabular}{|c|c|c|c|}
  \hline
  Size $n$ & $\mathrm{SepScore}_{500}$& $\mathrm{SepScore}_{1000}$& $\mathrm{SepScore}_{1500}$ \\ \hline
  $5$& $-0.869\pm 0.036$& $-0.073\pm 0.019$& $-0.057\pm 0.017$\\
  $10$& $-0.827\pm 0.022$& $0.136\pm 0.009$& $-0.309\pm 0.016$\\
  $20$& $-0.459\pm 0.015$& $-0.163\pm 0.012$& $-0.378\pm 0.014$\\
  $40$& $-0.490\pm 0.015$& $-0.205\pm 0.012$& $-0.305\pm 0.012$\\ 
  $60$& $-0.483\pm 0.015$& $-0.238\pm 0.012$& $-0.279\pm 0.012$\\ 
  $80$& $-0.423\pm 0.013$& $-0.306\pm 0.012$& $-0.272\pm 0.012$\\ 
  $100$& $-0.411\pm 0.013$& $-0.264\pm 0.011$& $-0.325\pm 0.012$\\ 
  $120$& $-0.497\pm 0.013$& $-0.289\pm 0.011$& $-0.214\pm 0.011$\\ 
    \hline
  \end{tabular}
\caption[SepScores for varying $n$]{The separation scores for the three $m_L$ hypotheses for SOM models of varying $n$ shown in Fig.~\ref{fig:Somgrids1}. Higher values
of score for a particular class indicate better separation of that class and occupation of a larger grid area. }
  \label{tab:sepscores_n}
\end{table}

We start with considering different combinations of $n$ (the size of the SOM). We fix $\sigma$ (the radius parameter) to $12$,  $L(t)=1.0$,
and the number of iterations to be $2000$. Figure~\ref{fig:Somgrids1} shows eight different SOM models with varying $n$.
Some separation is achieved in each case, and
as the grid size changes, the events have more {\it{space}} to spread into subclusters.
For the method we describe, there emerge two constraints on the grid size. A too small grid has large overlap of the different classes
of points, with no defined regions of separation. An extremely large SOM grid will have each input event occupy its own neuron,
which will erode the statistical power of judging the clustering.
Table~\ref{tab:sepscores_n} shows the separation scores for the models shown in Fig.~\ref{fig:Somgrids1}, with a statistical uncertainty
calculated from the population in each neuron.
One can see, for example, that while $n=10$ performs best for $m_L=1000\gev$, it performs especially poorly for $m_L=500\gev$.
For our analysis, we choose two example SOM grids, with $n=100$ and $n=40$. These show good separation between classes as well as
overall occupancy in grid. We note that choosing to work with two SOM grids for an analyst will also offer the advantage of
having a cross-check.

\begin{figure}[h!]
  \centering
  \begin{subfigure}[b]{0.3\textwidth}
    \centering
    \includegraphics[width=\textwidth]{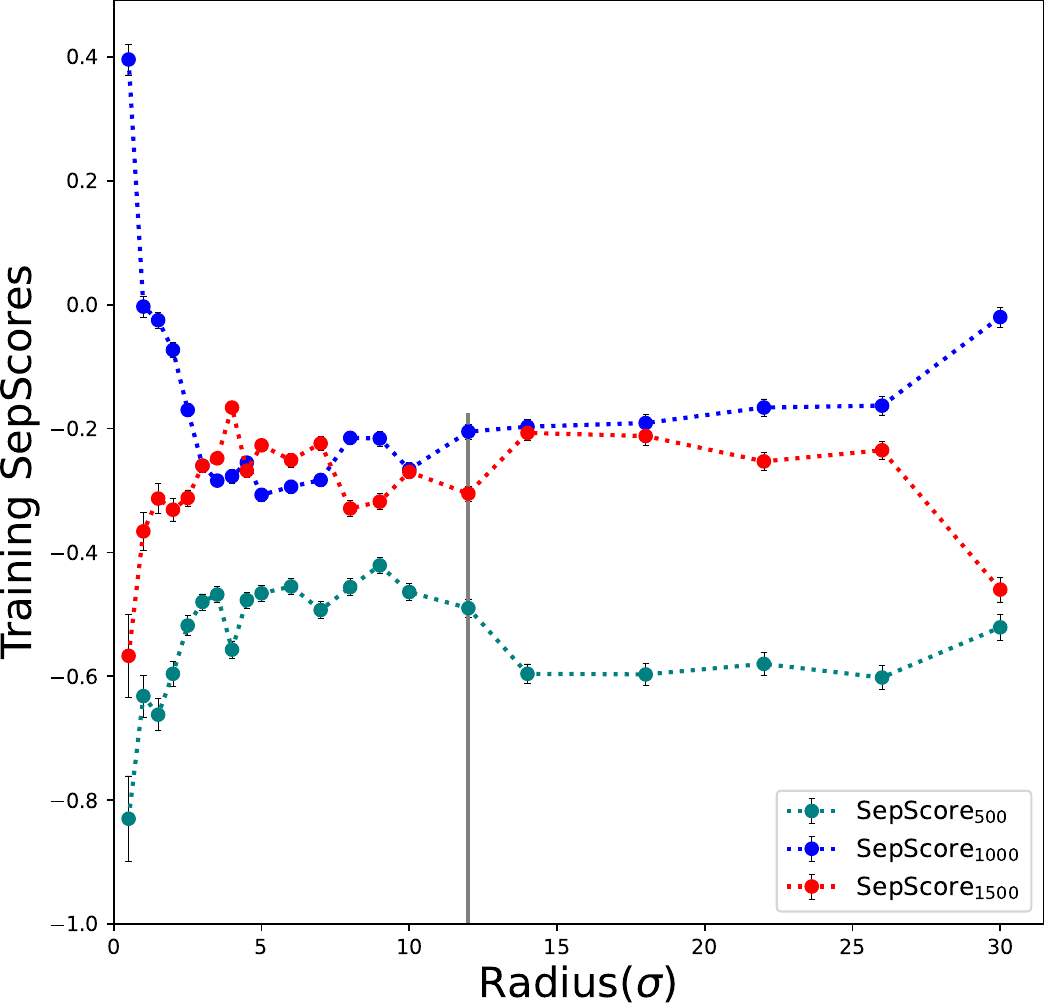}
    \label{subfig:(a1)}
  \end{subfigure}
  \hfill
  \begin{subfigure}[b]{0.3\textwidth}
    \centering
    \includegraphics[width=\textwidth]{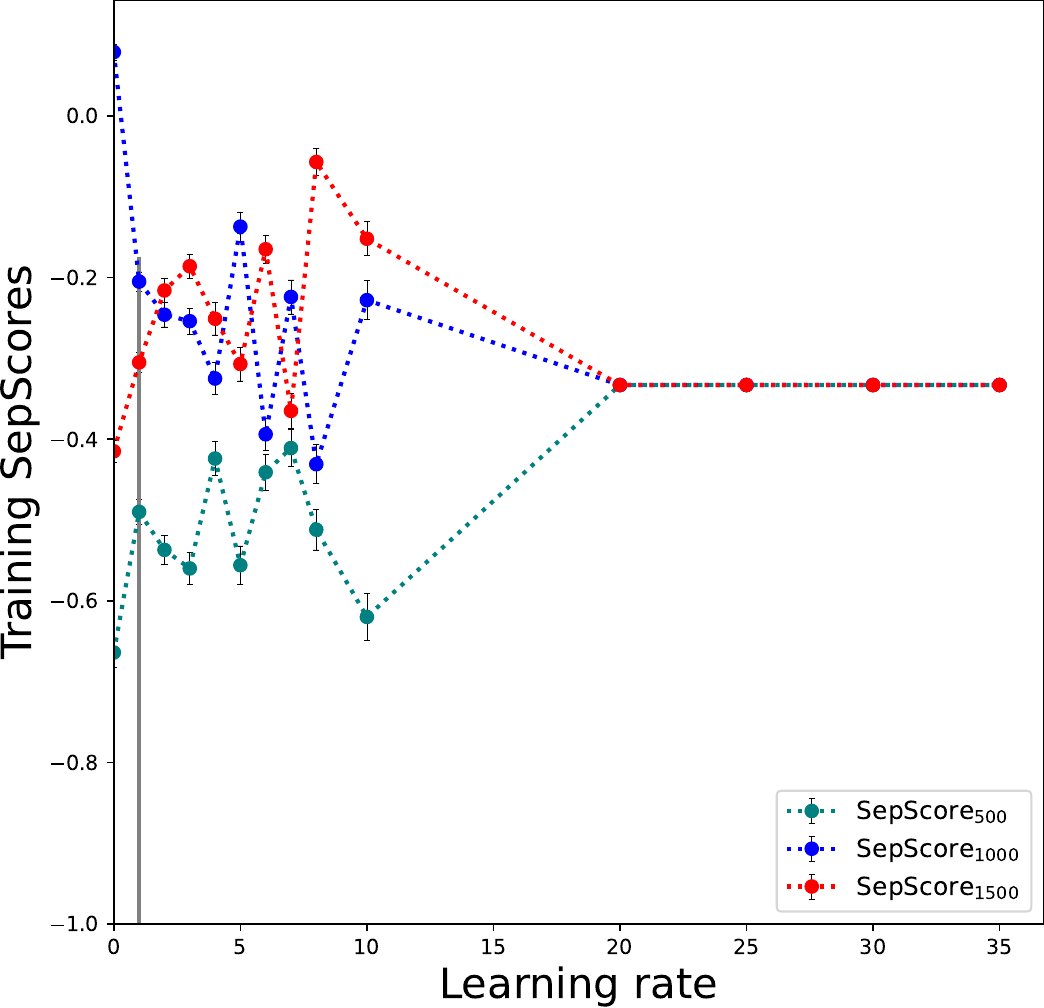}
    \label{subfig:(b1)}
  \end{subfigure}
   \hfill
  \begin{subfigure}[b]{0.3\textwidth}
    \centering
    \includegraphics[width=\textwidth]{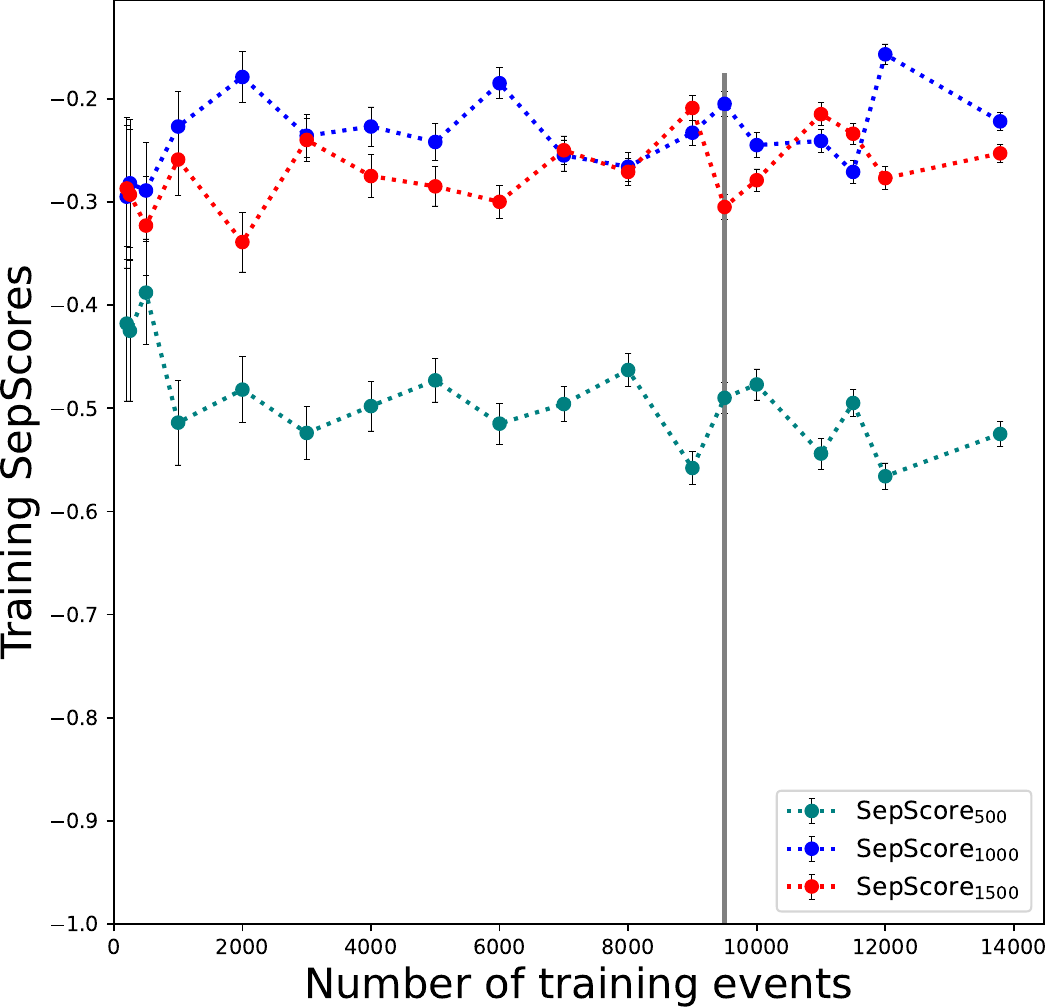}
    \label{subfig:(c1)}
  \end{subfigure}

  \hfill
   
  \begin{subfigure}[b]{0.3\textwidth}
    \centering
    \includegraphics[width=\textwidth]{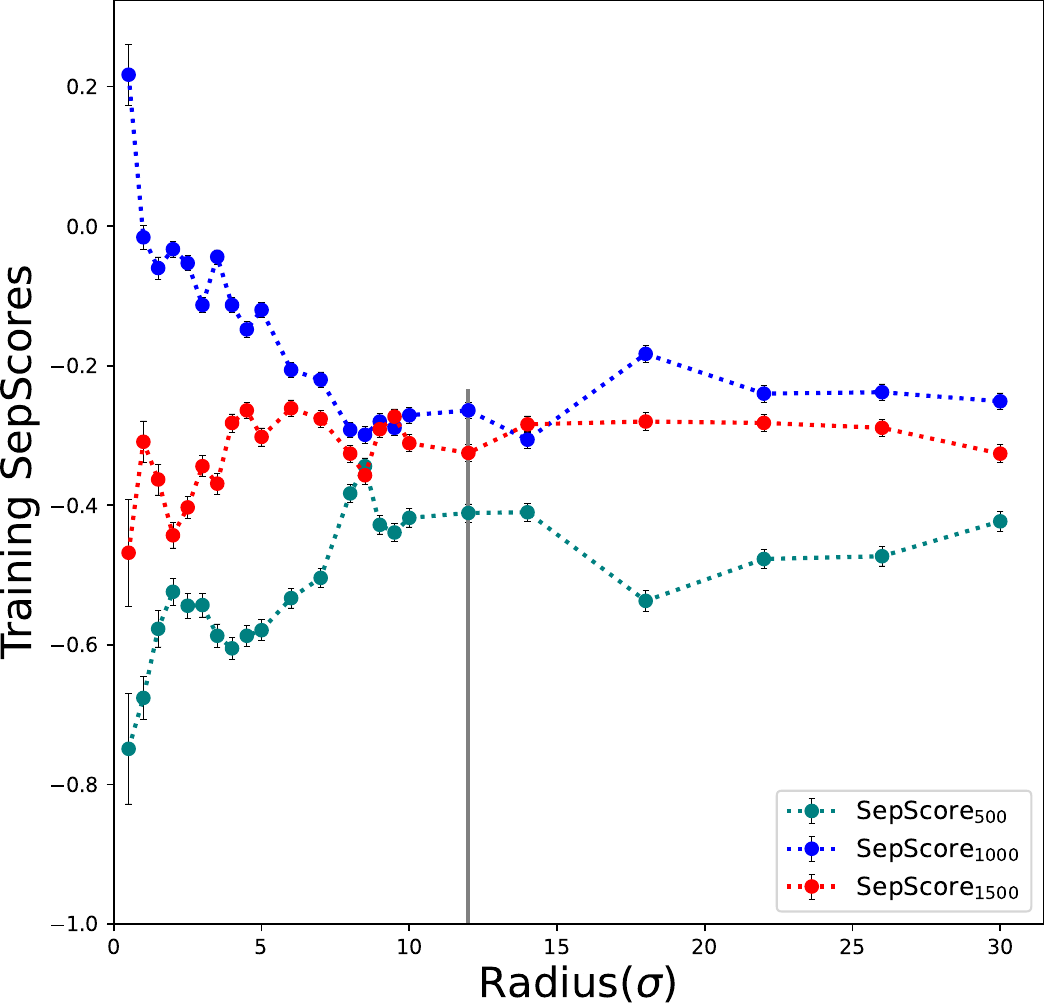}
    \label{subfig:(f1)}
  \end{subfigure}
   \hfill
  \begin{subfigure}[b]{0.3\textwidth}
    \centering
    \includegraphics[width=\textwidth]{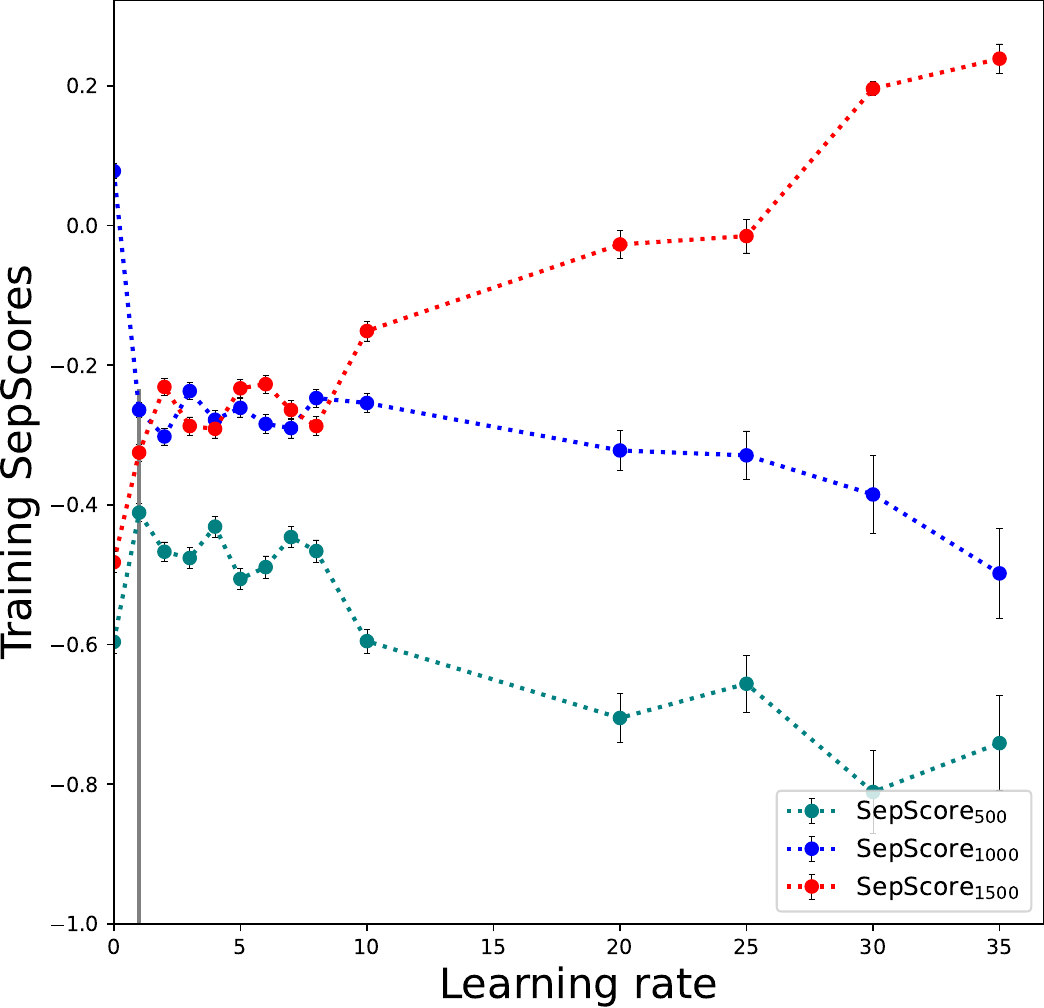}
    \label{subfig:(g1)}
  \end{subfigure}
   \hfill
  \begin{subfigure}[b]{0.3\textwidth}
    \centering
    \includegraphics[width=\textwidth]{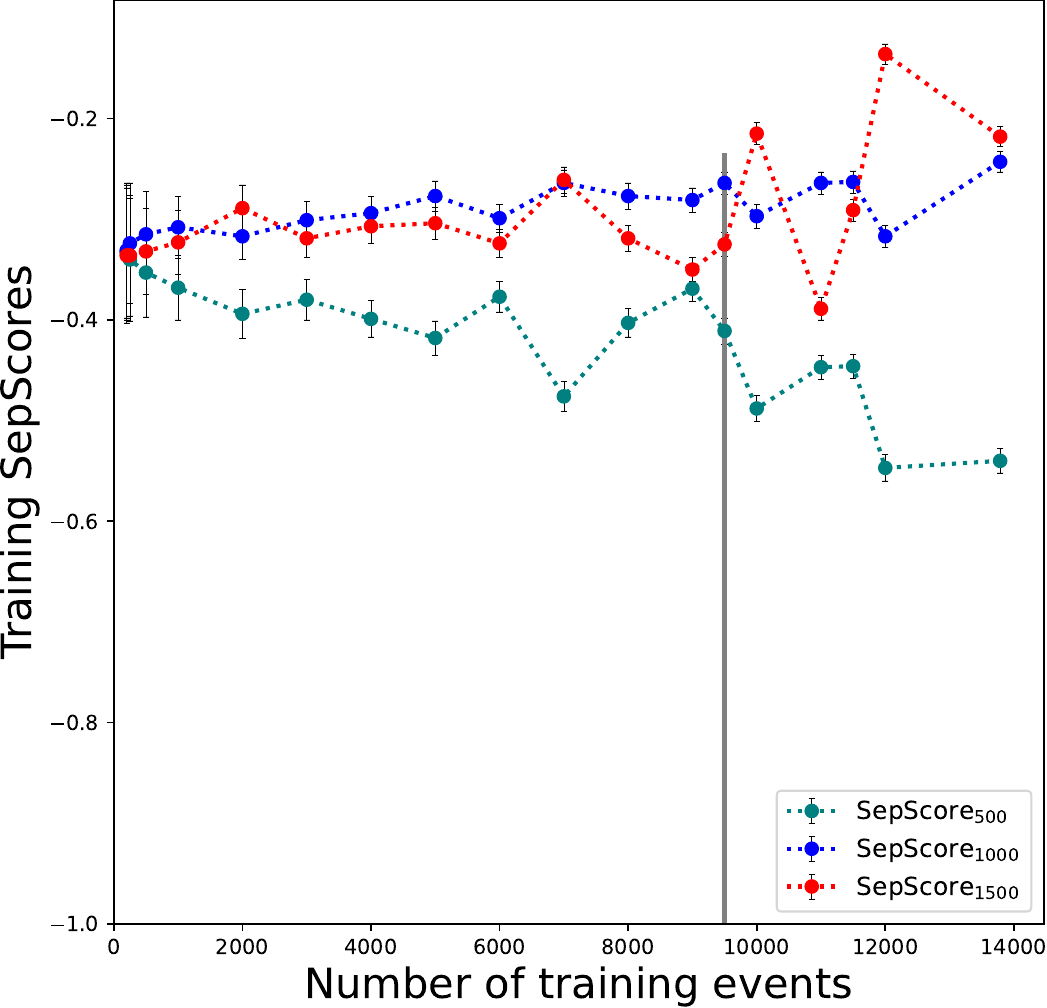}
    \label{subfig:(h1)}
  \end{subfigure}
  \caption[Variation of SOM hyperparameters]{Variation in separation scores with varying SOM hyperparameters: radius (left column),
    learning rate (middle column), and training sample size (right column) for $n=40$(top row) and $n=100$ (bottom row).}
  \label{fig:SomPara1}
\end{figure}

The performance of the SOM is reasonably stable against variations in the other training hyperparameters. Figure~\ref{fig:SomPara1}
shows the variation of training separation scores ($\mathrm{SepScore}_{500}$, $\mathrm{SepScore}_{1000}$, $\mathrm{SepScore}_{1500}$)
with varying radius, learning rate, and training sample size.
We see that the performance remains approximately constant with number of training events above a $1000$.

\begin{figure}[h!]
  \centering
  \includegraphics[width=0.40\textwidth]{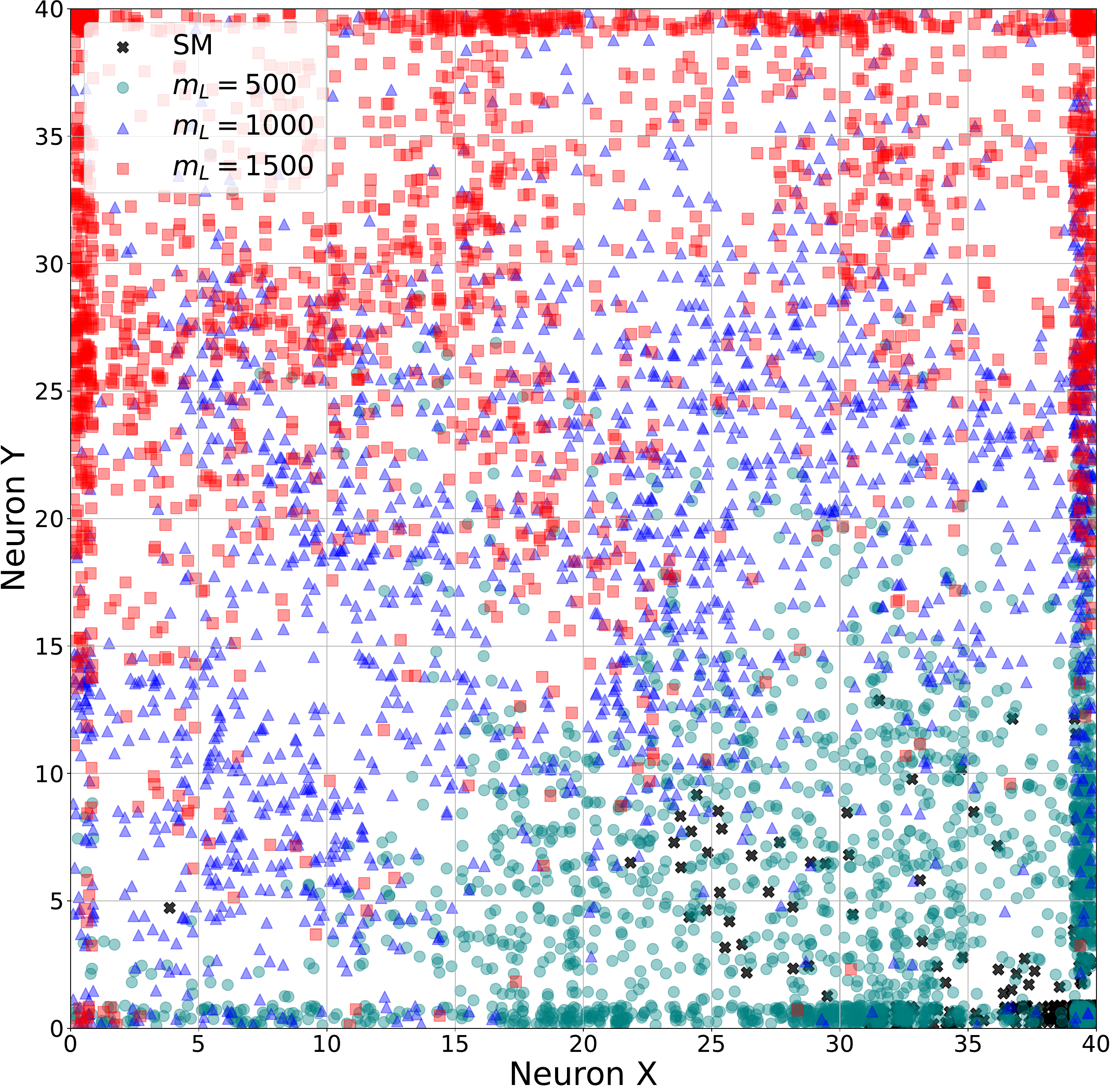}
  \includegraphics[width=0.40\textwidth]{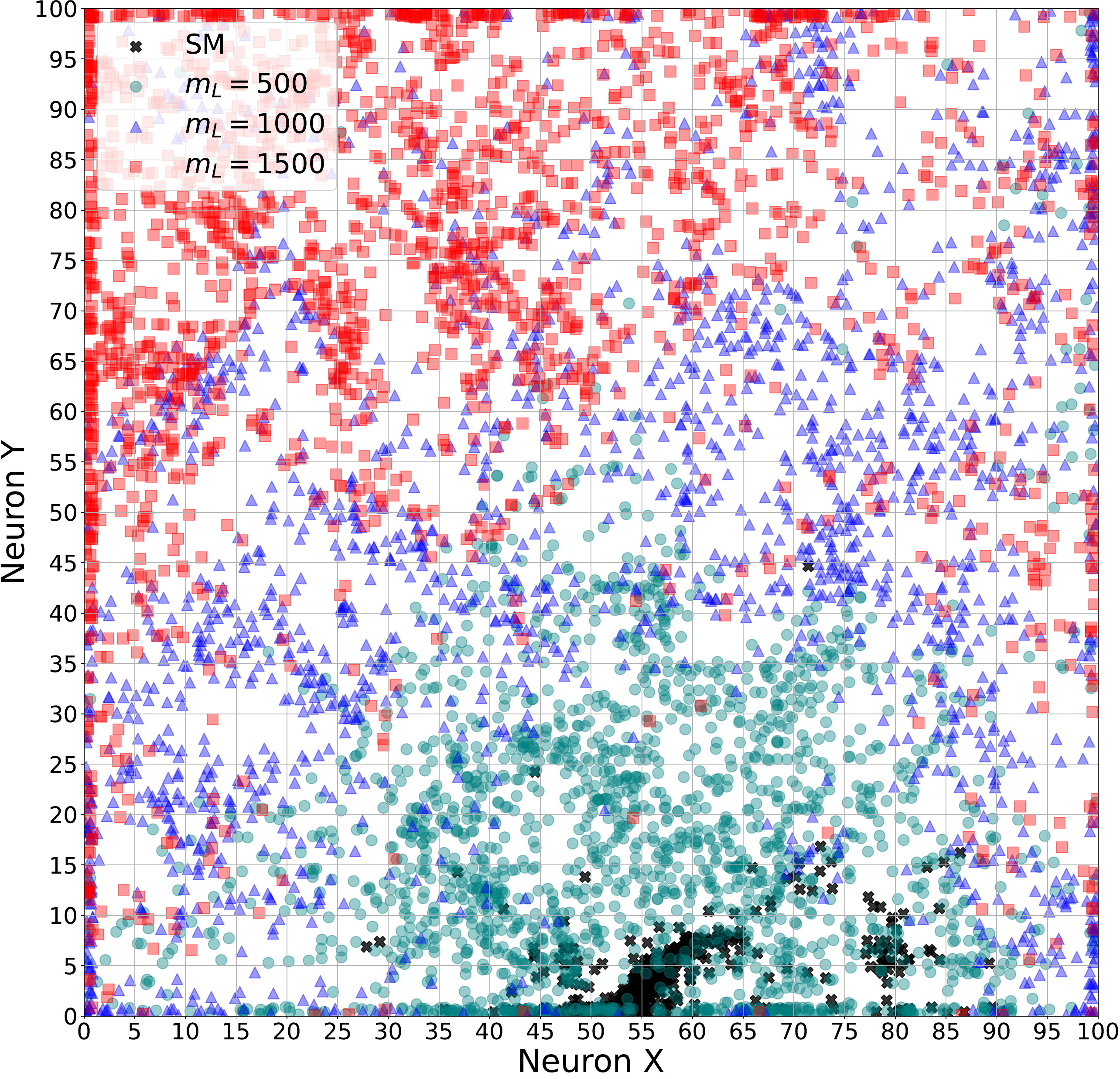}
  \caption[Testing SOM model.]{The distribution of BMUs for the testing dataset for SOM models with $n=40$ (left) and $n=100$ (right).}
  \label{fig:somtest}
\end{figure}

The first test involves testing the SOM model with a set of events independent of the training.
We construct a testing dataset consisting of $2000$ events each of $m_L=500,1000,1500\gev$, and $1000$ events each of $WZ$ and
$t\bar{t}Z$. The $WZ$ and $t\bar{t}Z$ events are collectively labelled as SM in the figures. 
The BMUs are calculated for the testing dataset using the weights of the trained SOM, and
the distribution of these events in the SOM grid are compared to that from the training events. Figure~\ref{fig:somtest} shows
the distribution of these BMUs for the various processes for $n=40$ and $n=100$ SOMs. These can be compared to Figs.~\ref{subfig:(d)} and
\ref{subfig:(g)}.
We find that the events of the testing dataset populate the same regions as that of the training set, indicating that the training
has happened in an optimal way. We also see that the SM processes cluster in a specific part of the SOM.
The trained SOM model populated with the testing dataset is now designated as the testSOM and is used in the subsequent studies.


\subsection{\label{sec:testSOM} Individual cases using a SOM}

We now consider the various cases from Section~\ref{sec:setup}. For each case, we use a defined procedure involving the testSOM
to infer the mass hypothesis. We describe the procedure for the first case in detail. Here as well we perform a $100$ independent
trials for each case, and report on the performance of these trials.

~\\ \noindent {\bf Case 1:} The experiment has observed $10$ events (each event corresponds to $m_L=1000\gev$).

The BMUs for the $10$ events are calculated using the trained SOM weights. 
For each of these BMUs, we now consider an $m\times m$ region, that is centred on the BMU.
We calculate a regional separation score for each of the $10$ events, using the testSOM. The regional separation score is defined
along the same lines as the ${\mathrm{SepScore}}$, but here including the SM processes:
\begin{align} \mathrm{score^{reg}_{500}}^i &= {\frac{{\mathrm{N_{500} - N_{1000} - N_{1500} - N_{SM} }}}{{\mathrm{N_{500} + N_{1000} + N_{1500} + N_{SM} }}}} \\
  \mathrm{SepScore^{reg}_{500}} &= \frac{\Sigma_i {\mathrm{score_{500}^{i}}}}{{\mathrm{N_{non-empty}}}}
\end{align}
where $i$ is one of the $m^2$ neurons, and similarly for $\mathrm{SepScore^{reg}_{1000}}$, $\mathrm{SepScore^{reg}_{1500}}$, and $\mathrm{SepScore^{reg}_{SM}}$. Here
$\mathrm{N_{non-empty}}$ is the number of neurons in that $m\times m$ region populated by at least one event of any class. Figure~\ref{fig:TestSOM} (right)
shows the BMUs for the $10$ events. For two BMUs, as an illustration, a region with $m=5$ (labelled as $1$) and a region with $m=11$ (labelled as $2$) are also
shown. These regions are mapped on the testSOM as shown in Fig.~\ref{fig:TestSOM} (left) and the $\mathrm{SepScore^{reg}}$ are calculated.

\begin{figure}[h!]
  \centering
  \includegraphics[width=0.8\textwidth]{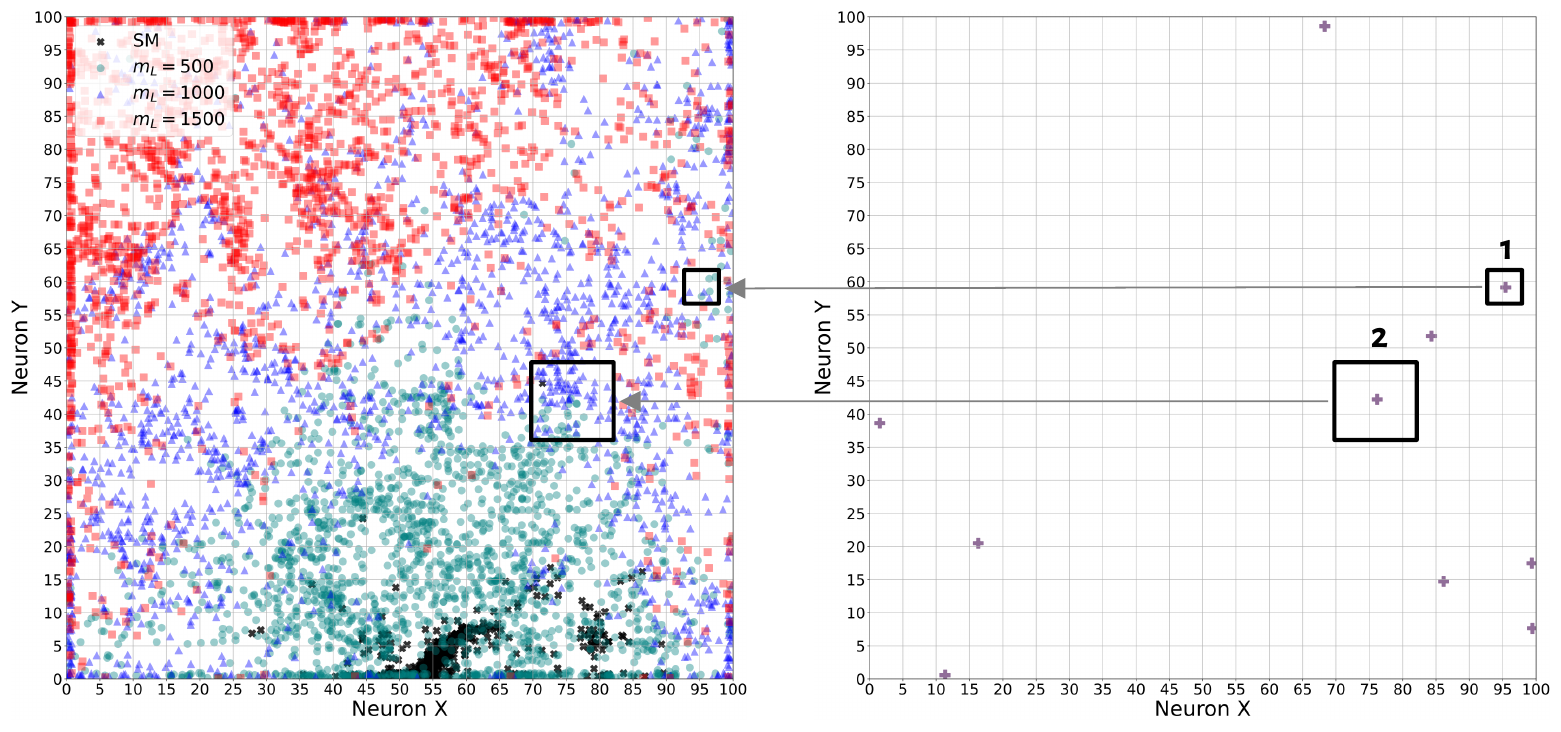}
  \caption[Testing SOM model with signal and background]{The testSOM (left) and the BMUs for the $10$ events of case 1 (right). For two BMUs, a region
    of size $m=5$, and one of size $m=11$ is shown as (1) and (2) respectively.}
  \label{fig:TestSOM}
\end{figure}

We will thus obtain $10$ separate $\mathrm{SepScore^{reg}_{500}}$ (and so on) corresponding to each of the $10$ events. 
Figure~\ref{fig:case1_hists} shows the histogram of the four regional separation scores overlaid on each other for two choices
of regions: $m=5$, and $m=11$. We consider the median of each distribution as we did in the DNN case, and we see that
the regional separation score corresponding to $m_L=1000\gev$ ($\mathrm{SepScore^{reg}_{1000}}$)
shows the highest median value. Based on these histograms, we would thus conclude that the observed excess corresponds to $m_L=1000\gev$.
In a $100$ independent trials, we obtain $m_L=1000\gev$ in $95$ trials for $m=11$, and in $91$ trials for $m=5$.

\begin{figure}[h!]
  \centering
  \includegraphics[width=0.40\textwidth]{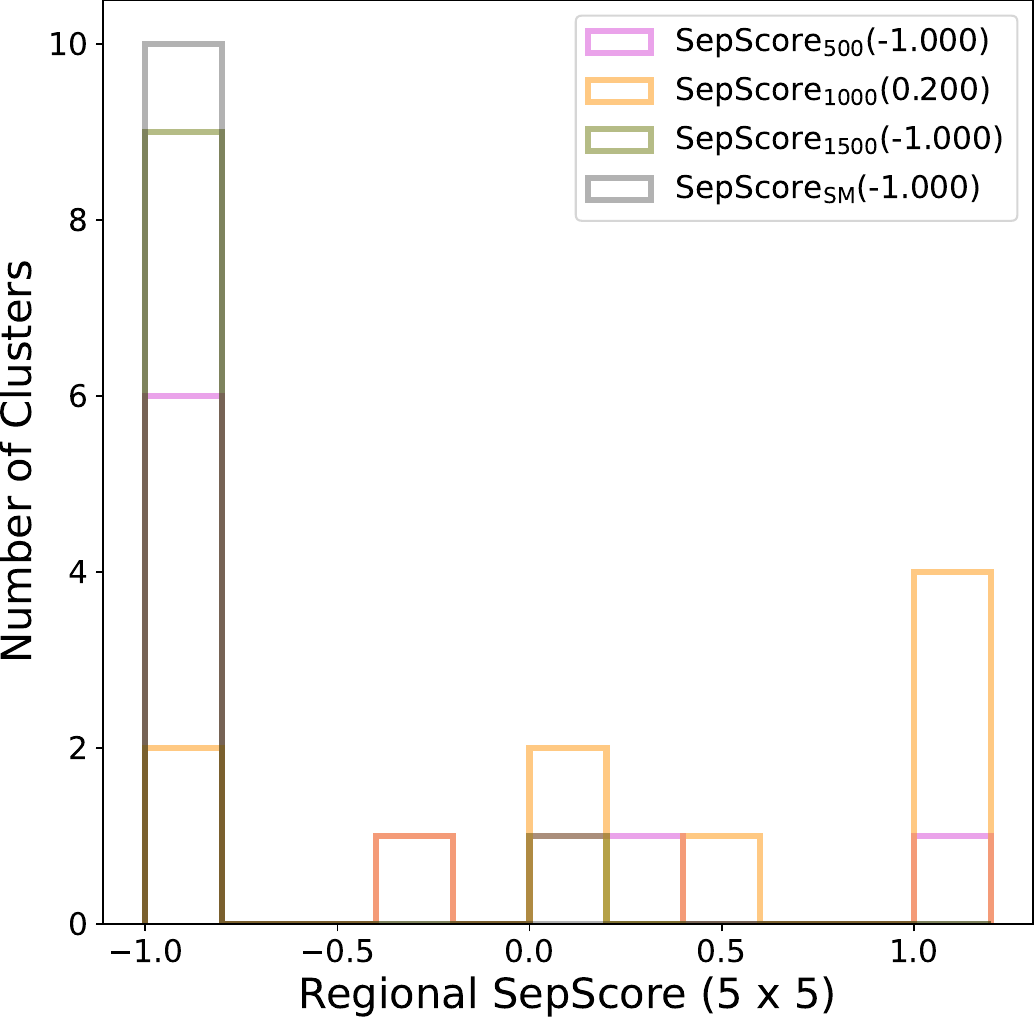}
  \includegraphics[width=0.40\textwidth]{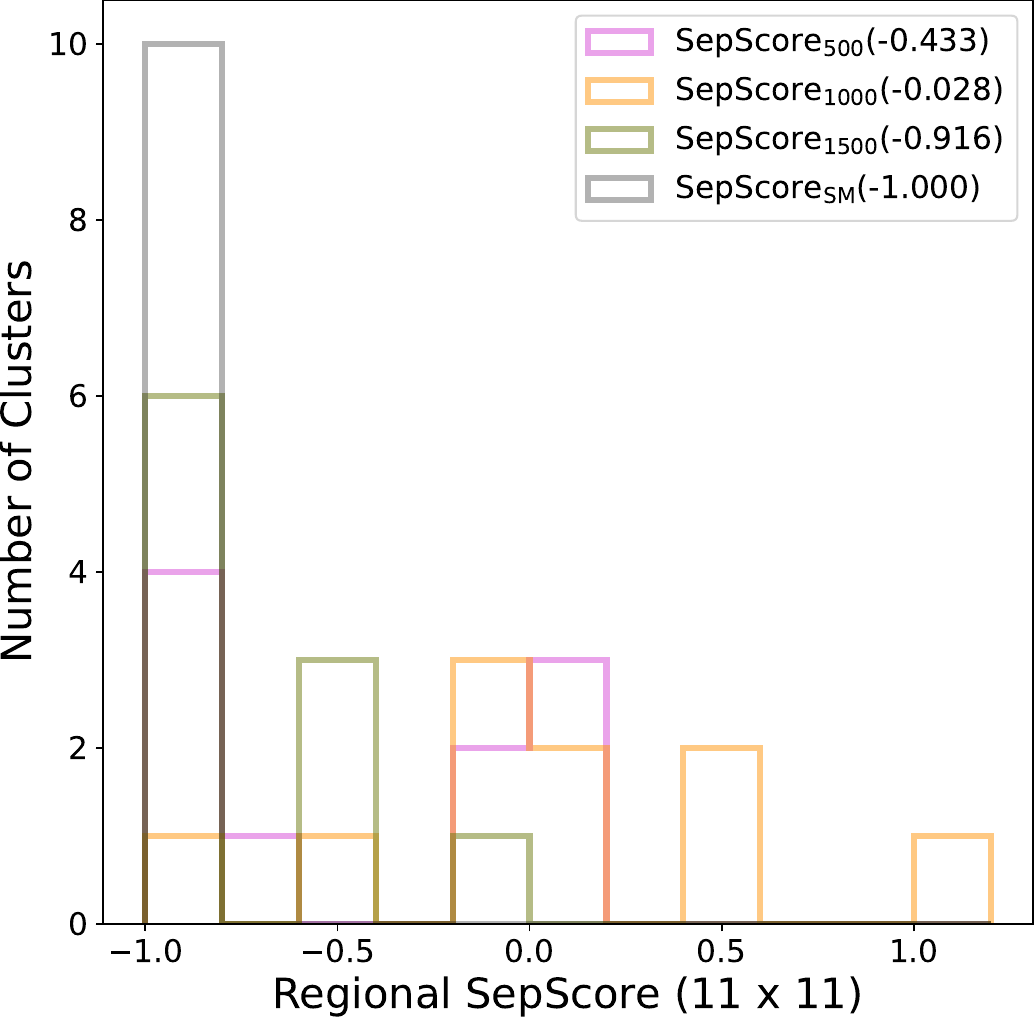}
  \caption[Testing Case 1: Histogram of regional separation scores]
          {The regional separation scores in a region of $m=5$ (left) and
            $m=11$ (right) for the BMUs of case 1. }
  \label{fig:case1_hists}
\end{figure}

~\\ \noindent {\bf Case 2:} The experiment has observed $10$ events (each event corresponds to $m_L=2500\gev$).

As in the previous case, we calculate the BMUs for the $10$ events using the trained SOM. We then define
regions with $m=5$ and $m=11$ around each of the BMUs of the $10$ events, and use the testSOM to calculate regional separation scores. Figure~\ref{fig:case2_hists}
shows the BMUs for the $10$ events, and the histograms of these regional separation scores. The highest median score is obtained for
$\mathrm{SepScore^{reg}_{1500}}$. Thus from the histograms, we would conclude that the observed excess corresponds to
$m_L=1500\gev$.
In a $100$ independent trials, we obtain $m_L=1500\gev$ in all trials for $m=11$, and in $99$ trials for $m=5$.
This erroneous conclusion is similar to that obtained with the DNNs and is anticipated.
The events corresponding to $m_L=2500\gev$ look most like the events of $m_L=1500\gev$.

\begin{figure}[h!]
  \centering
  \includegraphics[width=0.30\textwidth]{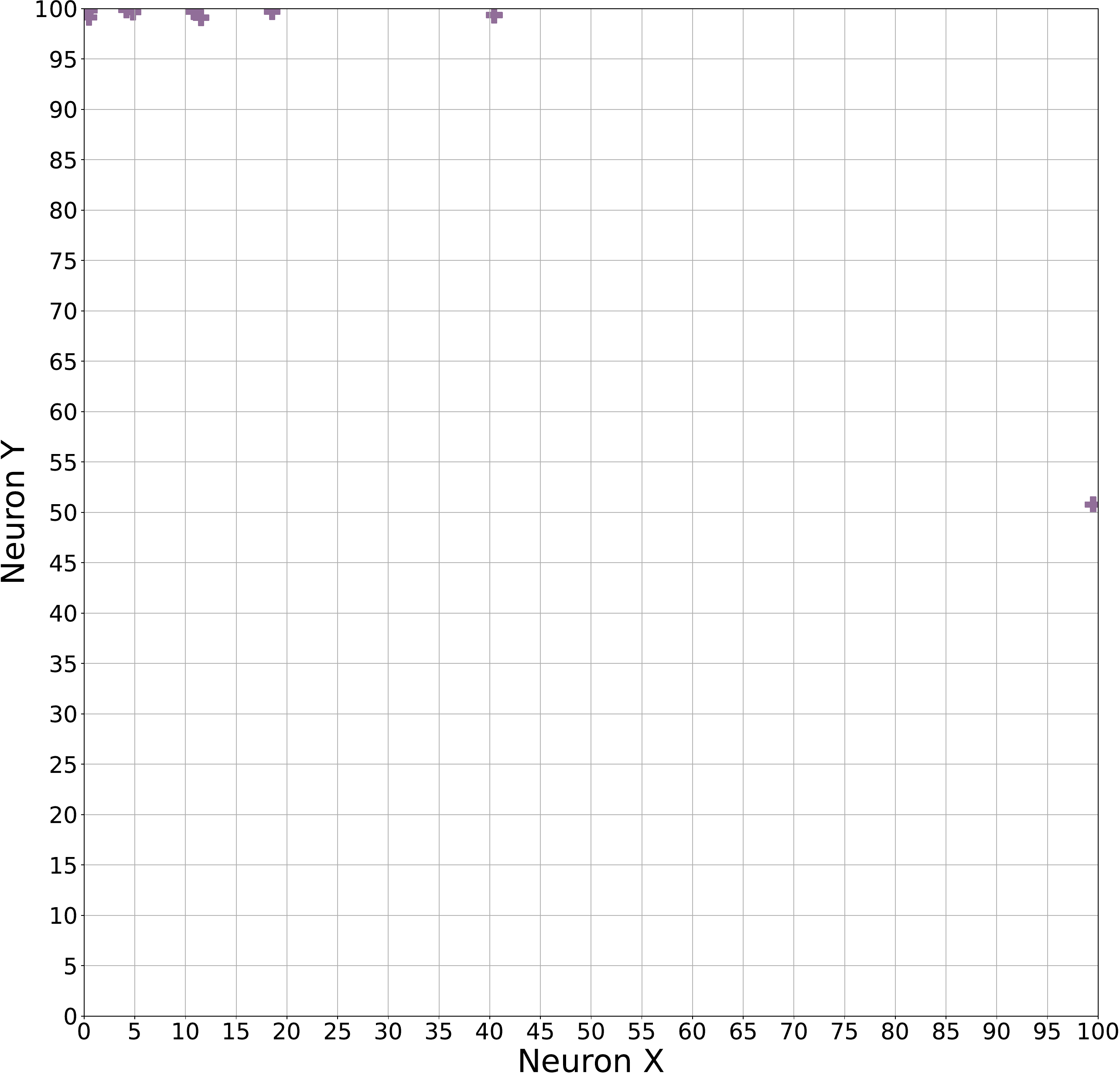}
  \includegraphics[width=0.30\textwidth]{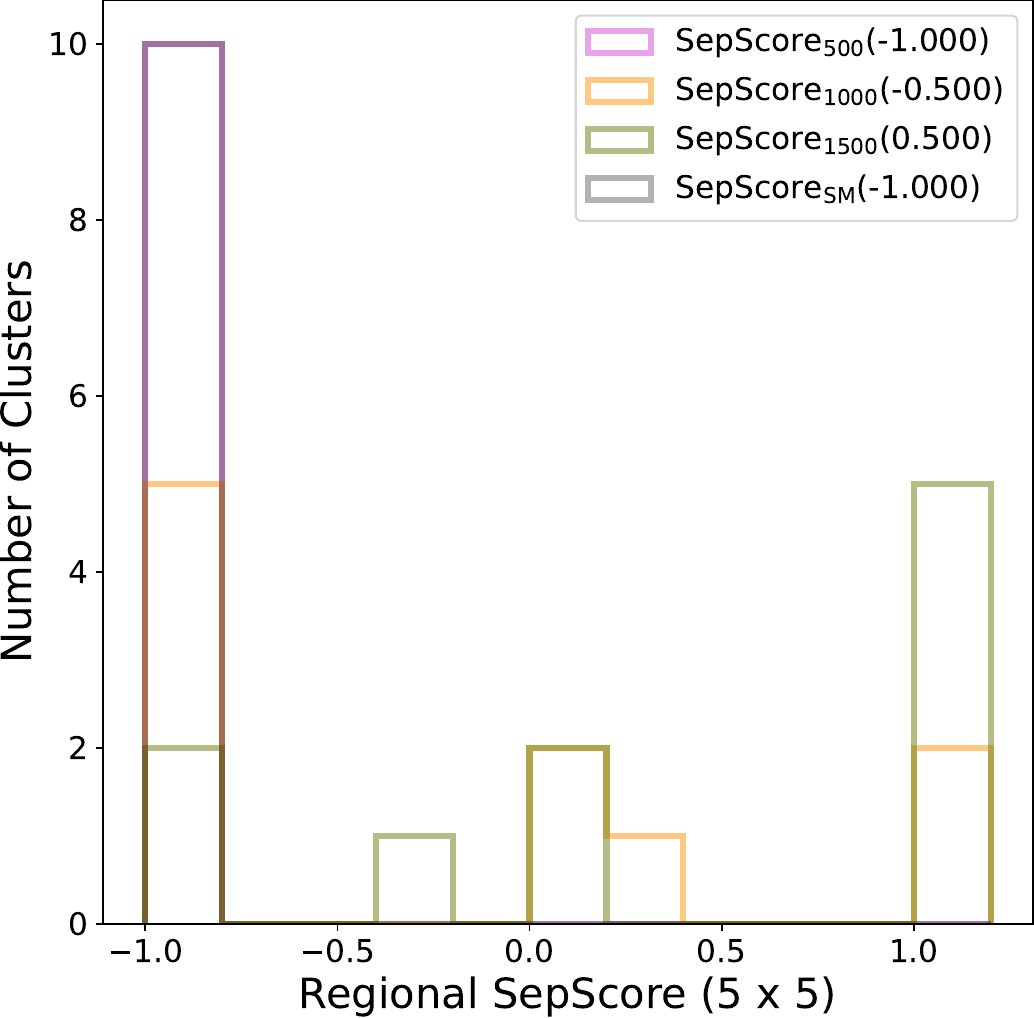}
   \includegraphics[width=0.30\textwidth]{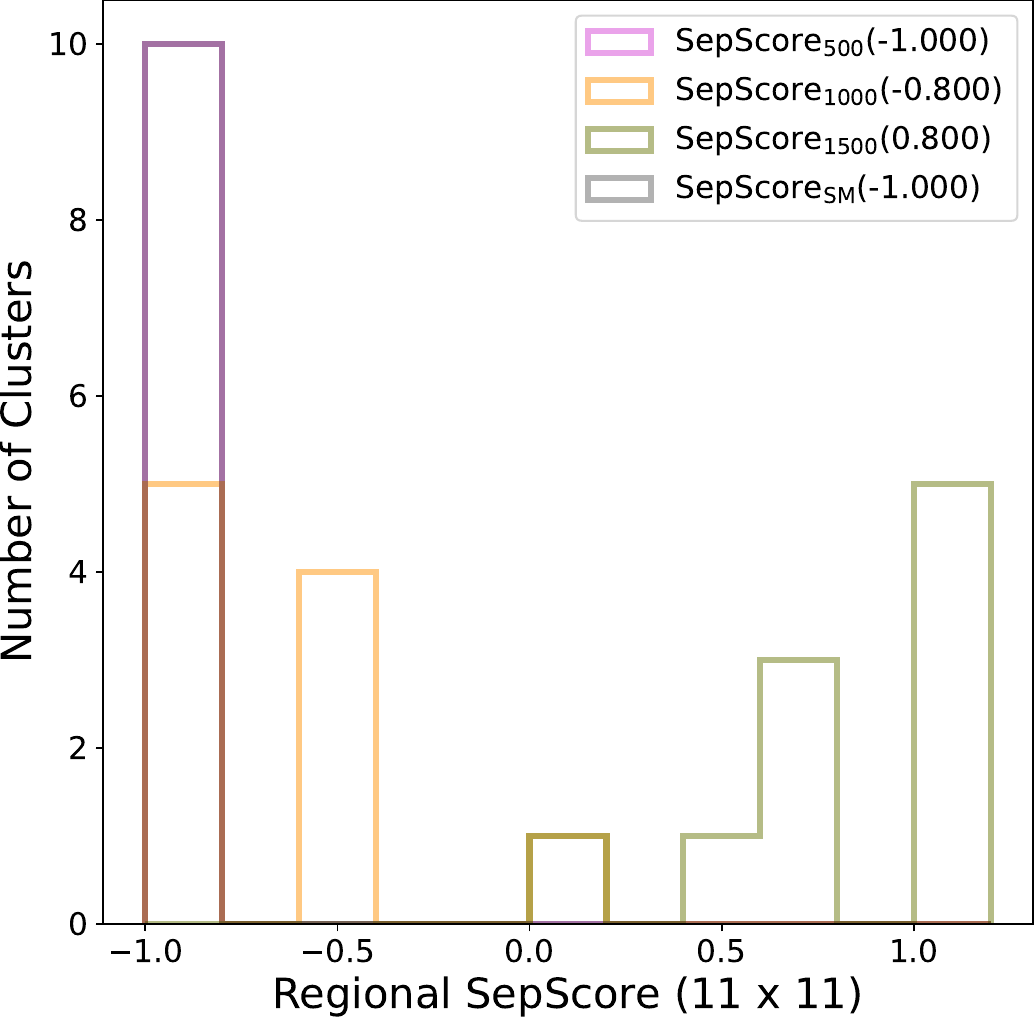}
   \caption[Testing Case 2]{The distribution of BMUs (left) for the $10$ events of case 2, and the regional separation scores for
     $m=5$ (middle) and $m=11$ (right) regions for the BMUs of case 2.}
  \label{fig:case2_hists}
\end{figure}

~\\ \noindent {\bf Case 3:} The experiment has observed $20$ events, with an expected background of $10$ events. (The composition of the observed events is $10$ from the SM
processes, and $10$ from the $m_L=500\gev$ process.)

As in the previous case, we set $m=5$ to define regions around the BMUs of each of these $20$ events, and obtain
the regional separation scores. From the histograms, we see that several events have a large $\mathrm{SepScore^{reg}_{SM}}$, and are
thus most likely events from SM processes. To eliminate the SM events, we select a subset of events that satisfy $\mathrm{SepScore^{reg}_{SM}}<0.6$.
This selection is expected to reject approximately 50\% of the SM events. This selection is less stringent than the DNN case since here the SOM
has not been trained on SM events, and an aggressive selection may not be justified.
The regional
separation scores for the $12$ surviving events can now be considered. Figure~\ref{fig:case3_hists} shows the BMUs, and the regional
separation scores before and after the requirement on $\mathrm{SepScore^{reg}_{SM}}$. From the regional separation score distribution, we can clearly
conclude that the excess corresponds to the $m_L=500\gev$ hypothesis. 
In a $100$ independent trials, we obtain $m_L=500\gev$ in $96$ trials for $m=11$, and in $99$ trials for $m=5$.

\begin{figure}[h!]
  \centering
  \includegraphics[width=0.30\textwidth]{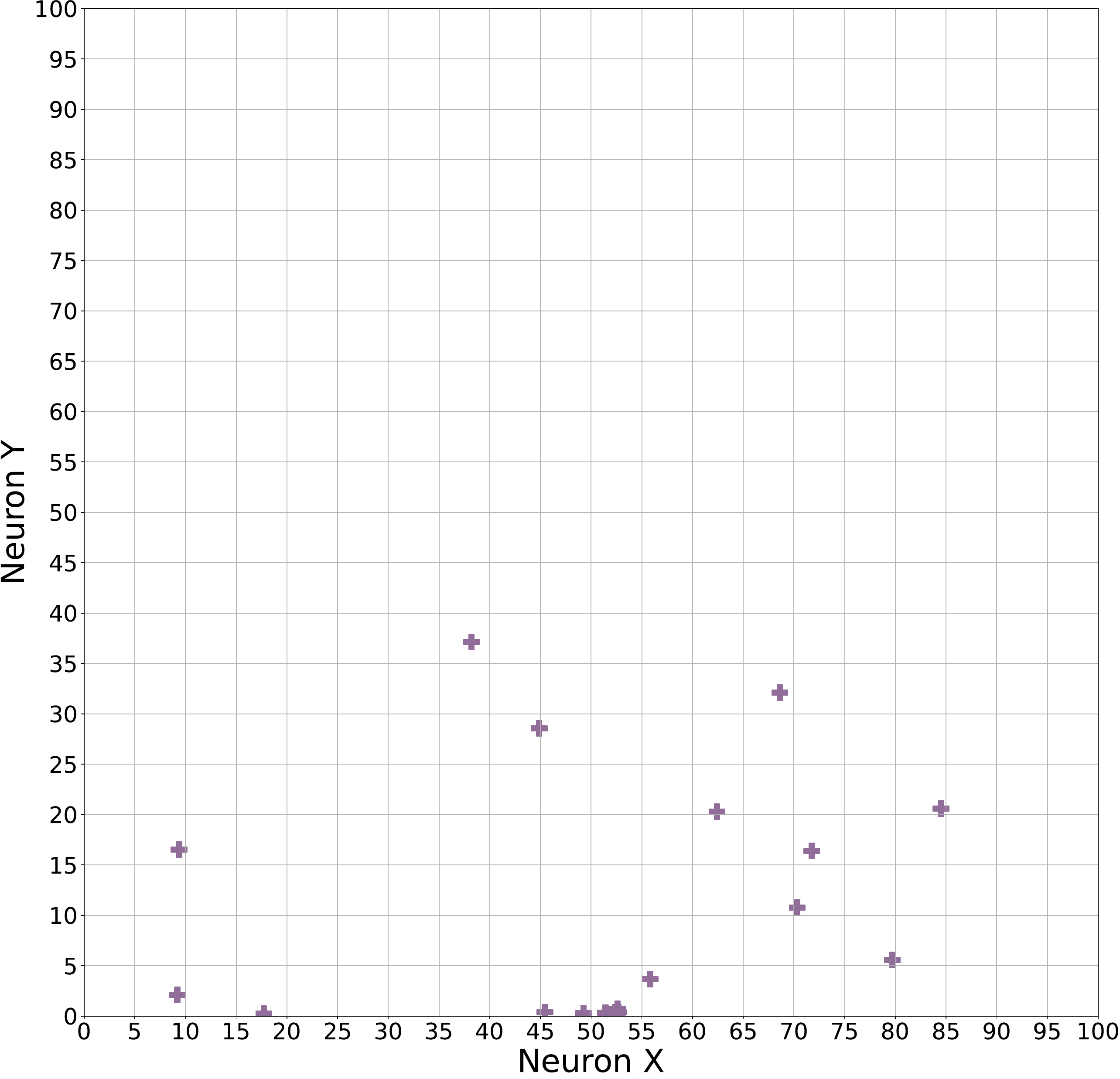}
  \includegraphics[width=0.30\textwidth]{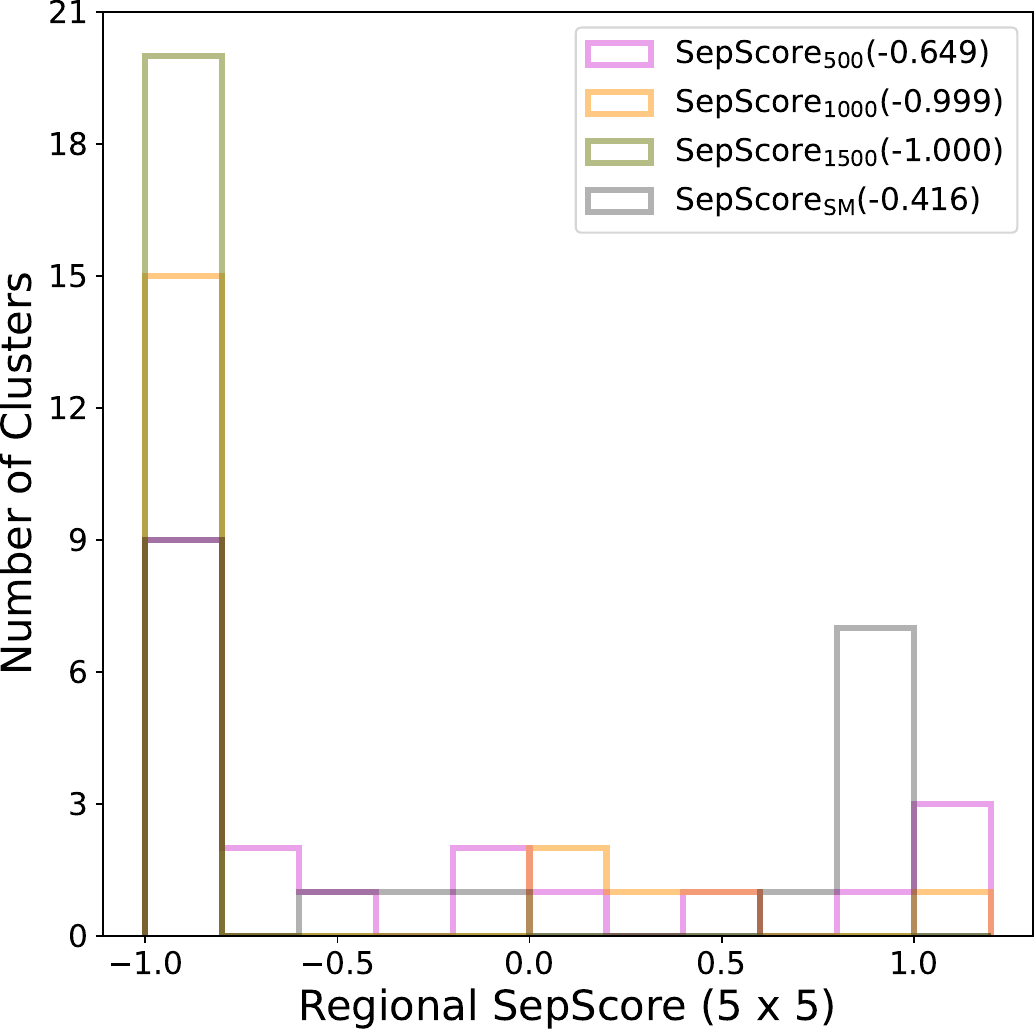}
  \includegraphics[width=0.30\textwidth]{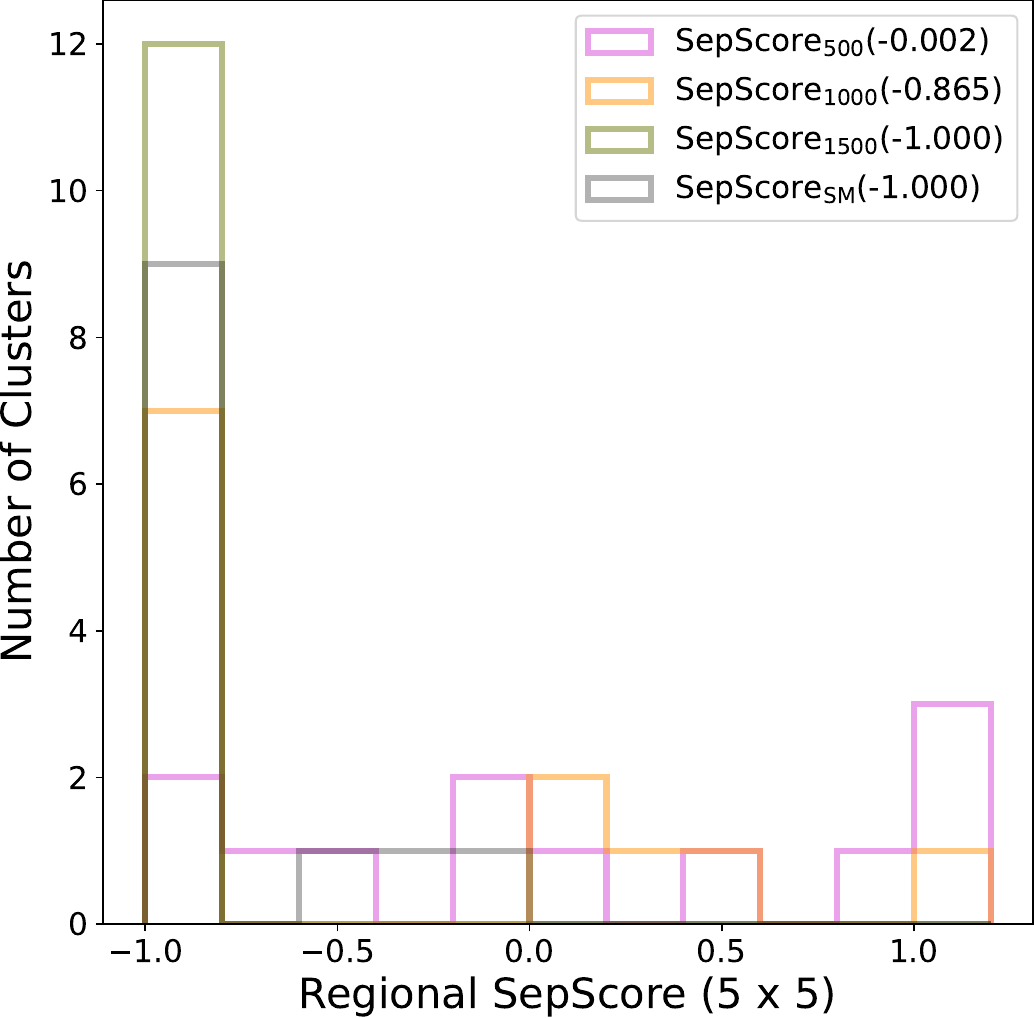}
  \caption[Testing Case 3]{The BMUs (left) for the $20$ events of case 3, and the regional separation scores for $m=5$
  around each BMU without (middle) and with (right) a cut of  $\mathrm{SepScore^{reg}_{SM}}<0.6$.}
  \label{fig:case3_hists}
\end{figure}

~\\ \noindent {\bf Case 4:} The experiment has observed $15$ events, with an expected background of $5$ events.
(The composition of the observed events is $5$ from the SM
processes, and $10$ from the $m_L=750\gev$ process.)

Figure~\ref{fig:case4_hists} shows the regional separation scores in $m=5$ regions around the BMUs of each of the $15$ events.
As in the previous case, we require that events satisfy $\mathrm{SepScore^{reg}_{SM}}<0.6$, and the figure also shows the separation scores
for the surviving $13$ events. We see that in this case, the distribution of $\mathrm{SepScore^{reg}_{1000}}$ shows the highest
median, and would lead us to conclude that the excess corresponds to $m_L=1000$.
In a $100$ independent trials, we obtain $m_L=500\gev$ in $57$ trials and $m_L=1000\gev$ in $43$ trials for $m=11$. For $m=5$, these
numbers are $67$ trials and $33$ trials respectively. As with the DNN, here as well it is possible to rule out the $m_L=1500\gev$ hypothesis in
all trials.

\begin{figure}[h!]
  \centering
  \includegraphics[width=0.30\textwidth]{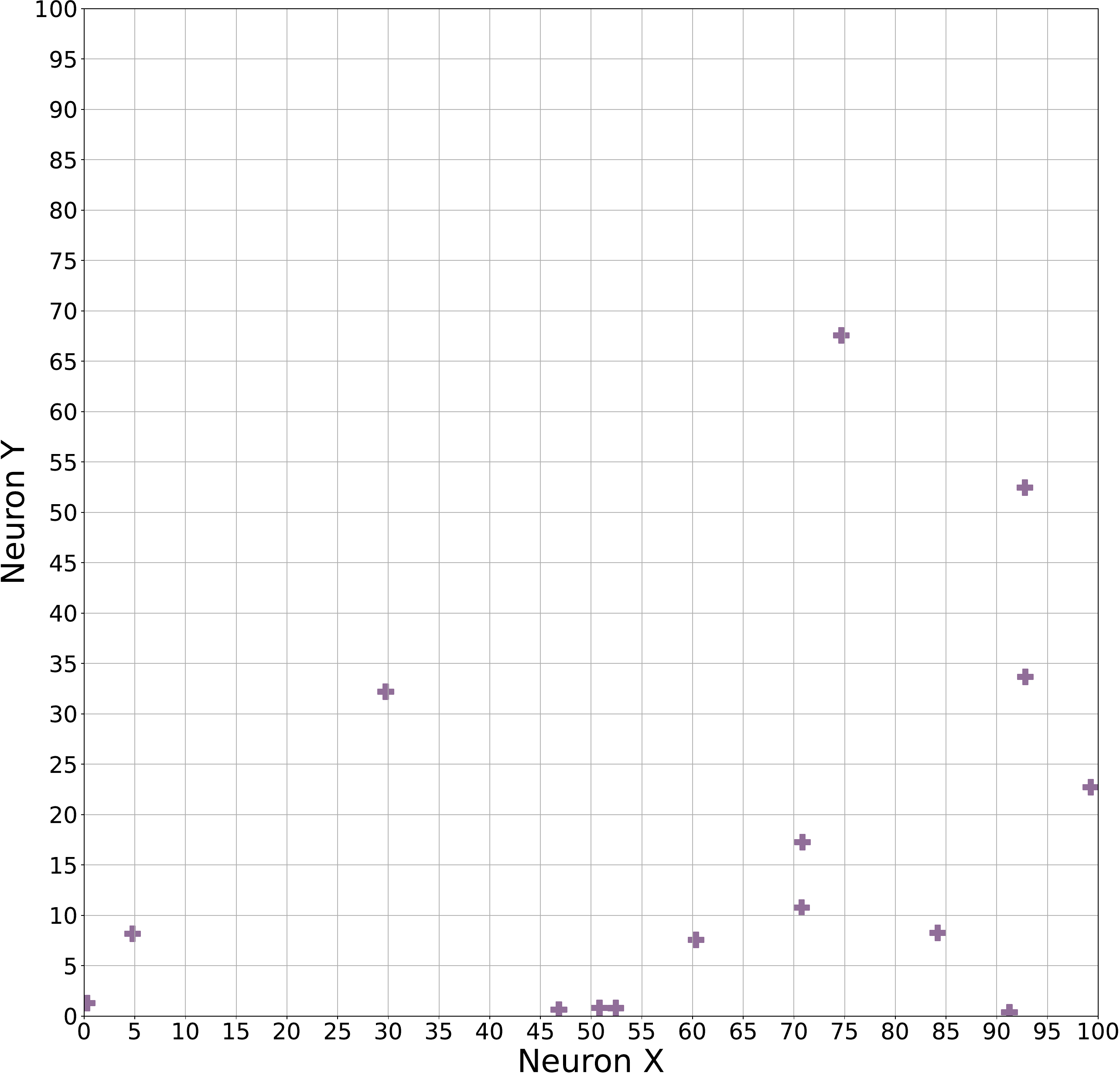}  
   \includegraphics[width=0.30\textwidth]{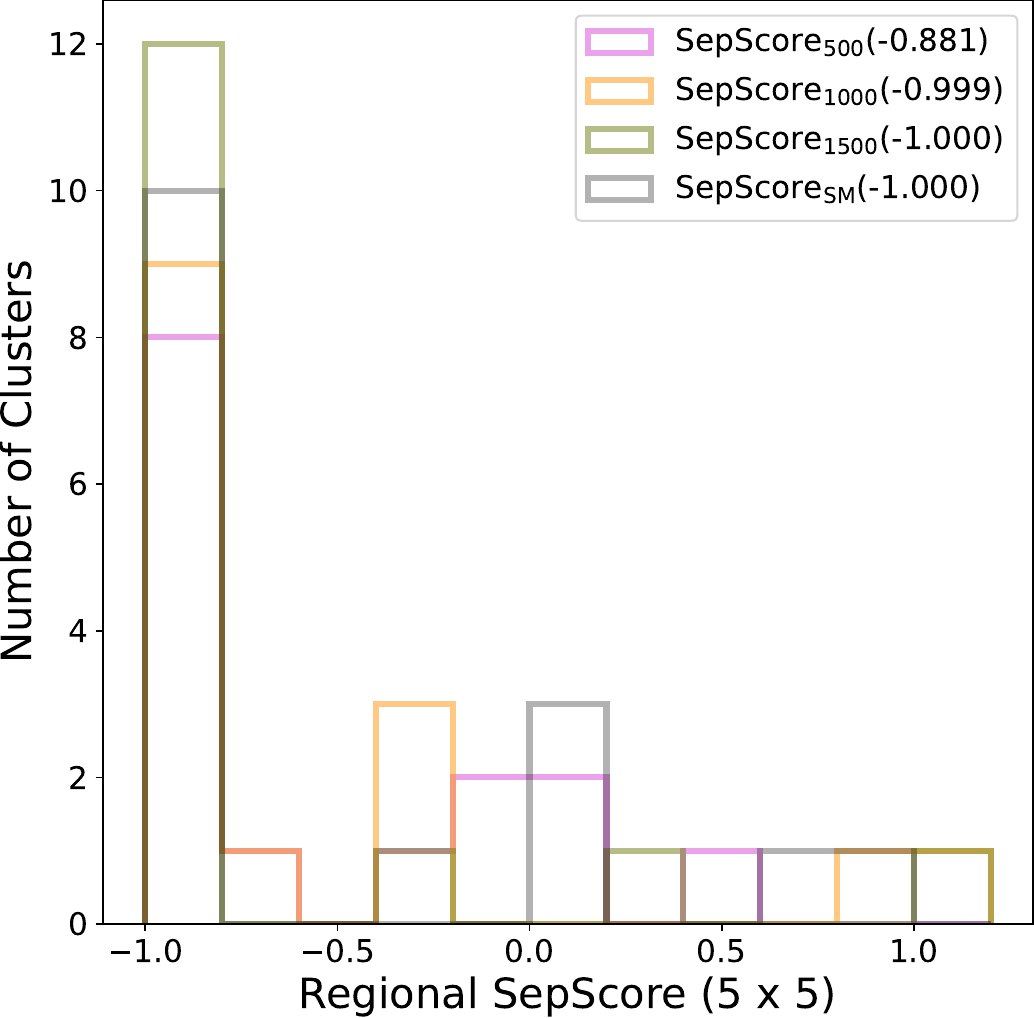}
  \includegraphics[width=0.30\textwidth]{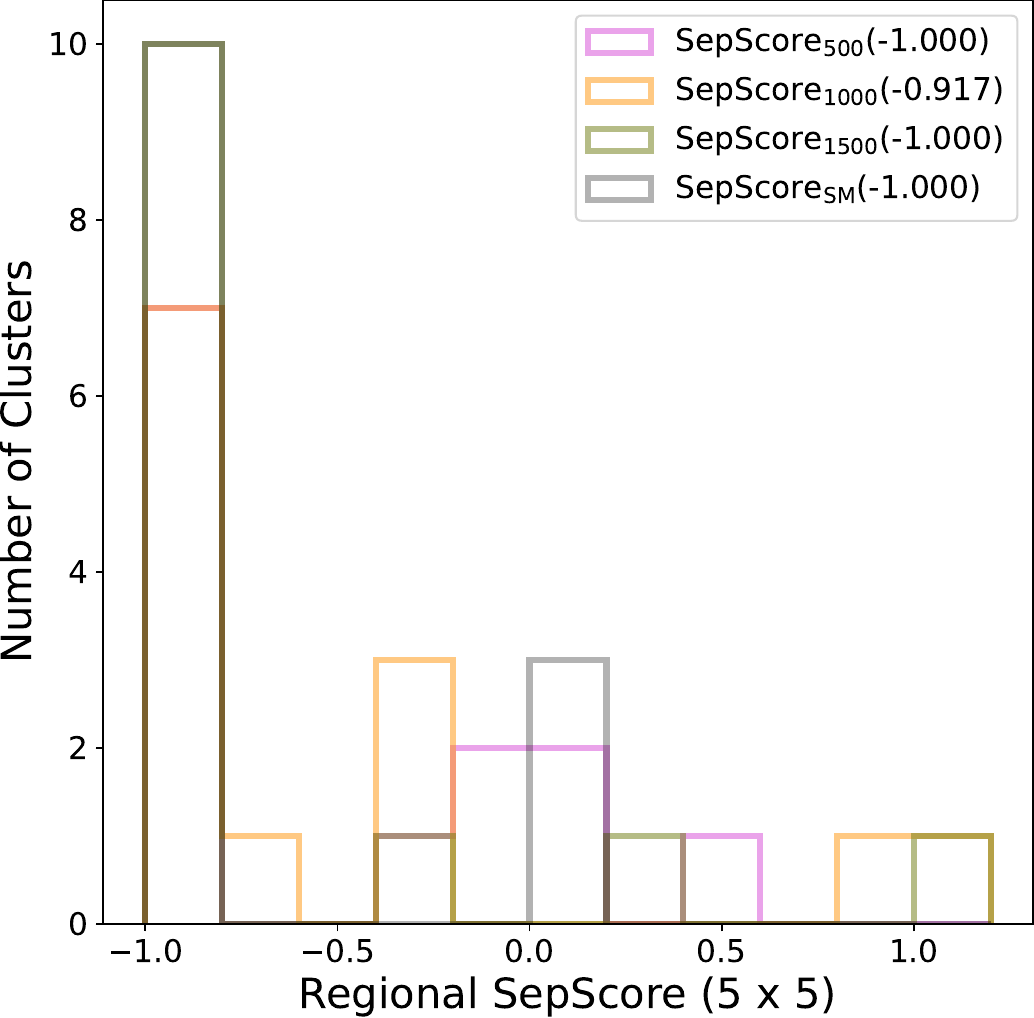}
  \caption[Testing Case 4]{The BMUs (left) for the $15$ events of case 4, and the regional separation scores for $m=5$
  around each BMU without (middle) and with (right) a cut of  $\mathrm{SepScore^{reg}_{SM}}<0.6$.}
  \label{fig:case4_hists}
\end{figure}

~\\ \noindent {\bf Case 5:} The experiment has observed $15$ events, with an expected background of $5$ events.
(The composition of the observed events is $5$ from the SM
  processes, and $10$ from the $m_L=600\gev$ process.)
  
Figure~\ref{fig:case5_hists} shows the regional separation scores in $m=5$ regions around the BMUs of each of the $15$ events.
The figure shows the separation scores for all events as well as the events satisfying $\mathrm{SepScore^{reg}_{SM}}<0.6$. The
distribution of $\mathrm{SepScore^{reg}_{500}}$ shows the highest median. One would thus conclude the excess events to be
from $m_L=500\gev$, which is similar to the conclusion obtained using DNN.
In a $100$ independent trials, we obtain $m_L=500\gev$ in $96$ trials for $m=11$ and in all trials for $m=5$.

\begin{figure}[h!]
  \centering
  \includegraphics[width=0.30\textwidth]{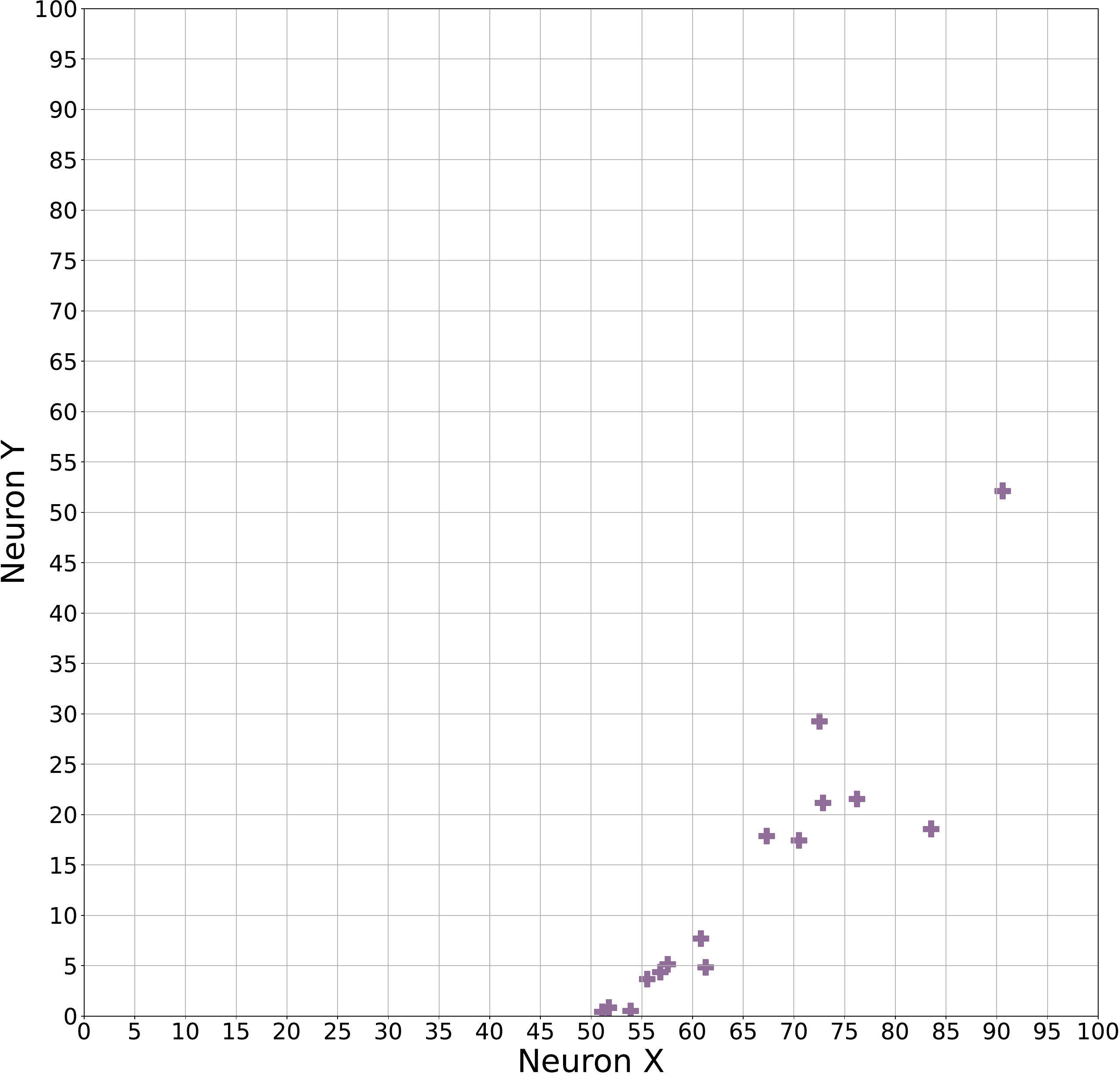}  
   \includegraphics[width=0.30\textwidth]{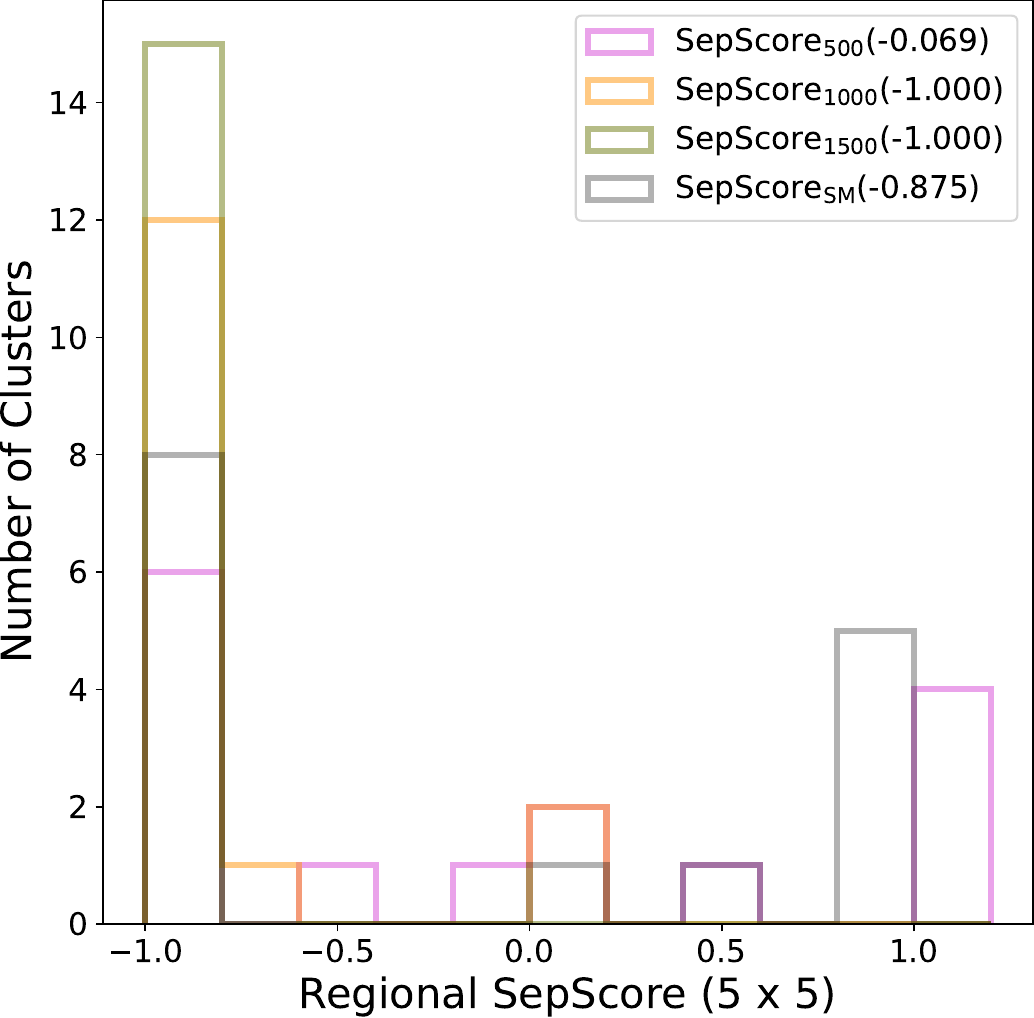}
   \includegraphics[width=0.30\textwidth]{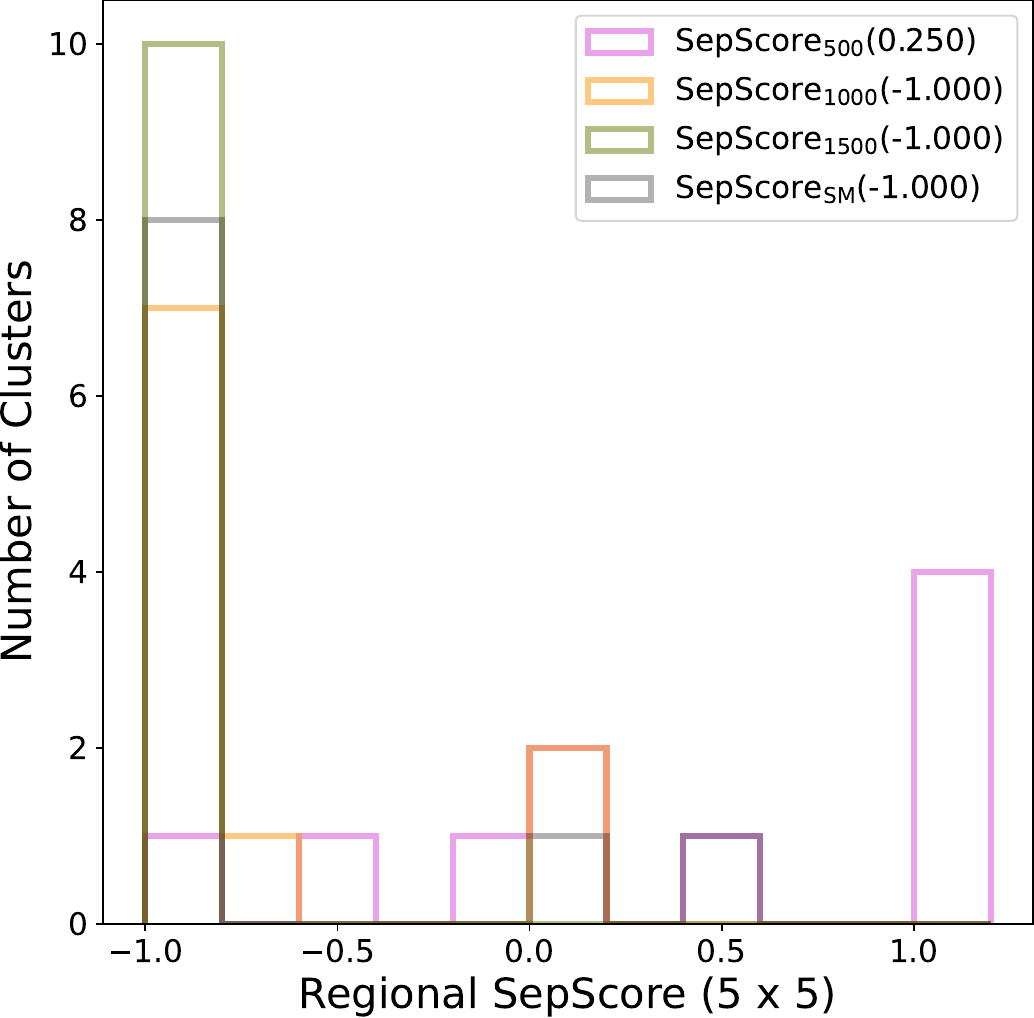}
  \caption[Testing Case 5]{The BMUs (left) for the $15$ events of case 5, and the regional separation scores for $m=5$
  around each BMU without (middle) and with (right) a cut of  $\mathrm{SepScore^{reg}_{SM}}<0.6$.}
  \label{fig:case5_hists}
\end{figure}

\subsection{Summarizing the cases}

We find that in cases where the excess corresponds to a mass hypothesis that we have trained on, the procedure very readily
allows us to unambigiously identify the hypothesis that the excess corresponds to. In case 1 and case 3, we show that irrespective
of the presence of SM processes, we can identify the events corresponding to the signal, and the $m_L$ value. In case 2, we see
that our procedure identifies the excess as $m_L=1500\gev$, despite it actually being $m_L=2500\gev$.
In fact, we see that events for any mass $m_L>1500\gev$ will be identified in the same way. The proposed fix to this is straightforward.
One repeats the process by choosing a different range of the $m_L$ hypotheses to train the SOM. Doing the process iteratively will allow
the definite $m_L$ hypothesis to be correctly identified. This is straightforward in BSM searches, given that a grid of signal simulation
samples is usually already created for interpretation of the result. 
In case 4 and case 5, the SOM is confronted with two classes of events, neither of which it has been trained to separate.
We see however that we are still able to isolate the non-SM events.
In the three cases, case 2, 4 and 5, one can take the events that survive the requirement on $\mathrm{SepScore^{reg}_{SM}}$, and compare
the distribution of their kinematic properties with those of the training data.
Figure~\ref{fig:somsum} shows the $m_{\ell\ell\ell}$ distribution for case 2 (left), case 4 (middle) and
case 5 (right). For case 2, one can see that the observed events seem to arise from a $m_L$ hypothesis much larger than $1500\gev$.
For case 4, the ambiguity between $m_L=500$ and $1000\gev$ is seen here as well, while in case 5 the $m_{\ell\ell\ell}$ distribution favors $m_L=500\gev$.
These observations will prompt further scrutiny, and by training
on a different range of $m_L$, one could narrow down the final answer.

\begin{figure}[h!]
  \centering
  \includegraphics[width=0.3\textwidth]{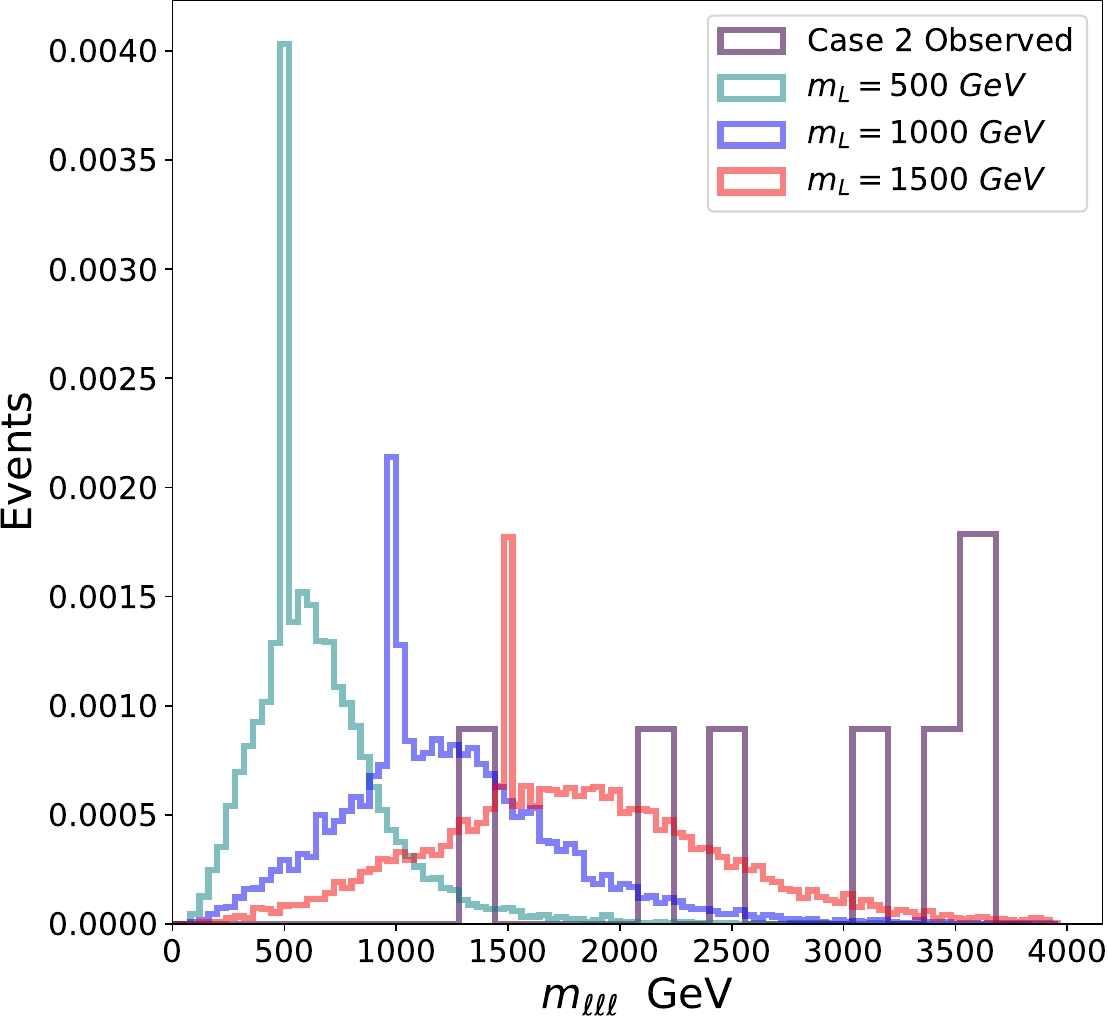}
  \includegraphics[width=0.3\textwidth]{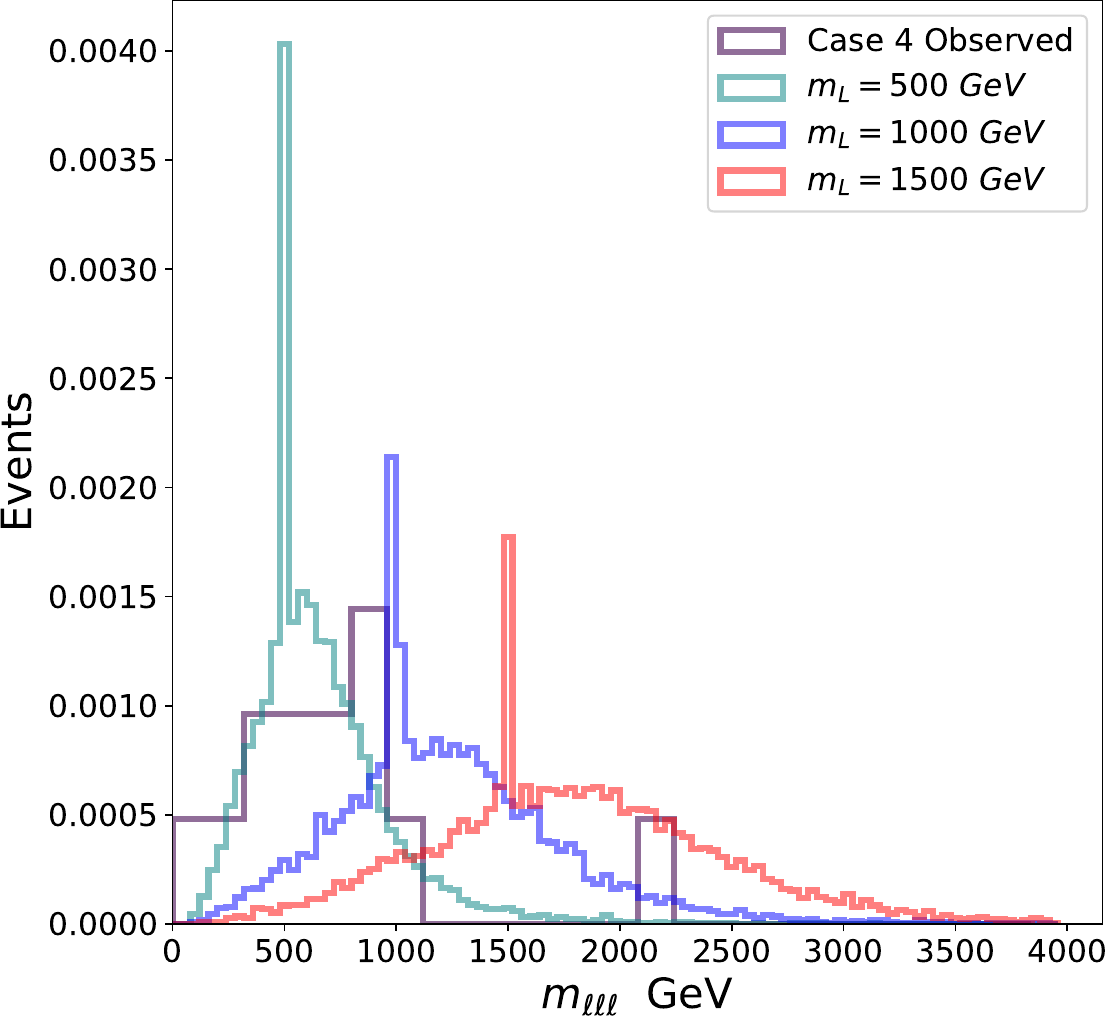}
  \includegraphics[width=0.3\textwidth]{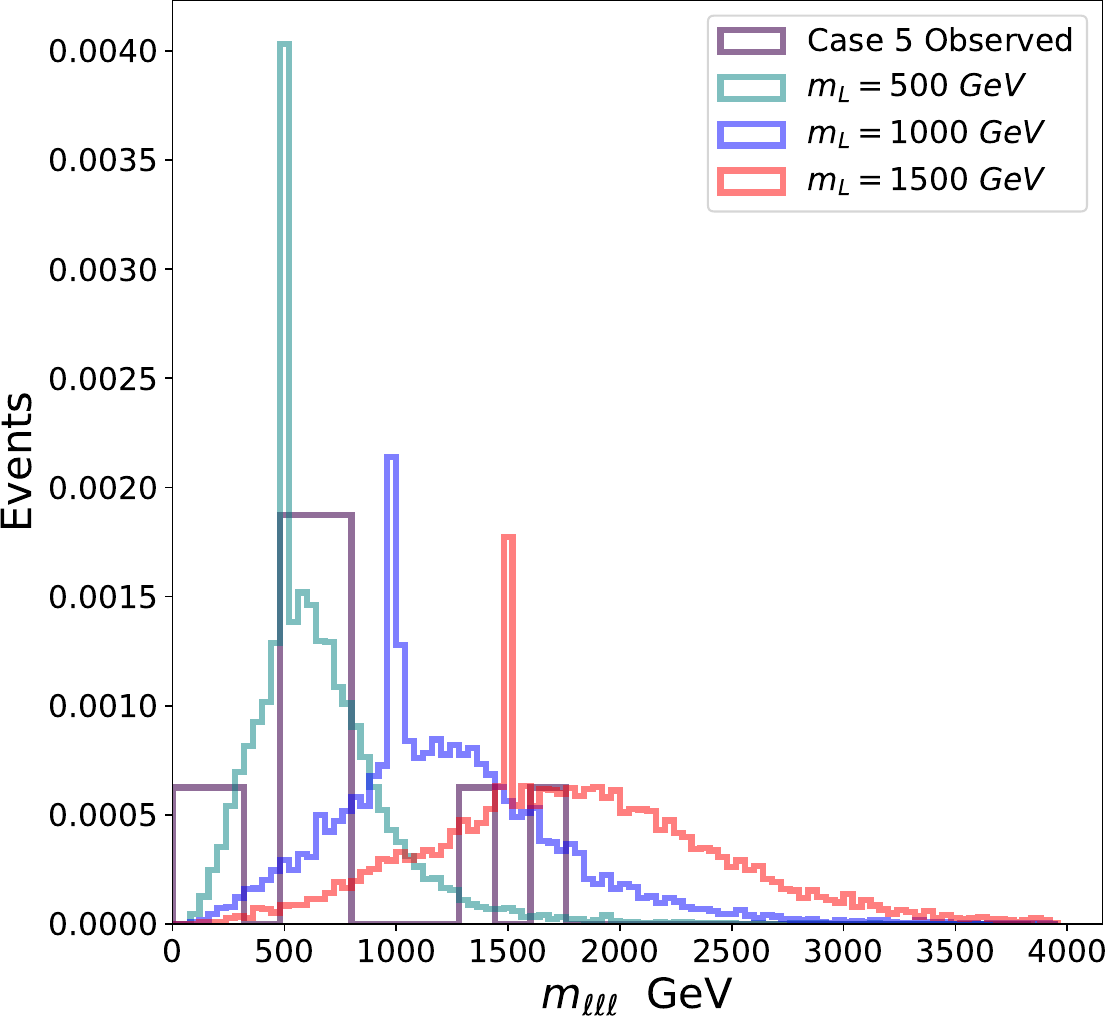}
  \caption[Summary mass comparison]{The $m_{\ell\ell\ell}$ distribution for the events of case 2 (left), case 4 (middle) and case 5 (right)
    are compared to the full distribution of the training
  dataset. In both cases, only the events satisfying $\mathrm{SepScore^{reg}_{SM}}<0.6$ are shown.}
  \label{fig:somsum}
\end{figure}


\subsection{Comparing the performance of SOM and DNN}\label{subsec:somdnnroc}

The definition of the regional separation scores, and the subsequent distributions can be exploited to define a ROC
curve for SOMs. To do this, we construct an independent dataset consisting of $500$ events each of
$m_L=500, 1000, 1500\gev$, $WZ$ and $t\bar{t}Z$ processes. We find the BMUs for these $2500$ events. For each of these,
as in the above cases, we can calculate a regional separation score in $m\times m$ regions using the testSOM. We perform this
procedure for each $m_L$ hypothesis. For example, take the $500$ events of $m_L=1000$ and calculate $\mathrm{SepScore^{reg}_{1000}}$.
Now take the other $2000$ events and calculate the same for them. The former distribution is treated as `signal'
and latter as `background' to calculate an ROC curve.
We show these ROC curves for the SOM model with $n=100$ and $m=11$ in Fig.~\ref{fig:somroc}, where
we also show the curves from the DNN. Here as well the uncertainty on the ROC curve is estimated using the bootstrapping
method, similar to the DNNs. To make a comparison with a much smaller set of events, we also define an `accuracy'. We
define accuracy to be the number of times {\it our procedure} is able to identify the correct answer from a $100$
independent trials using $10$ randomly sampled events. The accuracies for the various algorithms are shown in Table~\ref{tab:accs}.

\begin{figure}[h!]
  \centering
   \includegraphics[width=0.45\textwidth]{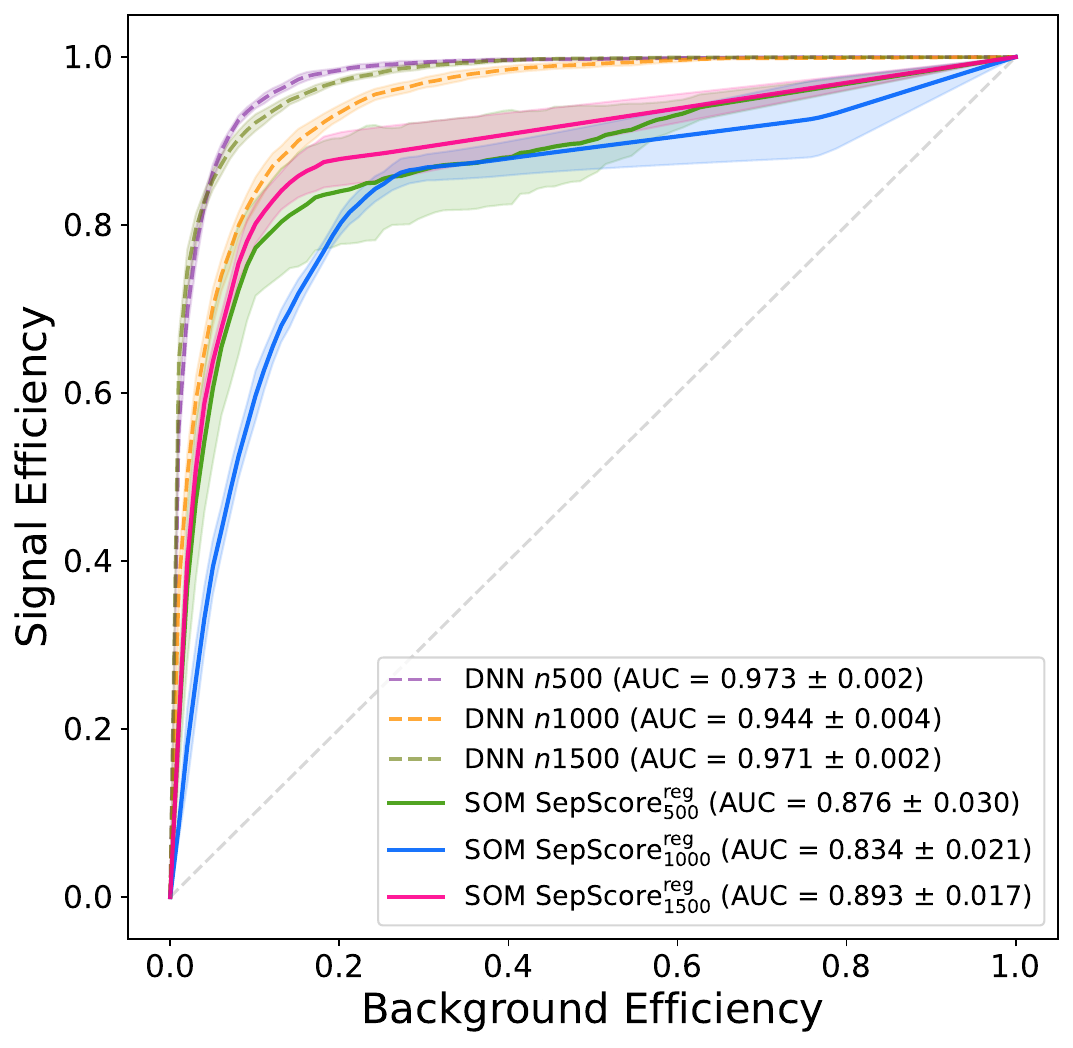}
  \caption[SOM ROC]{The ROC curves calculated using the SOM with $n=100$ and $m=11$ regions for the different VLL mass hypotheses are compared to the ROC curves from the DNN.}
  \label{fig:somroc}
\end{figure}

\begin{table}[!h]
  \centering
  \begin{tabular}{|l|c|c|c|}
    \hline
    Model &  $m_L=500\gev$ & $m_L=1000\gev$ & $m_L=1500\gev$ \\
    \hline
    DNN &        $100$& $100$& $99$\\
    SOM $n$=40, $m$=3 &   $100$& $96$& $98$\\
    SOM $n$=40, $m$=5 &   $100$& $97$& $97$\\
    SOM $n$=100, $m$=5 &  $100$& $91$& $93$\\
    SOM $n$=100, $m$=11 & $99$& $95$& $94$\\
    \hline
  \end{tabular}
  \caption[Accuracy for SOM and DNN]{The accuracy, i.e the number of trials yielding the correct answer from a total of $100$ independent trials done using $10$ events, is shown
  for the various algorithms.}
  \label{tab:accs}
\end{table}

\section{Conclusion and Outlook}
\label{sec:conc}

We addressed the problem of identifying a unique signal hypothesis from a small excess of events by approaching it 
with two methods, one using a multiclassifying DNN, and the other using
SOMs. Here we develop a method to use SOMs for a categorization problem with labelled data, and outline a subsequent procedure to be used to identify
the category from the data.
Both methods perform well, with the multiclassifier having a clear numerical advantage. The AUCs for the different models are summarized
in Table~\ref{tab:aucs}. The various SOM grids have a comparable performance.

\begin{table}[!h]
  \centering
  \begin{tabular}{|l|c|c|c|}
    \hline
    Model & ${\mathrm{Score_{500}}}$ AUC & ${\mathrm{Score_{1000}}}$ AUC & ${\mathrm{Score_{1500}}}$ AUC \\
    \hline
    DNN & $0.973 \pm 0.002$ & $0.944 \pm 0.004$ & $0.971 \pm 0.002$ \\
    SOM $n$=40, $m$=3 & $0.897 \pm 0.019$ & $0.868 \pm 0.008$ & $0.917 \pm 0.011$ \\
    SOM $n$=40, $m$=5 & $0.859 \pm 0.046$ & $0.885 \pm 0.012$ & $0.933 \pm 0.010$  \\
    SOM $n$=100, $m$=5 & $0.853 \pm 0.011$ & $0.776 \pm 0.025$ & $0.835 \pm 0.018$ \\
    SOM $n$=100, $m$=11 & $0.876 \pm 0.030$ & $0.834 \pm 0.021$ & $0.893 \pm 0.017$ \\
    \hline
  \end{tabular}
  \caption[AUCs for SOM and DNN]{The AUCs for the ROCs plotted for SOM are compared to that of the DNN. The $n$ is the size of the SOM model,
    and $m$ is the square region chosen around the BMU of each event. }
  \label{tab:aucs}
\end{table}

Despite the higher AUC score of the DNNs,
the SOMs have additional features that can be used. The clustering of the classes differs with the size of the grid - each cluster will have slightly
different kinematic properties. Thus the same observed events can be probed in different ways. Additionally, varying the size of the neighborhood of each
observed event can also be exploited to confirm or reject hypotheses. Here we probed the ability of SOMs to provide useful output when not trained on
a particular class of events (viz. the SM). This ability may be particularly useful when the dominant background for a search is an instrumental one,
and thus not easily amenable to \emph{a priori} simulation.

For the problem we consider, we find that pinpointing a specific hypothesis using the SOMs will require an iterative approach. Initial iterations
can be used to definitively rule out hypotheses. Both the DNN and SOM could be used to narrow down the observed events into SM vs BSM, following
which the BSM events can be probed further. The clustering nature of a SOM along with our defined separation score also allows for a possible
approach of training for a much larger number of classes.

Typical searches at the LHC also employ DNNs to define signal regions for searches. A useful strategy could be to train a SOM on the events that
pass the requirements on a background-supressing DNN. Given that SOMs work well even with a smaller training dataset, such a strategy would be useful
to probe any excess that may be found, even if the excess is not significant enough to claim a discovery.
Overall we find that SOMs could be a versatile tool in searches for BSM phenomena, and offer niche advantages over other multivariate techniques in terms
of abilities and simplicity.

\section*{Acknowledgements}

We would like to thank Riya~Sharma, and Prashant~Gaikwad for discussions of results
and comments on the manuscript. S.~D. and V.~T. are grateful to Nilanjana~Kumar
for help with model files for VLLs. N.~K. acknowledges the financial support
provided by the Indian Institute of Science Education and Research (IISER), Pune.


\bibliographystyle{unsrt}
\bibliography{VLL_SOM_DNN_refs}

\begin{thebibliography}{10}

\bibitem{ParticleDataGroup:2024cfk}
S.~Navas et~al.
\newblock {Review of Particle Physics}.
\newblock {\em Phys. Rev. D}, 110:030001, 2024.

\bibitem{cmspublic}
{CMS collaboration}.
\newblock Cms collaboration publications and results.
\newblock
  \url{https://cms-results.web.cern.ch/cms-results/public-results/publications/},
  2026.

\bibitem{atlaspublic}
{ATLAS collaboration}.
\newblock Atlas collaboration publications and results.
\newblock \url{https://twiki.cern.ch/twiki/bin/view/AtlasPublic}, 2026.

\bibitem{Shmakov:2023kjj}
Alexander Shmakov, Kevin Greif, Michael Fenton, Aishik Ghosh, Pierre Baldi, and
  Daniel Whiteson.
\newblock End-to-end latent variational diffusion models for inverse problems
  in high energy physics.
\newblock 2023.

\bibitem{Bornhauser:2012iy}
Nicki Bornhauser and Manuel Drees.
\newblock {Mitigation of the LHC Inverse Problem}.
\newblock {\em Phys. Rev. D}, 86:015025, 2012.

\bibitem{CMS:2022nty}
{CMS collaboration: A. Tumasyan} et~al.
\newblock {Inclusive nonresonant multilepton probes of new phenomena at $\sqrt
  s$=13{\,}{\,}TeV}.
\newblock {\em Phys. Rev. D}, 105:112007, 2022.

\bibitem{Kohonen2004SelforganizedFO}
Teuvo Kohonen.
\newblock Self-organized formation of topologically correct feature maps.
\newblock {\em Biol. Cybern.}, 43:59--69, 1982.

\bibitem{ATLAS:2024mrr}
{ATLAS collaboration: G. Aad} et~al.
\newblock {Search for vector-like leptons coupling to first- and
  second-generation Standard Model leptons in pp collisions at $ \sqrt{s} $ =
  13 TeV with the ATLAS detector}.
\newblock {\em JHEP}, 05:075, 2025.

\bibitem{Kumar:2015tna}
Nilanjana Kumar and Stephen~P. Martin.
\newblock {Vectorlike leptons at the Large Hadron Collider}.
\newblock {\em Phys. Rev. D}, 92(11):115018, 2015.

\bibitem{Bhattiprolu:2019vdu}
Prudhvi~N. Bhattiprolu and Stephen~P. Martin.
\newblock {Prospects for vectorlike leptons at future proton-proton colliders}.
\newblock {\em Phys. Rev. D}, 100(1):015033, 2019.

\bibitem{Alwall_2014}
J.~Alwall, R.~Frederix, S.~Frixione, V.~Hirschi, F.~Maltoni, O.~Mattelaer,
  H.-S. Shao, T.~Stelzer, P.~Torrielli, and M.~Zaro.
\newblock The automated computation of tree-level and next-to-leading order
  differential cross sections, and their matching to parton shower simulations.
\newblock {\em JHEP}, 2014:1--157, July 2014.

\bibitem{pythia83}
Christian Bierlich, Smita Chakraborty, Nishita Desai, Leif Gellersen, Ilkka
  Helenius, Philip Ilten, Leif Lönnblad, Stephen Mrenna, Stefan Prestel,
  Christian~T. Preuss, Torbjörn Sjöstrand, Peter Skands, Marius Utheim, and
  Rob Verheyen.
\newblock {A comprehensive guide to the physics and usage of PYTHIA 8.3}.
\newblock {\em SciPost Phys. Codebases}, page~8, 2022.

\bibitem{deFavereau2014}
J.~de~Favereau et~al.
\newblock {DELPHES 3: a modular framework for fast simulation of a generic
  collider experiment}.
\newblock {\em JHEP}, 2014(2):057, 2014.

\bibitem{Brun:1997pa}
Rene Brun and Fons Rademakers.
\newblock {ROOT {\textemdash} An object oriented data analysis framework}.
\newblock {\em Nucl. Instrum. Meth. A}, 389(1-2):81--86, 1997.

\bibitem{Cacciari:2008gp}
Matteo Cacciari, Gavin~P. Salam, and Gregory Soyez.
\newblock {The anti-$k_t$ jet clustering algorithm}.
\newblock {\em JHEP}, 04:063, 2008.

\bibitem{Cacciari:2011ma}
Matteo Cacciari, Gavin~P. Salam, and Gregory Soyez.
\newblock {FastJet user Manual}.
\newblock {\em Eur. Phys. J. C}, 72:1896, 2012.

\bibitem{tensorflow2015-whitepaper}
Mart\'{i}n Abadi et~al.
\newblock {TensorFlow}: Large-scale machine learning on heterogeneous systems.
\newblock \url{https://www.tensorflow.org/}, 2015.

\bibitem{keras_chollet2015}
Fran\c{c}ois Chollet et~al.
\newblock Keras.
\newblock \url{https://keras.io}, 2015.

\bibitem{adam_optimizer}
Diederik~P. Kingma and Jimmy Ba.
\newblock Adam: A method for stochastic optimization.
\newblock 2017.

\bibitem{Chowdhury:2025mul}
Shreecheta Chowdhury, Amit Chakraborty, and Saunak Dutta.
\newblock {Probes of anomalous events at LHC with self-organizing maps}.
\newblock {\em Eur. Phys. J. C}, 85(9):964, 2025.

\bibitem{vettigliminisom}
Giuseppe Vettigli.
\newblock Minisom: minimalistic and numpy-based implementation of the self
  organizing map.
\newblock \url{https://github.com/JustGlowing/minisom/}, 2018.

\end{thebibliography}

\end{document}